\documentclass[traditabstract]{aa}
\input psfig.tex
\usepackage{graphicx}
\usepackage{natbib}
\usepackage{array}
\usepackage{graphics}
\usepackage{latexsym}
\usepackage{amssymb}
\usepackage{amsmath}
\usepackage{fancyhdr}
\usepackage{morefloats}
\bibpunct{(}{)}{;}{a}{}{,}
\include{hyphe}

\begin{document}
\title{Formation and Evolution of Early-Type Galaxies:
       Spectro-Photometry from Cosmo-Chemo-Dynamical Simulations}

\author{Rosaria Tantalo$^{1}$, Simonetta Chinellato$^{2}$, Emiliano Merlin$^{1}$,
Lorenzo Piovan$^{1}$, Cesare Chiosi$^{1}$}

\institute{$^1$ Department of Astronomy, Padova University, Vicolo dell'Osservatorio 3, I-35122, Padova, Italy\\
$^2$ Padova Astronomical Observatory, Vicolo dell'Osservatorio 5, I-35122, Padova, Italy\\
\email{{rosaria.tantalo\char64unipd.it, simonetta.chinellato\char64oapd.inaf.it,
        emiliano.merlin\char64unipd.it, lorenzo.piovan\char64gmail.com, cesare.chiosi\char64unipd.it}}}
\date{Received: June 2009; Revised: November 2009; February 2010; Accepted: *** ***}

\titlerunning{Spectro-photometric models of early type galaxies}
\authorrunning{R. Tantalo et al.}

\abstract{
\textsc{Context}. One of the major challenges in modern
astrophysics is to understand the origin and the evolution of
galaxies, the bright, massive early type galaxies (ETGs) in
particular. There is strong observational evidence that massive ETGs
are already in place at redshift $z \sim 2-3$ and that they formed most
of their stars well before $z=1$. Therefore, these galaxies are
likely to be good probes of galaxy evolution, star formation and,
metal enrichment in the early Universe.

\noindent
\textsc{Aims}. In this context it is very
important to set up a diagnostic tool able to combine results from
chemo-dynamical N-Body-TSPH (NB-TSPH) simulations of ETGs with those
of spectro-photometric population synthesis and evolution so that
all key properties of galaxies can be investigated. These go from
the integrated spectrum and magnitudes in any photometry, both in
the rest-frame and as a function of the redshift, to present-day
structural properties. The main goal of this paper is to provide
a preliminary validation of the software package before applying
it to the analysis of observational data.

\noindent
\textsc{Methods}. The galaxy models in use where calculated by
the Padova group in two different cosmological scenarios: the standard
cold dark matter cosmology ($SCDM$), and the so-called Concordance
cosmology ($\Lambda CDM$, with $\Omega_\Lambda=0.762$). For these
template galaxies, we recover their spectro-photometric evolution
through the entire history of the Universe. This is done in particular
for two important photometric systems, the Bessell-Brett and the Sloan
Digital Sky Survey (SDSS) passbands.

\noindent
\textsc{Results}. We computed magnitudes and colors and
their evolution with the redshift along with the evolutionary and
cosmological corrections for the model galaxies at our disposal, and
compared them with data for ETGs taken from the COSMOS and the GOODS
databases. Finally, starting from the dynamical simulations and photometric
models at our disposal, we created synthetic images in a given photometric system,
from which we derived the structural and morphological parameters.
In addition to this, we address the question of the scaling
relations, and in particular we examine the one by Kormendy.
The theoretical results are compared with observational data
of ETGs selected form the SDSS database.

\noindent
\textsc{Conclusions}. The simulated colors for the
different cosmological scenarios follow the general trend shown by
galaxies of the COSMOS and GOODS surveys at lower redshifts and are in
good agreement with the data up to $z \sim 1$, where the number of
early-type galaxies observed falls abruptly. In conclusion, within the
redshift range considered, all the simulated colors reproduce the
observational data quite well. Looking at the structural parameters
derived from the surface imaging, the luminosities and effective radii
(Kormendy relation) measured for our model galaxies are consistent
with the archival data from the SDSS.}

\keywords{early-type galaxies - elliptical galaxies - galaxy formation
and evolution -integrated spectro-photometry}
\maketitle

\section{Introduction}\label{Introduction}

The origin and evolution of early-type galaxies (ETGs), the bright
massive ETGs in particular, are two of the major challenges in modern
astrophysics, and it is still a very controversial subject
\citep{Chiosi00}. Spheroidal systems are of interest in their own
right because they contain more than half of the total stellar mass in the
local Universe \citep{Fukugita98}. Giant ETGs appear to define a
homogeneous class of objects that predominantly consists of uniformly
old and red populations, which implies that they must have formed at
high redshift and that they have negligible amounts of gas and very little star
formation \citep{Bressan94}.

There is strong observational evidence that old, massive, red, and
metal-rich proto-ETGs are already in place at $z \sim$2-3 and that
the present-day early-type galaxies formed most of their stars well
before redshift $z=1$
\citep{Searle73,Brinchmann00,Treu05,vanderWel05}. Moreover, the
current rates of star formation in these systems are quite low,
whereas the rates increase sharply into the past
\citep{Butcher78,Dressler80}. Therefore, these ETGs are likely good
probes of galaxy assembly, star formation, and metal enrichment in
the early Universe.

\noindent
\textsf{The cosmological background.}
In a Universe dominated by cold dark matter (CDM), some kind of dark
energy in form of the cosmological constant $\Lambda$, and containing
a suitable mix of baryons and photons, the cosmic structures are
formed by the gravitational collapse of dark matter and are organized
in a hierarchy of halos inside which baryons dissipate their energy
and collapse to form luminous systems. The formation of ETGs can be
reduced to the following schemes \citep{Peebles02,Schade99}:

\noindent
i) {\it Early, monolithic-like aggregation}. This scenario of galaxy
formation predicts that all ETGs form at high redshift ($z \gg 1$) as
a result of rapid and dissipation-less collapse of a large mass of gas
soon transformed into stars. In the model, first proposed by
\citet{Eggen62} and then refined and improved by
\citet{Larson75},\citet{Arimoto87},\citet{Bressan94},and \citet{Chiocar02}, ETGs undergo a
single and short, but intense, burst of star formation, followed ever since
by the passive evolution of their stellar populations to the
present day. This simple model naturally accounts for the old ages
($\sim 12$\,Gyr) of spheroidal galaxies, their high densities, and
the weak temporal evolution of their stellar content.

In favor of this scheme are the observational data that convincingly
hint at old and homogeneous stellar populations
\citep[see][ for a review of the subject]{Chiosi00}. It is worth
mentioning, however, that \citet{Kauffmann93} and \citet{Barger99}
argue for some recent evolution in the stellar
populations of elliptical galaxies. This scenario reproduces the
optical properties of ETGs remarkably well, and successfully
explains the tightness of the fundamental scaling relations that
ETGs obey, like the color-magnitude relation and the fundamental
plane, as well as the evolution of these relations as a function of
redshift.

The monolithic formation mechanism fails to explain some recent
observational evidence that has become available with the advance of
more detailed data from present-day surveys. These indicate that the
star formation histories of at least some ETGs, and perhaps the
early-type population as a whole, deviate strongly from the
expectations of the monolithic collapse paradigm, both in terms of
their structural evolution and star formation experienced by them
over the whole Hubble time. It is less successful at explaining the
detailed luminosity dependence of their dynamical properties, the
apparent scarcity of very large star-bursts in the high-redshift
universe, and the origin of dynamical peculiarities indicating some
recent accretion events. This scenario does not fit the currently
accepted $\Lambda CDM$ picture of galaxy formation whose bottom line
is that massive dark matter halos are assembled by mergers of low-mass
halos and therefore the mass of a galaxy is thought to accumulate over
the lifetime of the Universe.

To cope with some of the above difficulties, a hybrid scenario named {\it
revised monolithic} has been proposed by \citet{Schade99} and
confirmed by NB-TSPH simulations by \citet{Merlin06,Merlin07}, who
suggest that a large number of the stars in massive galaxies are formed
very early-on at high redshift ($z \sim$1-2) and the remaining few
at lower $z$. The revised monolithic ought to be preferred to
the classical monolithic, as some evidence of star formation at
0.2$\leq z\leq$2 can be inferred from the emission
line of [OII], and also the number frequency of ETGs up to
$z\simeq$1 seems to be nearly constant. Recently,
\citet{PerezGonzalez08}, whom analyzing a huge sample of galaxies,
confirmed a scenario where most massive objects assemble their mass
very early, whereas the smallest galaxies evolve more slowly
building up their mass at lower redshift.

\noindent
ii) {\it The hierarchical aggregation.} This scenario instead suggests
that massive ETGs are the end product of subsequent violent mergers of
preexisting smaller subunits, on time scales almost equal to the
Hubble.  In this scenario, the epoch of assembly of ETGs differs
markedly from the epoch of formation of their constituent stars, and
the high density of elliptical galaxies is ascribed to the effects of
dissipation during the formation of the progenitor disks. As the
look-back time increases, the density in comoving space of bright
(massive) ETGs should decrease by a factor 2 to 3 \citep[see
e.g.][]{White78,Kauffmann93}.

This model accounts naturally for the scarcity of very bright
elliptical progenitors at high redshift, for the rapid evolution of
the galaxy population with look-back time, and for
dynamical peculiarities. In favor of this view is some
observational evidence that the merger rate likely increases with
$\sim(1+z)^{3}$ \citep{Patton97}, together with some hint for a
color-structure relationship for E \& S0 galaxies: the color
becomes bluer at increasing complexity of a galaxy structure. This
could indicate some star formation associated to the merger event.
Finally, there are the many successful numerical simulations of
galaxy encounters, mergers, and interactions
\citep[e.g.][]{Barneshern96}.

It is, on the other hand, less successful in explaining the apparent
old ages of stars in elliptical galaxies and their uniformity in
dynamical properties. Nevertheless, contrary to the expectation from
this model, the number density of ellipticals do not seem to
decrease with the redshift, at least up to $z\simeq 1$ \citep{Im96}.
A significant population of massive and passive ETGs up to $z
\approx 2.5$ and some hints about massive ETGs at redshift $z>3$
\citep{Cimatti09}, clearly do not agree with the classical
hierarchical scenario, because we need to have big objects already
in place at higher and higher $z$.

There is a companion scheme named \textit{dry merger}, in which bright
ETGs form by encounters of quiescent, no star-forming galaxies. This
view is advocated by \citet{Bell04}, who find that the B-band
luminosity density of the red peak in the color distribution of
galaxies shows mild evolution starting from $z\simeq$1. As old stellar
populations would fade by a factor 2 or 3 in this time interval, and
the red color of the peak tells us that new stars are not being
formed in old galaxies, this mild evolution hints for a growth in the
stellar mass of the red sequence, either coming from the blue-peak
galaxies in which star formation is truncated by some physical
process, or by "dry mergers" of smaller red, gas-poor
galaxies. However, according to \citet{Bundy05,Bundy06}, dry mergers
cannot be the leading mechanism in the history of galaxy assembly
because of the weak dependence on the environment, in contrast to what
expected. Indeed the majority of quiescent galaxies seem to be
assembled by a mechanisms that depends on their mass rather than the
environment, as the merger rate does not seem to increase with
environment density \citep{Bundy06}.

\textsf{Putting Dynamics and Photometry together.} How can we
disentangle the above scenarios? Comparing theoretical predictions
to observational data concerning the light and hence mass profiles,
velocity, SEDs, magnitudes, colors, line strength indices, and
associated gradients. In this context, spectro-photometric models of
galaxies have long been the key tool to investigate how galaxies
formed and evolved with time.  Consequently, an impressive number of
chemo-spectro-photometric models for ETGs have been proposed. To
mention a few among the recent ones, we recall
\citet{Bressan94},\citet{Vazdekis97},\citet{Tantalo96a},\citet{Gibson97},
\citet{Kodama97},\citet{Fioc97b},\citet{Tantalo98b},\citet{Pipino04}, and \citet{Piovan06,Piovan06b}.
Nearly all these models simulate a galaxy and its evolution adopting
the point source approximation, in which no morphological structure
and no dynamics are considered.

In parallel to this line of work, many fluid-dynamical models in
n-dimensions (from 1 to 3) and with multi-phase descriptions of the
gaseous component were developed for galaxies of different
morphological type. To mention a few, we recall \citet{Theis92},\citet{Ferrini93},\citet{Samland97},\citet{Boissier99a,Boissier99b},
\citet{Samland00},\citet{Samland01},\citet{Berczik03}, and \citet{Immeli04}. For some of
them the spectro-photometric aspect of the models was also
investigated with successful results.

Finally, there are the  NB-TSPH simulations with dark and baryonic
matter, the hydrodynamic treatment of the baryonic component, and
even multi-phase descriptions of the gas. The NB-TSPH simulations
are one of the best tools to infer the 3-D structure of ETGs, to
follow the temporal evolution of the dynamical structure, the
stellar content, and the chemical elements. Recent models of this
type at different level of complexity are by
\citet{Chiocar02},\citet{Kobayashi04a,Kobayashi04b,Kobayashi05},
\citet{Merlin06,Merlin07}, and \citet{Scannapieco06a,Scannapieco06b}. So
far the corresponding photometric properties of the models are
either left aside or treated in a very rudimentary way.

It follows from these considerations that the ideal tool to develop
would be the one folding together NB-TSPH simulations and
chemo-spectro-photometry to generate 3-D chemo-dynamical,
spectro-photometric models of galaxies. This would allow us to
simultaneously predict and discuss both the structural properties
related to dynamical formation process and the spectro-photometric
ones related to the stellar content in a self consistent fashion, and
hopefully to cast light on the above issues.
Therefore, we have taken the cosmo-chemo-dynamical models of galaxies
calculated with the Padova Code [G\textsc{al}D\textsc{yn}] (see details below)
from which we get the star formation (SFH), the chemical enrichment
(Z(t)) histories, and the structure of the simulated galaxy. The output
of these models is fed into the Padova photometric code [S\textsc{pe}C\textsc{ody}],
which generates the spectral energy distribution (SED) of the whole
galaxy. From this SED we derive the absolute magnitudes, colors,
indices, etc., in a chosen photometric system. This allows us to
determine the rest-frame and cosmological evolution of magnitudes and
colors for the set of models at our disposal.

\textsf{Aims and plan of the paper.} The purpose of this study is to
validate the whole procedure before applying it to a set of
simulations under preparation and/or to extensive study of
observational data. The outline of the paper is as follows.
Section~\ref{dynsim} describes the dynamical NB-TSPH simulations of
ETGs. Section~\ref{specody} describe the photometric package used to
get the rest-frame magnitudes and colors of the galaxy models in a
photometric system (some details are given for two of those, namely
the Bessell-Brett and the Sloan Digital Sky Survey (SDSS). We also
we present a study of the color-magnitude diagram (CMD) of the
stellar populations of the model galaxies. Section~\ref{cosmev}
presents a multi-wavelength study of the optical and near-IR
high-$z$ photometric properties of the ETGs and compares the results
with a sample of galaxies selected from COSMOS and GOODS surveys. In
Section~\ref{chapim} we describe the  method followed to derive 2-D
artificial images, starting from the 3-D model galaxies. These
images resemble observational data and can be analyzed in a similar
manner. Isophotal analysis with aid of the Fourier and S\'ersic
technique is applied to derive some structural properties of the
model galaxies. We obtain the morphological and structural
parameters and compare them with the data for a sample of
elliptical galaxies selected from  SDSS. This allows us to
establish the consistency of the models with photometric data. In
Section~\ref{scalrel} we use the parameters derived from the surface
photometry to investigate  the Kormendy scaling relation. Finally,
in Section~\ref{concl}, we discuss some unsettled issues that
require future work and present some general, conclusive remarks.

\section{Dynamical Models of ETGs}\label{dynsim}

For our analysis we have considered three numerical simulations,
calculated by \citet{Merlin06,Merlin07} using
[G\textsc{al}D\textsc{yn}] the cosmo-chemo-dynamical evolutionary
code developed by \citet{Merlin06,Merlin07}. The code stems from the
original NB-TSPH code developed in Padova by \citet{Carraro98}. It
combines the Oct-Tree algorithm \citep{Barneshut86} for the
computation of the gravitational forces with the SPH
\citep{Lucy77,Benz90} approach to numerical hydrodynamics of the gas
component. It is fully-Lagrangian, three-dimensional, and highly
adaptive in space and time owing to individual smoothing lengths and
individual time-steps. It includes self-consistently a number of
non-standard physical processes: radiative and inverse Compton
cooling, star formation, energy feedback, and metal enrichment by type
Ia and II SN{\ae} \citep{Lia02}. The numerical recipe for star
formation, feedback, and chemical enrichment along with all the other
physical processes considered, the improvements to the initial
conditions and the multi-phase description of the interstellar medium
are described in \citet{Merlin06,Merlin07}. No details are given here
but for a few key points.

Particles, representing dark matter and baryons both in form of gas
and stars, evolve in the dynamical phase space under the action of
cosmological expansion, self-gravity, and (in the case of gas)
hydrodynamical forces. In the single-phase description (only one type
of gas), the gas-particles are turned into star-particles as soon as
they satisfy three physical requirements: (i) to be denser than a
threshold value; (ii) to belong to a convergent flow; and (iii) to
cool efficiently. There is also an additional statistical criterium
\citep[as described in details in ][]{Lia02} to be fulfilled. In the
multi-phase description (hot-rarefied and cool-dense gas),
gas-particles that become colder and denser than suitable thresholds
are subtracted from the SPH scheme and turned into sticky particles;
in this case, star formation can take place only within this
cold and dense phase. Star-particles then refuel the interstellar
medium with energetic and chemical feedbacks, ultimately quenching
star formation when the gas heated by SN explosions is hot enough to
leave the galaxy potential well (galactic winds).

Two cosmological scenarios are adopted to calculate the galaxy models:
the so-called standard-$CDM$ ($SCDM$) and the {\it concordance}
$\Lambda CDM$ as inferred by WMAP3 data \citep{Spergel03}. One galaxy
model is calculated with $SCDM$ and the one-phase description, and two
with $\Lambda CDM$. These latter in turn differ for the treatment of
the interstellar medium: (i) one-phase medium; the model is shortly
indicated as $\Lambda CDM$ and (ii) multi-phase medium; the models is
named $\Lambda CDM_{mp}$ \citep[see ][ and below for more
details]{Merlin06,Merlin07}.

All the models are constructed as follows: we start from a realistic
simulation of a large region of the Primordial Universe carried out
with a given cosmological scenario. At certain value of the redshift,
typically $z_{ini} \simeq 50-60$ (the precise value changes from model
to model), a spherical, over-dense, galaxy-sized proto-halo is
selected, detached from its surroundings, and let evolve with void
boundary conditions, after that an outwards radial initial velocity
has been added to simulate the Hubble flow. Initially, the proto-halo
continues to expand but, reached a maximum extension, it turns around
and collapses toward higher and higher densities. In the meantime,
baryons (gas) collapse too and start forming stars, at the beginning
very slowly and then at increasing rate. The redshift at which
significant star formation begins is in between 50-60 and 5, but close
to about 5. In all models at redshift about 2 the conversion of gas
into stars is nearly complete \citet[see ][ for all
details]{Merlin06,Merlin07}.  In this picture, there is no sharp value
of the redshift at which star formation is supposed to start,
$z_{ini}$ is simply the redshift at which the proto-halo, inside which
a galaxy will later be formed, is singled out from the cosmological
tissue. For a similar choice of $z_{ini}$ see also
\citet{Li06a,Li06b,Li06c,Li07}. Table~\ref{cosmo} provides a summary
of the relevant cosmological parameters.

\begin{table}
\begin{center}
\caption[Cosmological parameters]{Cosmological parameters adopted in
our simulations} \label{cosmo}
\begin{tabular}{|c | c | c| c|}
\hline \multicolumn{1}{|c|}{Model} & \multicolumn{1}{c|}{$SCDM$} &
\multicolumn{1}{c|}{$\Lambda CDM$} &
\multicolumn{1}{c|}{$\Lambda CDM_{mp}$} \\
\hline
$h_0$              & 0.5  & 0.730 & 0.730 \\
$\Omega_M$         & 1    & 0.238 & 0.238 \\
$\Omega_{\Lambda}$ & 0    & 0.762 & 0.762 \\
$\sigma_8$         & 0.5  & 0.740 & 0.740 \\
\hline
\end{tabular}
\end{center}
\end{table}

Inside each proto-halo (proto-galaxy), the baryonic component is
initially in the gaseous phase and follows the dark matter
perturbations until it is heated up by shocks and mechanical
friction. When radiative cooling becomes efficient, the first cold
clumps begin to form, and the gas is finally turned into
star-particles. Because of the mass resolution of the models, a
star-particle is so massive that it can be thought of to correspond to
an assembly of real stars, which in turn distribute in mass according
to some initial mass function over the mass interval $m_{l}$ to
$m_{u}$, i.e. 0.1 to 100 $M_{\odot}$.  At the present time, each
star-particle contains living stars, from $m_{l}$ to a maximum mass
$m_{max}(t)$ that depends on the age, and remnants (black holes,
neutron stars, and white dwarfs) generated by all stars in the mass
interval $m_{max}(t) < m < m_{u}$. Therefore, in a star-particle,
SN-explosions may occur (their rate can be easily calculated), thus
releasing energy that cause evaporation of the nearby clouds which
quenches the star formation.

\begin{table*}
\begin{center}
\caption[Initial dynamical parameters]
        {Initial dynamical and computational parameters for the three
        model galaxies}
\label{inipar}
\footnotesize{
\begin{tabular*}{146.1mm}{|l r| c| c| c|}
\hline
\multicolumn{2}{|c|}{\rule{0pt}{10pt} Model} &
\multicolumn{1}{c|}{$SCDM$} &
\multicolumn{1}{c|}{$\Lambda CDM$} &
\multicolumn{1}{c|}{$\Lambda CDM_{mp}$}\\[1mm]
\hline\rule{0pt}{10pt}
Initial Total Mass  [$M_{\odot}$]             & $M_{tot}$        & $1.625 \times 10^{12}$ &  $8.830 \times 10^{11}$ & $2.449 \times 10^{11}$ \\[1mm]
\hline\rule{0pt}{10pt}
Initial Baryonic Mass Fraction                & $f_B$            & $0.1$                  &  $0.176$                & $0.176$              \\[1mm]
\hline\rule{0pt}{10pt}
Initial Baryonic Mass [$M_{\odot}$]           & $M_B$            & $1.624 \times 10^{11}$ & $1.413 \times 10^{11}$  & $4.164 \times 10^{10}$ \\[1mm]
\hline\rule{0pt}{10pt}
Initial Sphere Radius [kpc]                   & $R$              & $33.2$                 & $28$                    & $19$                 \\[1mm]
\hline\rule{0pt}{10pt}
Initial Number of Dark Matter Particles       &                  & $13685$                & $13657$                 & $6966$               \\[1mm]
\hline\rule{0pt}{10pt}
Initial Number of Gas Particles               &                  & $13719$                & $13707$                 & $6995$               \\[1mm]
\hline\rule{0pt}{10pt}
Dark Mass per particle [$M_{\odot}$]          & $M_{DM,p}$       & $1.069 \times 10^8$    & $5.435 \times 10^7$     & $2.919 \times 10^7$      \\[1mm]
\hline\rule{0pt}{10pt}
Baryonic Mass per particle [$M_{\odot}$]      & $M_{B,p}$        & $1.184 \times 10^7$    & $1.031 \times 10^7$     & $5.954 \times 10^6$      \\[1mm]
\hline\rule{0pt}{10pt}
Initial Redshift of the Proto-Halo            & $z_{ini}$        & $50$                   & $60$                    & $58$                 \\[1mm]
\hline\rule{0pt}{10pt}
IMF                                           &                  & $Kroupa$               & $Kroupa$                & $Salpeter$           \\[1mm]
\hline\rule{0pt}{10pt}
Softening length for Dark Matter [kpc]        & $\epsilon_{DM}$  & $2$                    & $2$                     & $2$                  \\[1mm]
\hline\rule{0pt}{10pt}
Softening length for Baryonic Matter [kpc]    & $\epsilon_{b}$   & $1$                    & $1$                     & $1$                  \\[1mm]
\hline
\end{tabular*}}
\begin{minipage}{0.96\textwidth}
\end{minipage}
\end{center}
\end{table*}

The three galaxy models we have considered differ in important aspects
that deserve some comments:

i) For the $SCDM$ model (standard cold dark matter cosmology)
calculated with the one-phase description of the interstellar medium,
the cosmological parameters are chosen in accordance with the kind of
model that cosmologists classify as the reference case. This explains
why we have chosen the normalized Hubble constant $h_0=0.5$ although
nowadays $h_0=0.7$ ought to be preferred.  Nevertheless, since testing
cosmology is beyond the scope of this study which simply aims to test
the ability of our code in predicting the photometric properties of
the model galaxy, the exact choice of $h_0$ is not particularly
relevant here.

ii) The models $\Lambda CDM$ and $\Lambda CDM_{mp}$ refer to the
standard {\it concordance} cosmology in presence of dark energy.
However, they have different assumptions concerning the treatment of
the interstellar medium: only one phase for the first and two phases
for the second.

iii) As a consequence of the different cosmological backgrounds, the
three models do not have the same initial total mass nor the same
ratio $f_B$ of baryonic to total mass, nor the initial redshift at
with the perturbations are singled out from the cosmological tissue.

iv) The evolution of the model galaxies is followed up to the present,
except for the $\Lambda CDM$ simulation that stops at $z
\sim 1$ and age of about 7\,Gyr in the adopted cosmology, due to
computational difficulties. However, for the purposes of this paper,
to consider also this truncated model is safe. Indeed, the model has
already relaxed to dynamical equilibrium so that its shape will not
change significantly during the remaining 5\,Gyr. The rate of star
formation has already decreased to very low levels like in $SCDM$
and $\Lambda CDM_{mp}$ models, and there is no reason to imagine
that it would strongly increase during the age interval from 7 to
13\,Gyr. From a photometric point of view, it is a galaxy in passive
evolution.

v) The numerical simulations track the metal content of each gas- and
star-particle. In brief, using the prescription for chemical evolution
by \citet{Lia02}, in each gas- and star-particle the evolution of the
mass abundance of 10 elements (He, C, O, N, Mg, Si, S, Ca, and Fe) and
the total metallicity $Z$ is followed in detail.  Each star-particle
carries its own age and chemical composition "tag". It is worth noting
that the pattern of abundances of the galaxy models is fully
consistent with the ones adopted to calculate the evolutionary tracks
and isochrones at the base of our SSPs.

vi) To conclude, all the three models can be used for the aims of this
study, i.e. to set up the photometric package suited to NB-TSPH
simulations and to validate it.

In Table~\ref{inipar} we summarize the initial dynamical and
computational parameters for the three models. The total mass
$M_{tot}$ of the model galaxies within the initial sphere radius $R$
comprises both the dark $M_{DM}$ and baryonic matter $M_{B}$, in the
cosmological ratio $f_{B}$. We also list the initial total number of
DM- and B-particles, these latter in form of gas, together with the
mass value per particle shortly indicated as $M_{DM,p}$ and
$M_{B,p}$. The mass per particle depends on the number of particles
adopted for each species. In the course of evolution, the gas-
particles are turned into star-particles of the same mass, which in
the text is indicated as $m_{sp}$. We give the {\it softening length}
for DM- and B-particles that is used to calculate the gravitational
interaction. In brief, because of the time and mass resolution limits,
close encounters between particles may cause numerical errors. To cope
with this, it is common practice to {\it soften} the gravitational
force between close pairs of bodies: if the distance between them
becomes smaller than a suitable {\it softening length} $\epsilon$, the
force exerted on each body is corrected and progressively reduced to
zero with decreasing distance. Finally, we give the IMF that is used
to calculate chemical enrichment and energy feedback by the real stars
composing each star-particle (see below for more details).

\begin{figure*}
\begin{center}
\includegraphics[width=16truecm, height=12truecm]{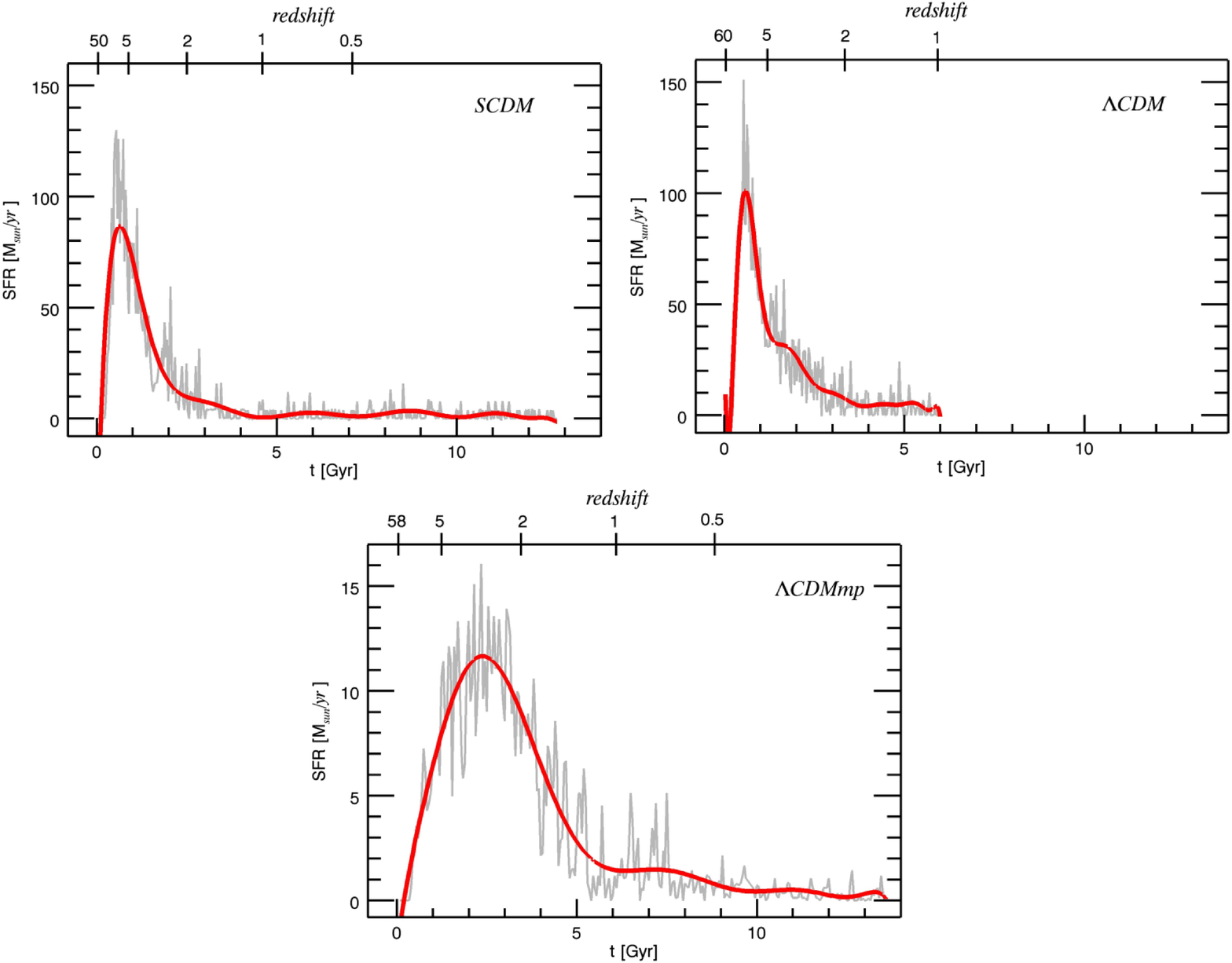}
\caption{Star formation rate, in $M_{\odot}/yr$, versus
      age, in Gyr, and/or redshift for the different model galaxies
      considered. Star formation is let start when the proto-galaxy
      gets gets into existence with respect to the surrounding medium,
      i.e.  $z_{ini} \sim 50$ for the $SCDM$ model (upper left),
      $z_{ini} \sim 60$ for the $\Lambda CDM$ model (upper right),
      $z_{ini} \sim 58$ for the $\Lambda CDM_{mp}$ model (lower).}
\label{sfr}
\end{center}
\end{figure*}

\subsection{Results for ETG models}
\label{dynres}

The final properties of the three galaxies considered are summarized
in Table~\ref{endpar} \citep[for other parameters regarding the
simulations see ][]{Merlin06,Merlin07}.

\begin{table*}
\begin{center}
\caption{End-product for the three model galaxies} \label{endpar}
\footnotesize{
\begin{tabular*}{129mm}{| l r| c| c| c|}
\hline
\multicolumn{2}{|c|}{\rule{0pt}{10pt} Model} &
\multicolumn{1}{c|}{$SCDM$} &
\multicolumn{1}{c|}{$\Lambda CDM$} &
\multicolumn{1}{c|}{$\Lambda CDM_{mp}$} \\[1mm]
\hline\rule{0pt}{10pt}
Final Redshift                 &                    & 0                    & 0.95                 & 0                  \\[1mm]
\hline\rule{0pt}{10pt}
Age of the last model [$Gyr$]  & $T_{gal}$          & 13                   & 7                    & 13                 \\[1mm]
\hline\rule{0pt}{10pt}
Final Number of Star Particles &                    & 7674                 & 8700                 & 5197               \\[1mm]
\hline\rule{0pt}{10pt}
Final Star Mass [$M_{\odot}$]  & $M_{star,tot}^{f}$ & 9.1 $\times 10^{10}$   & 8.9 $\times 10^{10}$   & 3.09 $\times 10^{10}$ \\[1mm]
\hline\rule{0pt}{10pt}
Final Gas Mass [$M_{\odot}$]   & $M_{gas,tot}^{f}$  & 7.2 $\times 10^{10}$ & 5.1 $\times 10^{10}$ & 1.05 $\times 10^{10}$ \\[1mm]
\hline\rule{0pt}{8pt}
Final Effective Radius of Stars [$kpc$] & $R_e$     &      7               &       3              &    4.6             \\[1mm]
\hline
\end{tabular*}}
\end{center}
\end{table*}

Fig.~\ref{sfr} shows the star formation history (SFH) versus time, in
Gyr, and/or redshift for the three models. Stars form in clumps of
cold gas that have collapsed on small scales so that the entire
process can be described as triggered by a number of early dissipative
gravitational collapses, followed by very early merging of stellar
substructures. As it is clearly shown, the galaxies form from an
initial star-burst comprised in the time interval 1 to 3\,Gyr in all
cases. As the cold gas is depleted, the SFR declines rapidly. Anyway,
at lower redshifts small amounts of the previously heated gas have
cooled down again at the center of the galaxy, so that SFR may
continue till the present epoch ($z = 0$), even if at much lower
rates. Although we suspect that to a great extent, this feature of the
models might be of numerical nature, there are no strong compelling
physical reasons to rule it out. Some residual star formation could
occur even at the present time in the very central regions of ETGs. If
so, some effects on the central colors are easy to foresee (see
below).

Fig.~\ref{massass} displays the stellar mass assembly process, i.e.
the growth with time of the total mass in form of stars, $M_{star}$.
The evolution turns the initial irregular proto-galaxy into a well
shaped spheroid, that quickly relaxes into the final configuration,
closely resembling a real ETG. The final stellar mass is essentially
fixed at $z \sim 2$, since little gaseous material is added to the
galaxy afterwards and stars age passively for the remaining time of
the evolution. Due to the early star-burst, the galaxy is already
old and very massive at a redshift of $z \sim$1-2 for the different
models.

\begin{figure*}
\begin{center}
\includegraphics[width=0.9\textwidth]{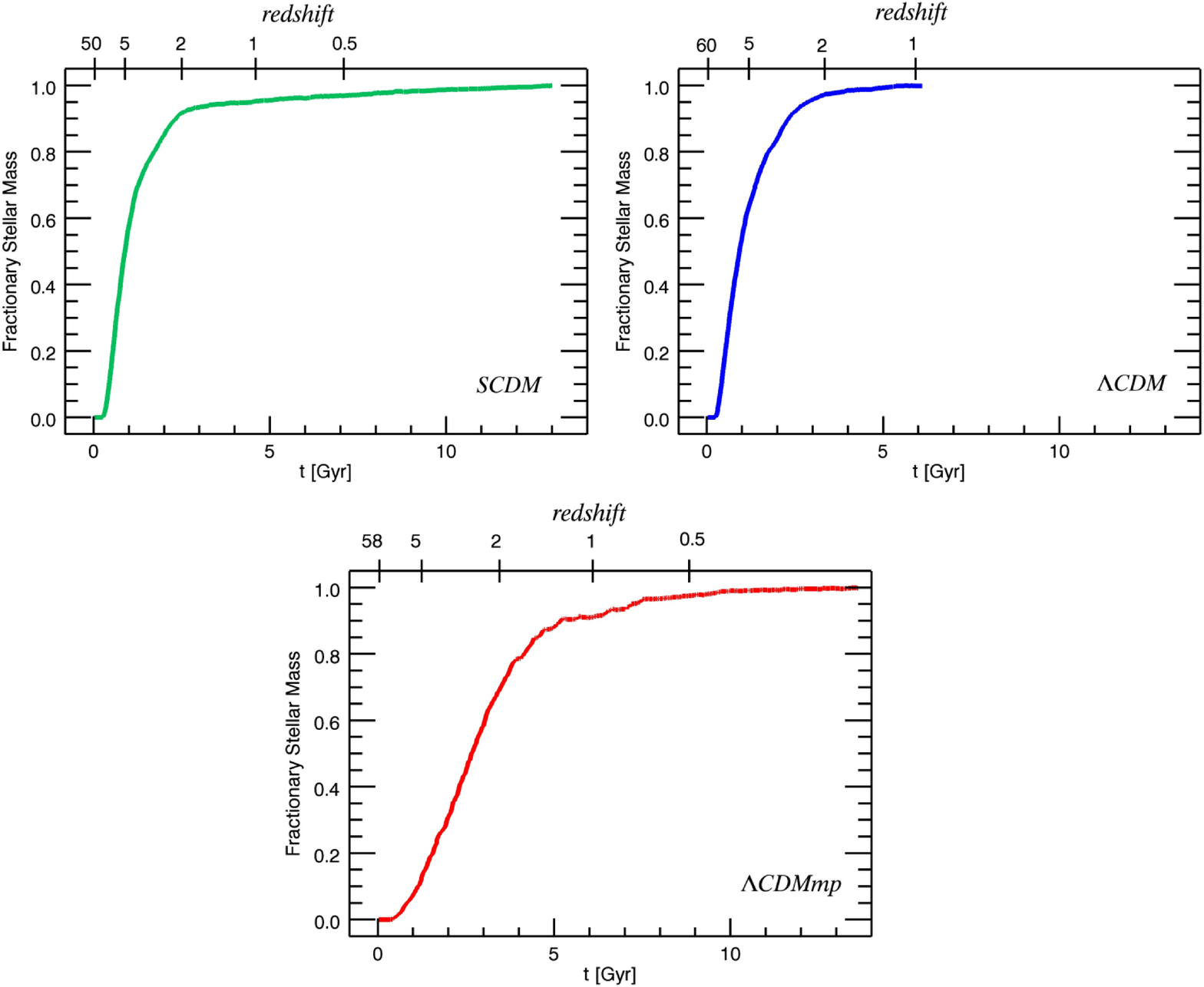}
\caption{Growth of the fractionary total star mass $M_{star}(t)
         /M_{star}^{f}$ as a function of the age, in Gyr, and/or
         redshift for the different models. The final value of
         $M_{star}^{f}$ is the quantity given in Table~\ref{endpar}.}
\label{massass}
\end{center}
\end{figure*}

\begin{figure*}
\begin{center}
{\includegraphics[width=8.0cm,height=8.0cm]{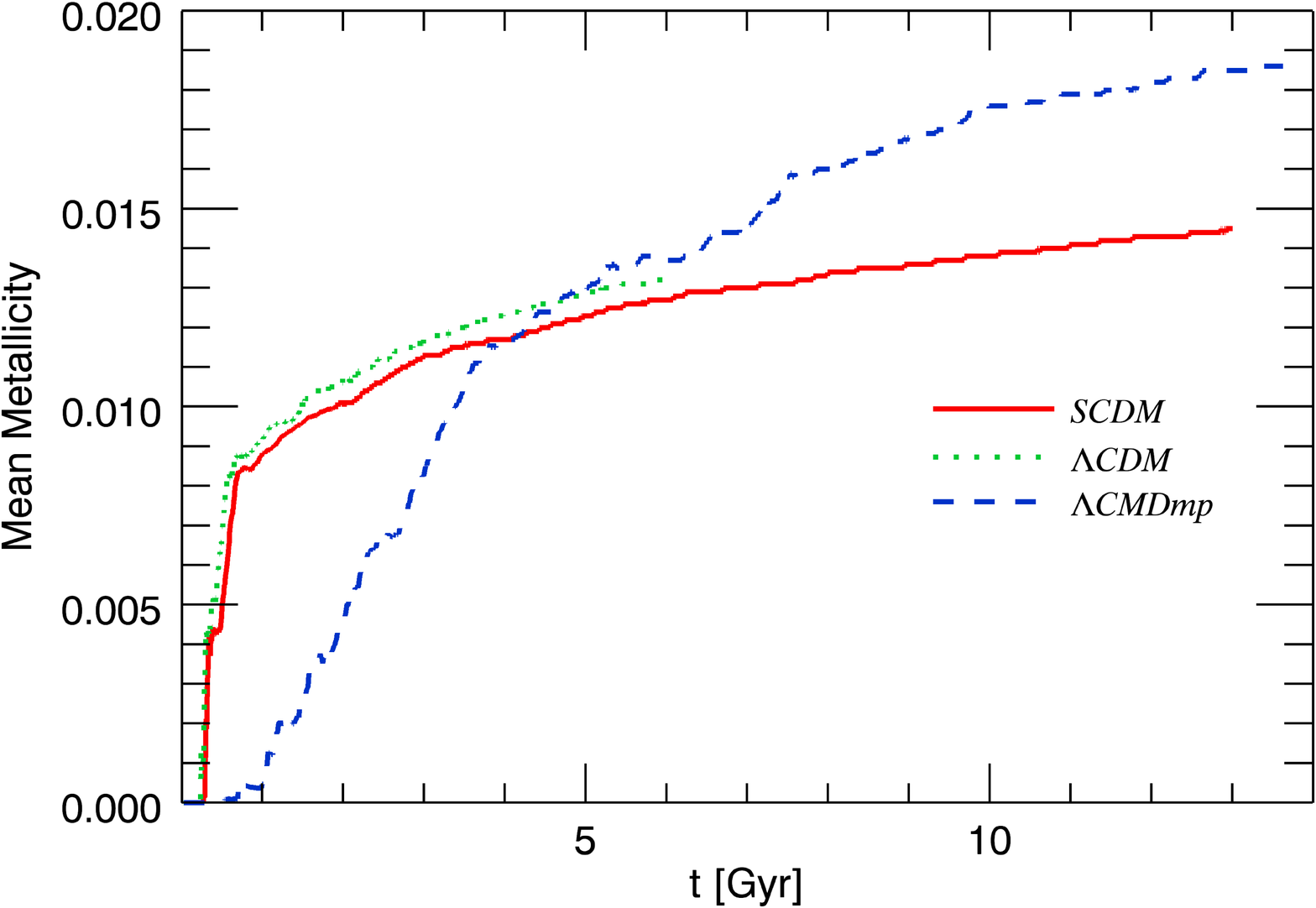}
\includegraphics[width=8.0cm,height=8.0cm]{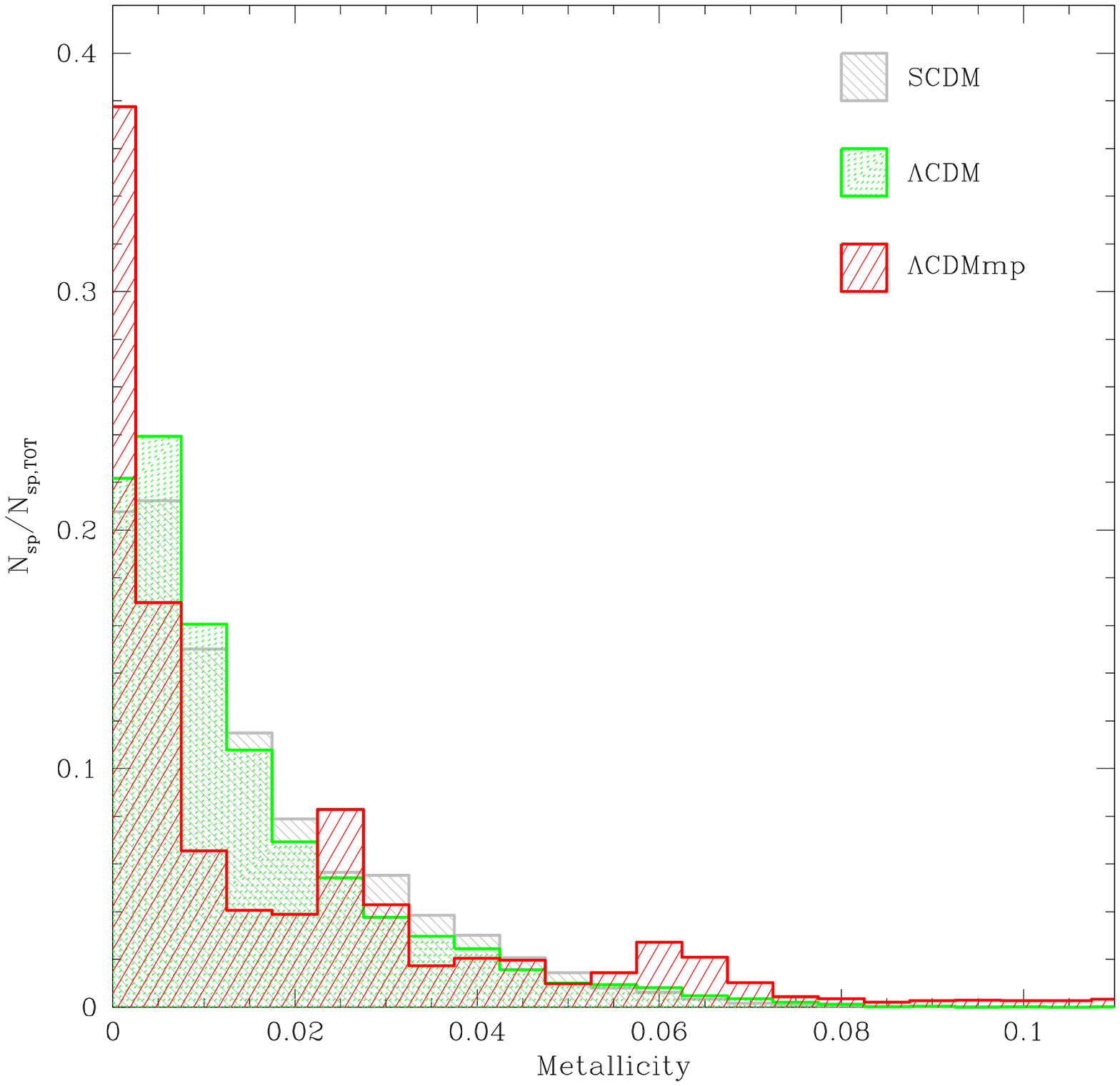}}
\caption{{\bf Left Panel}: mean metallicity $Z(t)$ versus
       time (in Gyr) for the three galaxy models, as indicated. {\bf
       Right Panel}: metallicity distribution (number of
       star-particles per metallicity bin) in the three models. The
       histograms labelled $SCDM$ and $\Lambda CDM_{mp}$ refer to the
       13\,Gyr age model, whereas the $\Lambda CDM$ one is for the
       7\,Gyr age model.}
\label{met}
\end{center}
\end{figure*}

Finally, in Fig.~\ref{met} we show the mean metallicity versus age
relationship for the three models (left panel). The mean metallicity
is simply the mean value of all star-particles evaluated at different
ages. The right panel shows the metallicity distribution (number of
star-particles per metallicity bin) in the three models.  The
histograms labelled $SCDM$ and $\Lambda CDM_{mp}$ refer to the 13\,Gyr
age models, whereas the $\Lambda CDM$ one is for the 7\,Gyr age
model. It is worth noting the long tail towards high
metallicities. The stars with these high metallicities are responsible
of a great deal of the ultraviolet excess in the SEDs via the
so-called AGB manqu\'e phase. The same consideration applies to the
very old stars in the lowest metallicity bin which may contribute to
the ultraviolet excess via the extended horizontal branch phase (see
below).

\section{The Spectral Energy Distribution of a NB-TSPH simulation: S{\footnotesize PE}{\normalsize C}{\footnotesize ODY}}
\label{specody}

As already mentioned, owing to the mass resolution of the dynamical
simulations fixed by the number of particles to our disposal, each
star-particle has the mass $M_{sp}\sim 10^{7} M_{\odot}$ or so (see
Table~\ref{inipar}), i.e. each star-particle represents a big assembly
of real stars which distribute in mass according to a given initial
mass function and are all born in a short burst of star formation,
therefore being homogeneous both in age and chemical composition. In
this way, each star-particle can be approximated to a SSP of mass
$M_{sp}$.

To derive the SED of our NB-TSPH galaxy model we start from the
definition of the integrated monochromatic flux generated by the
stellar content of a galaxy of age $T$

\begin{equation}
F_{\lambda}(T) = \int_{0}^{T} \int_{m_l}^{m_u} S(m,t,Z)\ f_{\lambda}(m,\tau',Z)\ dt\ dm
\end{equation}

\noindent
where $S(m,t,Z)$ denotes the stellar birth-rate and
$f_{\lambda}(m,\tau',Z)$ the monochromatic flux of a real star of mass
$m$, metallicity $Z$, and age $\tau'=T-t$. Separating $S(m,t,Z)$ into
the product of the star formation rate $\Psi(t,Z)$ (expressed in
suitable units) and the initial mass function $\phi(m)$, the above
integral becomes

\begin{equation}
F_{\lambda}(T) = \int_{0}^{T} \Psi(t,Z)\ f_{ssp,\lambda}(\tau',Z)\ dt
\end{equation}

\noindent
where

\begin{equation}
f_{ssp,\lambda}(\tau',Z) = \int_{m_l}^{m_u} \phi(m)\ f_{\lambda}(m,\tau',Z)\ dm
\end{equation}

\noindent is defined as the integrated monochromatic flux of a SSP,
i.e. of a coeval, chemically homogeneous assembly of stars with age
$\tau'$ and metallicity $Z$. The lower and upper mass limits of
integration in equations $m_{l}$ and $m_{u}$ respectively, define
the mass range within which stars are generated by each event of
star formation. Therefore, the stellar content of a galaxy can be
modelled as the convolution of many SSPs of different composition
and age, each of which is weighted by the rate of star formation at
the age at which it was born. Extensive tabulations of SSPs at
varying age, chemical composition, and initial mass function,
containing  magnitudes and colors for many photometric systems are
currently to our disposal in literature (see below for a very short
summary of the situation). Conventionally, the SED and
monochromatic flux, and hence luminosities in whatsoever passband
are calculated assuming that the SSP has a total mass equal to
1\,$M_{\odot}$. Having established the correspondence between the
star-particles of the NB-TSPH simulation with the classical SSP and
the consistency between the chemical parameters and IMF of the two
descriptions, the monochromatic flux and/or luminosity of a galaxy
with age T is given by

\begin{equation}
F_{\lambda}(T)= \sum_{i} f_{ssp,\lambda}(\tau_{i},Z_{i}) \times M_{sp,i}
\label{flamgal}
\end{equation}

\noindent
where $f_{ssp,\lambda}(\tau_{i},Z_{i})$ is the monochromatic flux of a
SSP of age $\tau_{i}$ and metallicity $Z_{i}$. To get the real
monochromatic flux emitted by a star-particle we must multiply
$f_{ssp,\lambda}(\tau_{i},Z_{i}) $ by $M_{sp}$ in solar
masses. Finally, the summation extends over all
star-particles. Extending the same procedure to the whole wavelength
interval we get the multi-wavelength SED, from which we can
immediately derive magnitudes and colors.

\subsection{Some details on the SSPs in use}
\label{ssp}

In this paper we have adopted the SSPs computed by
\citet{Tantalo05} and available online from the Padova Galaxies and
Single Stellar Population Models database (GALADRIEL) at {\it
http://www.astro.unipd.it/galadriel/}.

The spectral energy distributions (SEDs) of the SSPs have been
calculated following the method described by \citet{Bressan94}. To
this purpose we have adopted the stellar tracks calculated by
\citet{Girardi00}. These stellar models include modern physical
input as far as opacity, nuclear reaction rates, neutrino losses,
mixing schemes, etc. are concerned. The evolutionary sequences go
from the zero age main sequence to the latest evolutionary phases
and cover wide ranges of stellar masses and chemical compositions.
In particular, they include the planetary nebula phase, the
so-called AGB manqu\'e phase that may develop in low-mass stars when
the metallicity is higher than about three times solar, and the
extended horizontal branch typical of low-mass stars with very low
metallicity. Finally, the underlying isochrones are calculated by
means of the algorithm of "equivalent evolutionary points" described
in \citet{Bertelli94}.

In order to derive SEDs, magnitudes, and colors, corresponding to a
source of given luminosity, effective temperature, gravity, and
chemical composition, one needs a library of stellar spectra as
function of these parameters. The spectral library considered in
this paper was assembled by \citet{Girardi02} adopting the ATLAS9
release \citep{Kurucz93} of synthetic atmospheres: these latter are
those for the no-overshooting case calculated by \citet{Castelli97}
and subsequently extended by other authors.

For each SSP, GALADRIEL provides also large tabulations of
magnitudes and colors for the following photometric systems:

\begin{itemize}
\item Bessell-Brett
\item Hubble Space Telescope (NICMOS, WFPC2, ACS)
\item Sloan Digital Sky Survey (SDSS)
\item GAIA
\item GALEX
\end{itemize}

\noindent
All the details of the computational procedure adopted to derive the
synthetic magnitudes and colors are described in
\citet{Girardi02} to whom we refer.

In this study we have considered only two photometric systems: the
Bessell-Brett and SDSS for the VEGAmag and ABmag. The transmission
curves considered for first photometric system are from
\citet{Bessell90} for the $UBVRI$ passbands and from
\citet{Bessell88} for the $JHKLMN$ passbands. The SDSS photometric
system \citep{Fukugita96} comprises five non-overlapping passbands
that range from the ultraviolet cutoff at 3,000\AA\ to the
sensitivity limit of silicon CCDs at 11,000\AA. To interpret large
samples of galaxies of recent acquisition, such as COSMOS and GOODS,
we have also implemented the photometric systems used in these
campaigns (see below).

\subsection{SEDs of the galaxy models}

\begin{figure*}
\begin{center}
\includegraphics[width=0.9\textwidth]{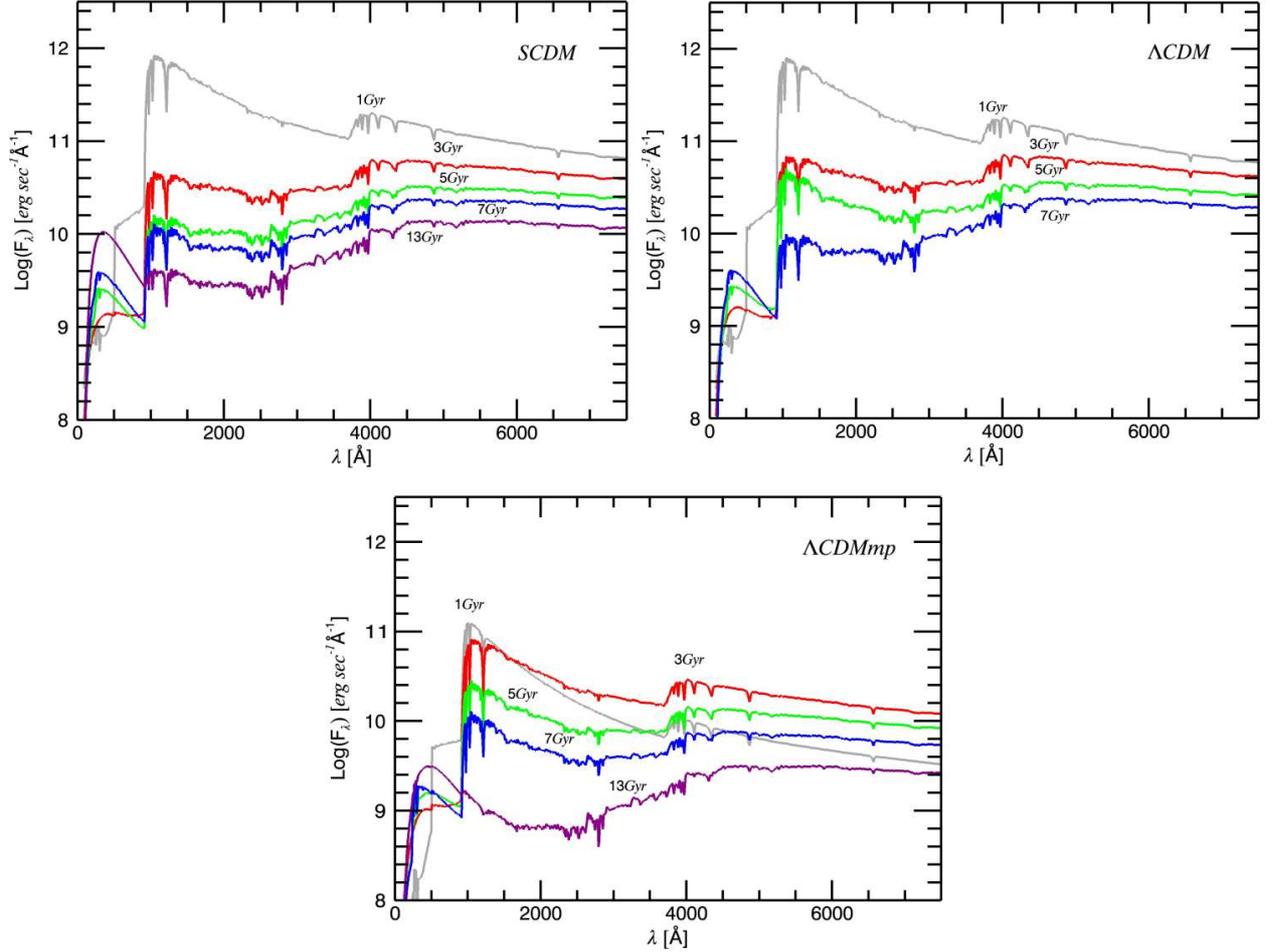}
\caption{SEDs of the model galaxies for the different
          cosmological scenarios, shown at different ages as
          indicated ($SCDM$ upper left, $\Lambda
          CDM$ upper right, $\Lambda CDM_{mp}$ lower panel). All
          curves refer to the rest-frame fluxes. Ages are in Gyr.}
\label{specgals}
\end{center}
\end{figure*}

Using the above technique and integrating over the SEDs of all
star-particles we can derive the SEDs of the model galaxies. The SEDs
are shown in Fig.~\ref{specgals} at different ages (in view of the
cosmological application of these results we remind the reader that
these are the SEDs seen in the rest-frame).

It is worth noting that starting from the age of about 5 Gyr the three
SEDs show an important ultraviolet excess, i.e. rising branch and a
peak in the flux short-ward of 2000\AA. At the last age in common,
7\,Gyr, the flux level is nearly comparable in the $SCDM$ and $\Lambda
CDM$ models and significantly lower in the $\Lambda CDM_{mp}$
case. For the two models arriving to 13\,Gyr age, namely $SCDM$ and
$\Lambda CDM_{mp}$, the ultraviolet excess in the first model is
significantly higher than in the second one. What is the source of
this excess of flux? There are several candidates: the short-lived
planetary nebulae, the AGB manqu\'e phase for stars with the
appropriate metallicity, the hot horizontal branch stars of low metal
content, and finally massive stars if star formation is going on. The
Planetary Nebulae are the descendants of low- and intermediate-mass
stars on their way from AGB to the White Dwarf regime. Although they
can be very bright and hot, they are too short-lived (a few $10^{4}$
yr). The AGB manqu\'e stars have a low-mass and a lifetime amounting
to a fraction of the core He-burning phase. These stars appear when
the age is older than approximately 5-6\,Gyr. Low-mass stars of very
low metal content during part of their core He-burning phase in a very
extended horizontal branch are bright and long lived so that they may
significantly contribute to the UV flux. Finally there are the young
stars if star formation goes on even at minimal levels of
activity. Owing to the much higher intrinsic luminosity of these
stars, if they are present even in small numbers they would
significantly contribute to the flux in the far
ultraviolet. Disentangling the contribution of each possible source is
a cumbersome affair. The fact that the flux in this wavelength
interval increases starting from about 5\,Gyr suggests a combination
of AGB manqu\'e stars and young stars, leaving planetary nebulae and
extreme horizontal branch objects in the background.

\subsection{Magnitudes and colors of galaxy models}
\label{magcolpart}

Given a photometric system, it is straightforward to derive the
temporal evolution of magnitudes and colors of each star-particle
and of the whole galaxy. For the sake of illustration we show
results only for the Bessell-Brett and the SDSS photometric systems.

\begin{figure*}
\begin{center}
{\includegraphics[width=6cm]{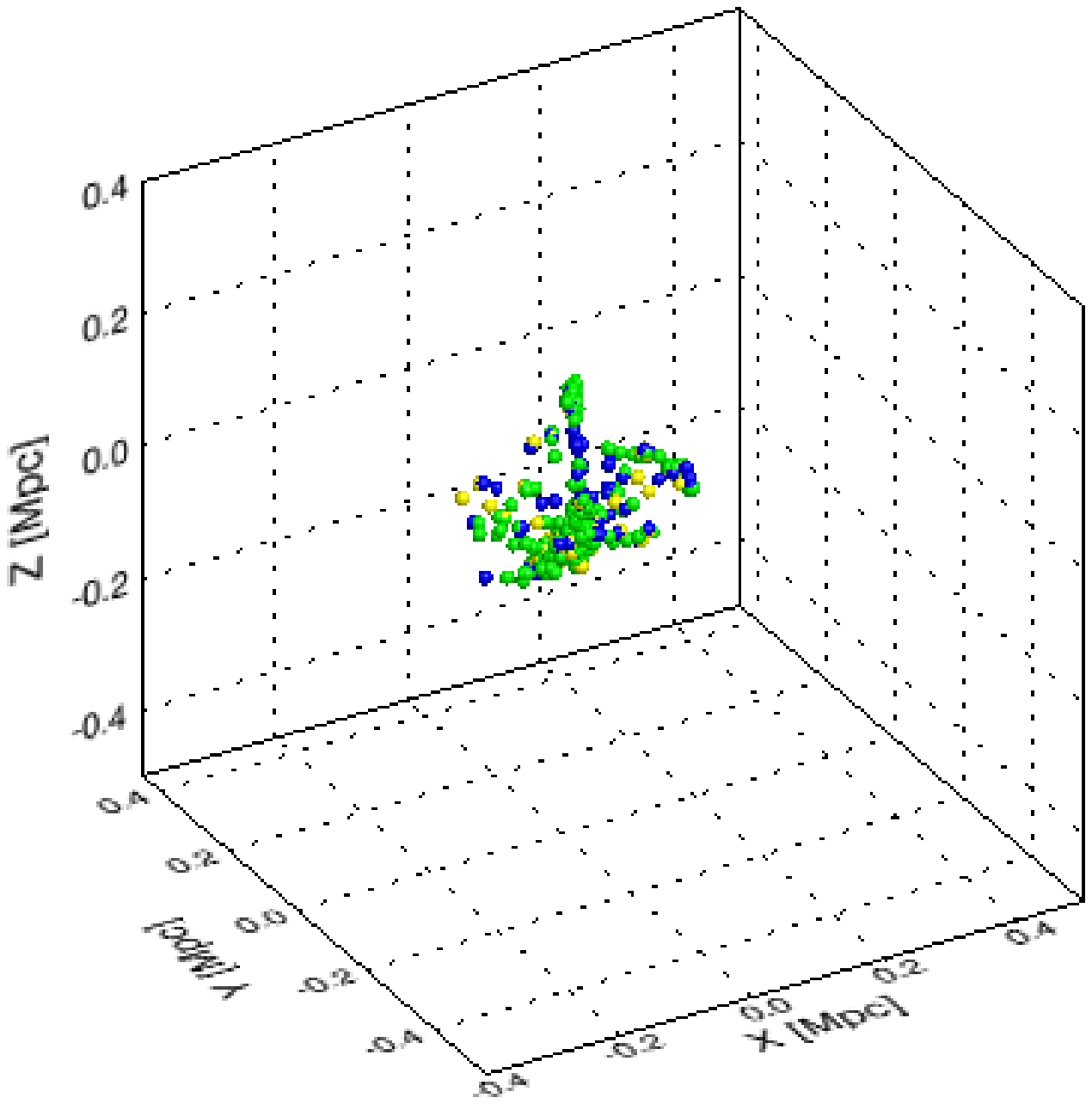}
 \includegraphics[width=6cm]{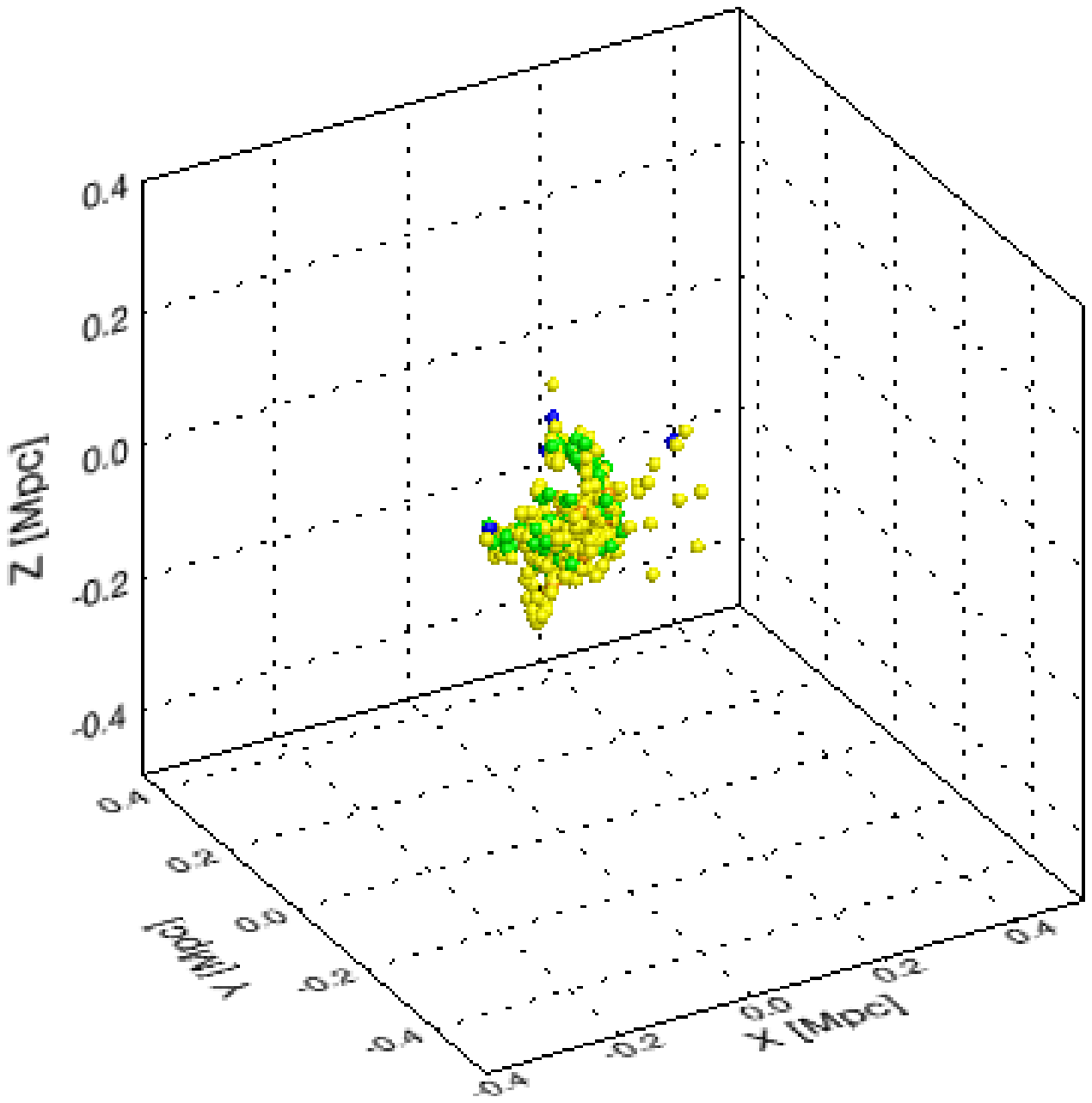}}
{\includegraphics[width=6cm]{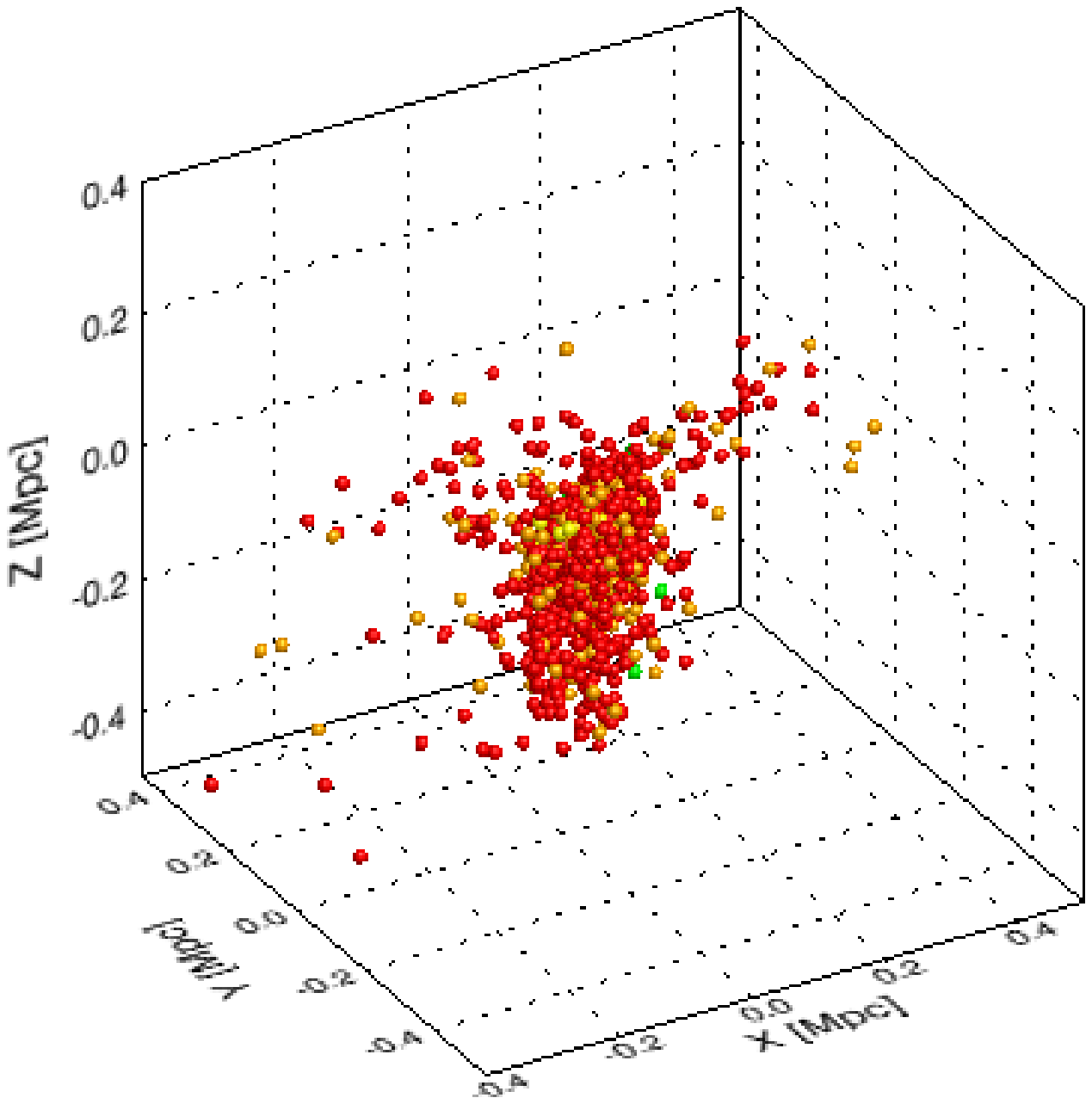}
 \includegraphics[width=6cm]{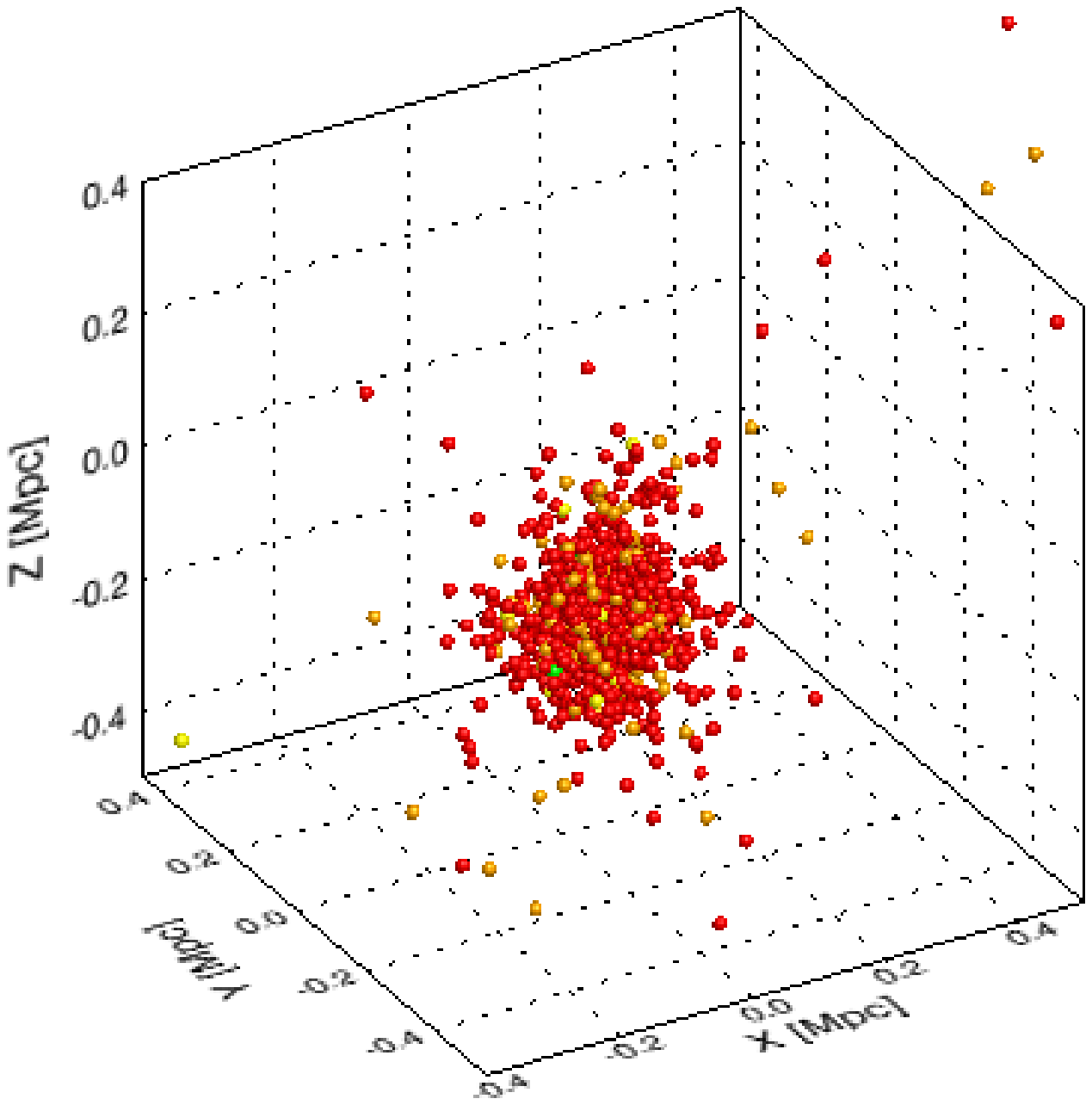}}
\includegraphics[width=0.7\textwidth]{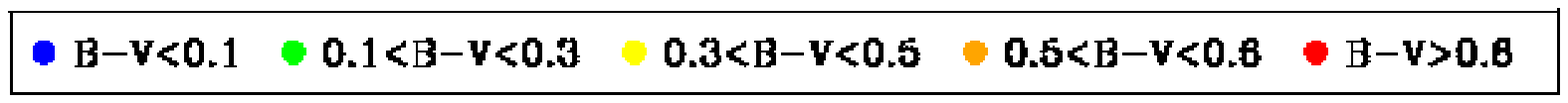}
\caption[3-D simulation]{Three-dimensional view of the $SCDM$
    simulation for four epochs (1\,Gyr upper left panel, 2\,Gyr upper
    right, 5.5\,Gyr lower left, and 13\,Gyr lower right). The color code
    of each star-particle corresponds to the value of the (B-V)
    color as indicated.}
\label{simpart}
\end{center}
\end{figure*}

\noindent
\textsf{Star-particle by star-particle view}. In
Fig.~\ref{simpart} we show the color evolution of the $SCDM$ model
galaxy for the Bessell-Brett system. The color code for each
star-particle in the 3D-view of the galaxy structure indicates the
color evolution of the corresponding SSP with same mass, age and
chemical composition. During the evolution, the galaxy starts from an
initial proto-galaxy to end up to a final stage where the structure
resembles that one of a spheroidal system. While the galaxy builds up
itself, young star-particles have blue colors.  Later, as the galaxy
ages, the stellar component gets redder and redder.

\begin{figure*}
\begin{center}
\centerline{
\includegraphics[width=8.0cm,height=8cm]{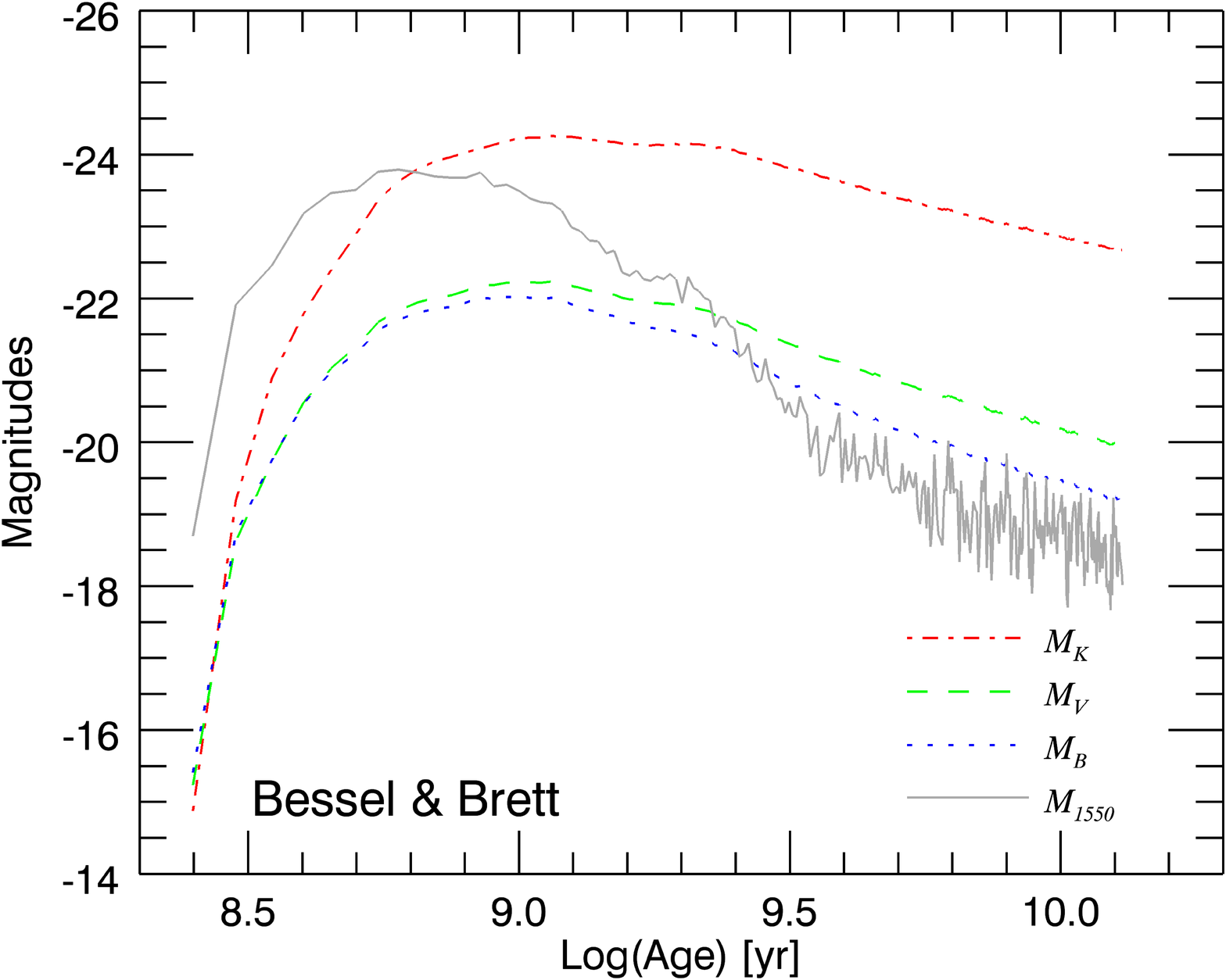}
\includegraphics[width=8.0cm,height=8cm]{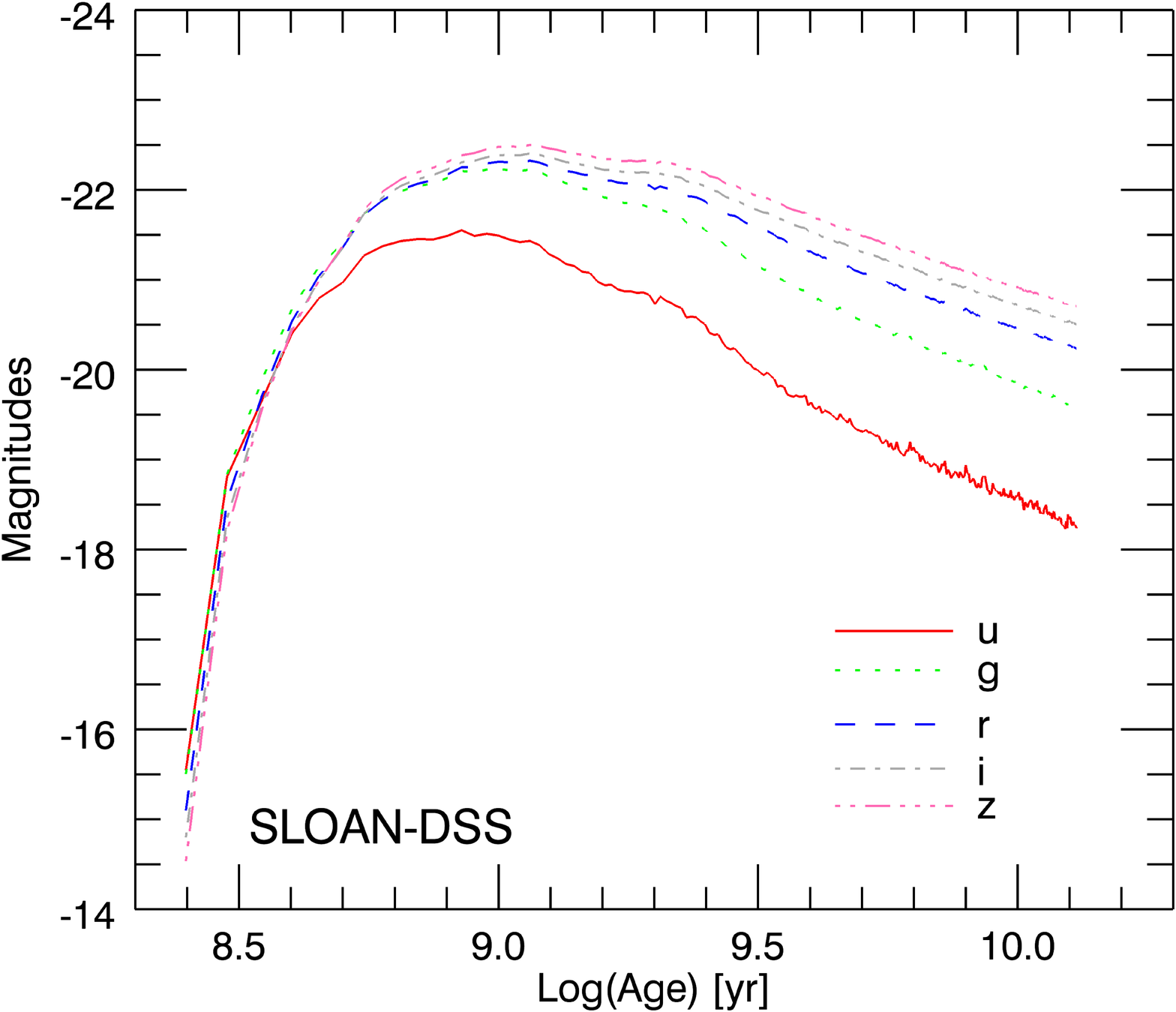}}
\caption[Rest-frame evolution of magnitudes in the Bessell-Brett and
SDSS systems]
   {{\bf Left Panel}: rest-frame evolution of the total absolute
    Bessell \& Brett magnitudes,  $M_{K}$, $M_{V}$, $M_{B}$,
    and $M_{1550}$, of the  $SCDM$ model. {\bf Right Panel}: the
    same as in the left panel but for the $u$, $g$, $r$, $i$, and
    $z$ passbands of the SDSS photometric system.}
\label{magevsb}
\end{center}
\end{figure*}

\noindent
\textsf{Integrated magnitudes and colors.} The evolution
of the integrated magnitudes in the rest-frame, for the $SCDM$ model
galaxy, is shown in Figs.~\ref{magevsb}, for the VEGAmag in the
Bessell-Brett photometric system (left panel) and ABmag in the SDSS
(right panel), respectively.

The temporal evolution of magnitudes mirrors the history of star
formation (see Fig.~\ref{sfr}): in brief the magnitudes decrease as
the galaxy gets the peak of SFR and hence becomes more luminous.
Afterwards, the magnitudes increase following the SFR that declines
rapidly. Fig.~\ref{magevsb} shows the magnitudes $M_B$, $M_V$, $M_K$,
and $M_{1550}$ of the SCDM case for the Bessell-Brett system (left
panel). The 1550-magnitude, that probes the UV region of the
spectrum, weights the star formation at each epoch and reproduces
the trend seen in the SFR. In the right panel of Fig.~\ref{magevsb}
we show the magnitudes in the SDSS photometric system of the same
model. Although to lower extent, the same effect is visible in the
$u$ and $g$ bands where the evolution is less smooth than for the
other three bands.

\begin{figure*}
\begin{center}
\centerline{
\includegraphics[width=8.0cm,height=8cm]{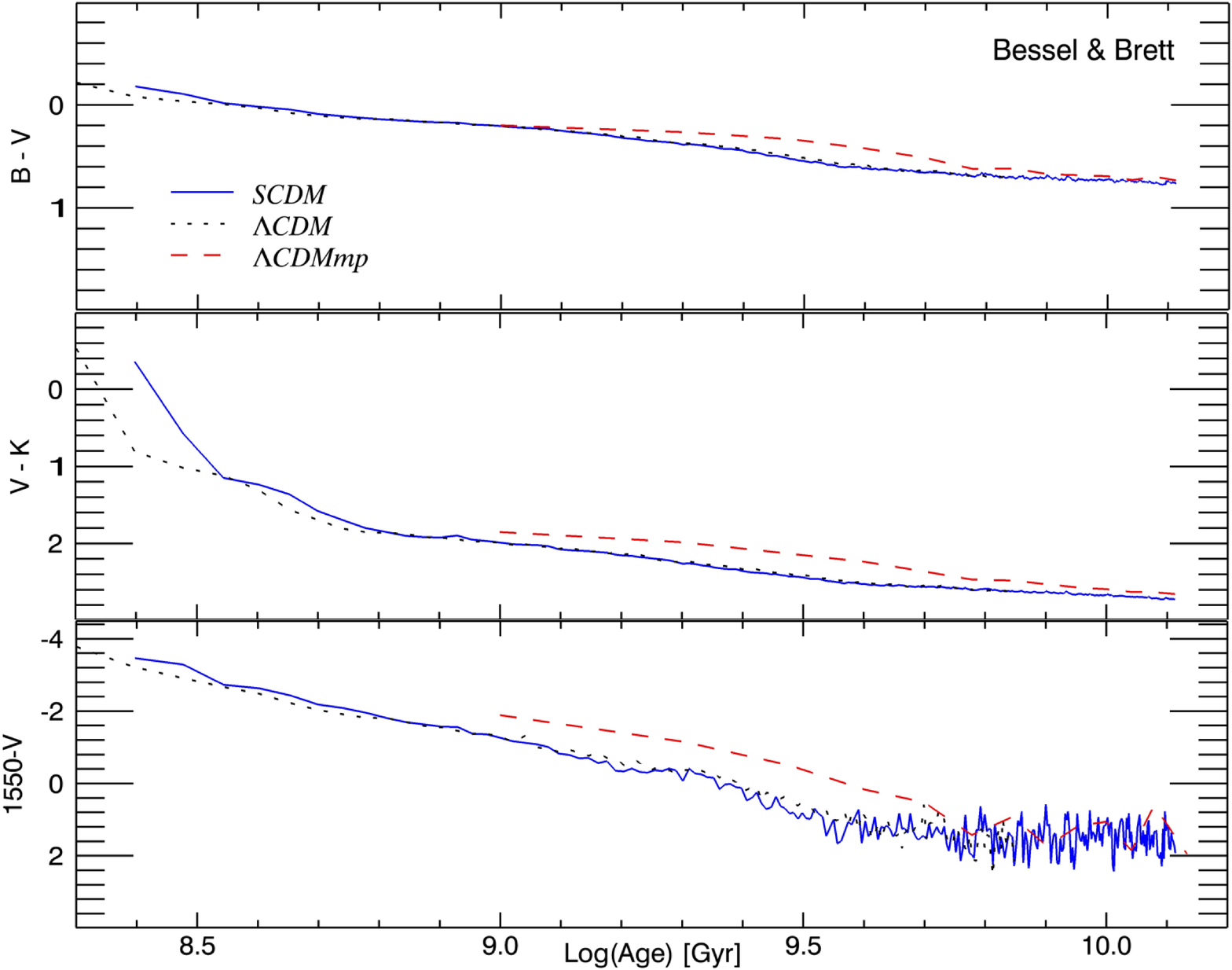}
\includegraphics[width=8.0cm,height=8cm]{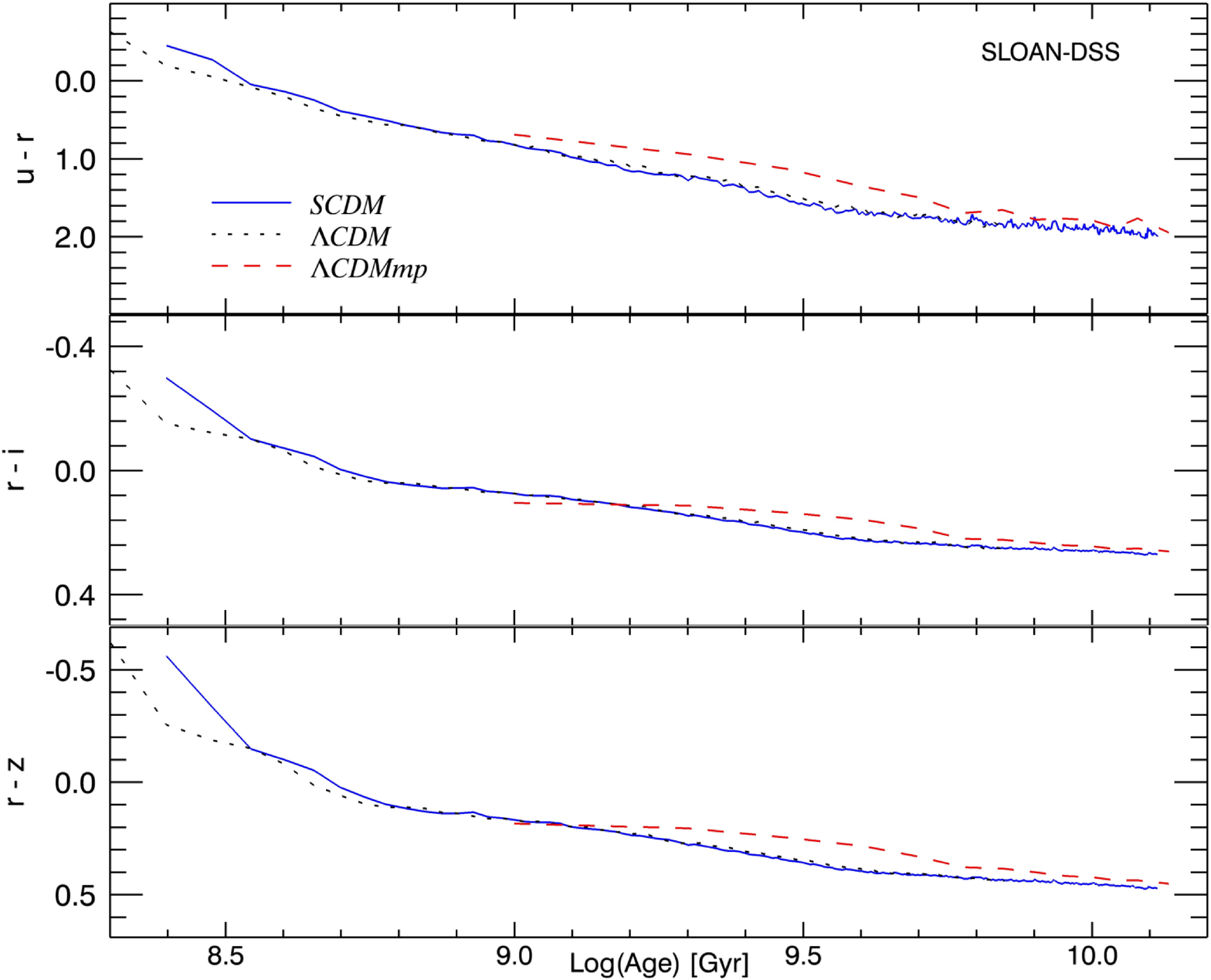}}
\caption[Rest-frame color evolution in the Bessell-Brett and SDSS system]
        {{\bf Left Panel}: rest-frame evolution of the B--V, V--K, and
        1550--V colors for the Bessell-Brett photometric system shown
        by our model galaxies as indicated. {\bf Right Panel}: the
        same as in the left panel but for SDSS colors $u-r$, $r-i$,
        and $r-z$. In both panels the $\Lambda CDM_{mp}$ case is
        plotted only for age older than 1\,Gyr for the sake of
        clarity. For younger ages it runs very close to the other two
        cases.}
\label{colevsb_vss}
\end{center}
\end{figure*}

\begin{figure*}
\begin{center}
\includegraphics[width=0.97\textwidth]{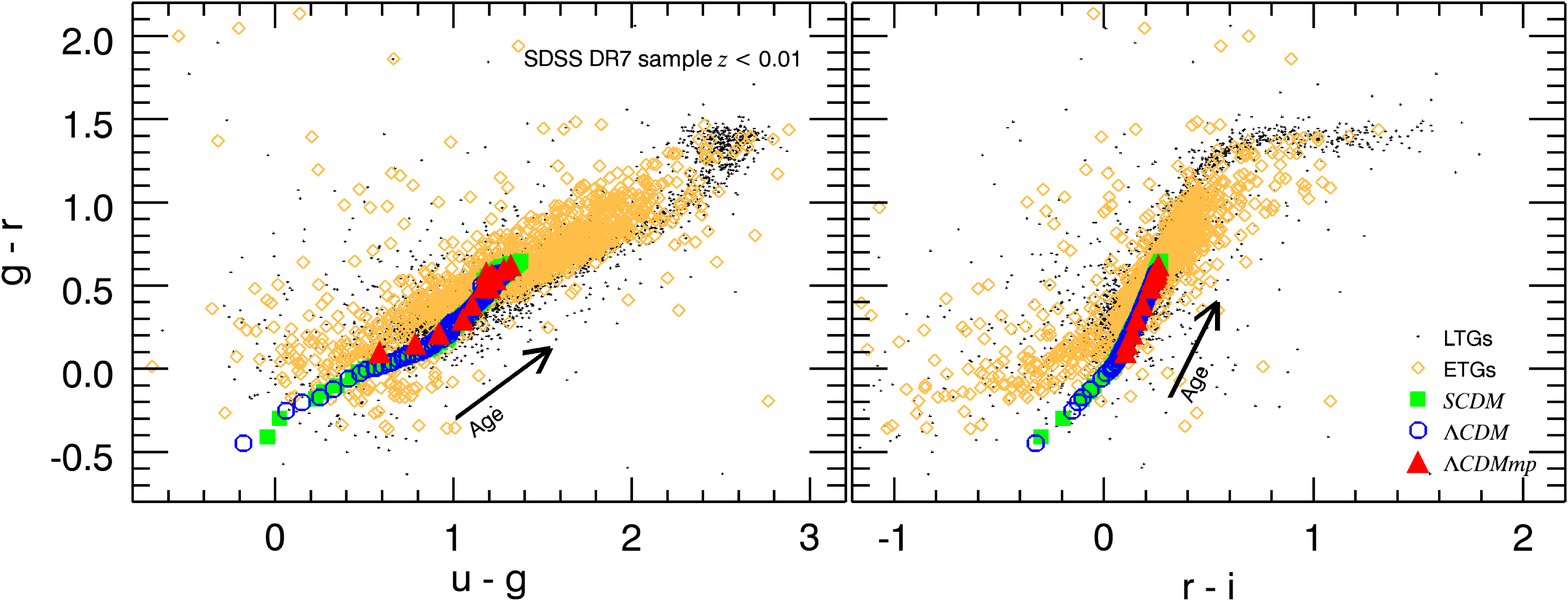}
\caption[Distribution of a SDSS Sample in the rest-frame color-color diagram]
    {Validation of the integrated colors of our models compared to
    the observational data for a sample of nearby galaxies. We show
    the color-color distribution of a sample of galaxies from DR7
    SDSS. The data are selected for $z \le 0.01$. Small dots and
    diamonds are for late type galaxies (LTGs) and ETGs, respectively.
    The evolution of $SCDM$ ({\it filled squares}), $\Lambda CDM$
    ({\it empty circle}), and $\Lambda CDM_{mp}$ ({\it filled
    triangles}) in the rest-frame are overlapped with age increasing
    as indicated.}
\label{2col}
\end{center}
\end{figure*}

The temporal evolution of colors is shown in the two panels of
Fig.~\ref{colevsb_vss}. Once again, the differences between models are
primarily due to their SFH. In brief, the $\Lambda CDM$ and $SCDM$
models have the same prescription for the SFR; therefore they have
similar SFHs (both qualitatively and quantitatively) and similar
color evolution. In the $\Lambda CDM_{mp}$ case with the multi-phase
ISM, star formation is more gradual and lasts longer, the peak of
activity is lowered and shifted to older ages as compared to the
$\Lambda CDM$ case. The total mass assembled by the models is almost
equal and the different behavior obtained with the two star formation
prescriptions is likely due to the different time scales required to
form new stars \citep[see ][ for details]{Merlin07}.

To assess the quality of the integrated colors of our models we
compare them to observed colors of a sample of nearby galaxies taken
from the SDSS-DR7 database. We select the sample with the following
criteria: the galaxies must have redshift $z \lesssim 0.01$, the
galaxy images should be taken far away from the CCD edges, they should
unsaturated, and finally, the photometric errors in each bands should
be smaller than 0.1 mag (good signal to noise ratios). With these
criteria we obtain a sample of 5986 galaxies. Fig.~\ref{2col} shows
the galaxies in a color-color diagram and compare them with the
rest-frame color evolution of the models. The age increases as
indicated by the arrow. Galaxies classified as ETGs, using the
exponential ($P_{exp}$) and de Vaucouleurs' ($P_{deV}$) profile
likelihoods \citep[see ][ for more details on the SDSS morphological
classification]{Shimasaku01,Strateva01}, are indicated with empty
circles, whereas late type galaxies (LTGs) are shown with black
dots. The sample has been corrected for the extinction. The color
evolution of our simulations is indicated with {\it filled squares},
{\it empty circles}, and {\it filled triangles} for the three models.
Remarkably, simulated colors and data nicely agree, in particular for
ages older than 5\,Gyr.

\subsection{Distribution of the Stellar Populations in the Color-Magnitude Diagram}

\begin{figure*}
\begin{center}
\centerline{
\includegraphics[width=8.0cm,height=8.0cm]{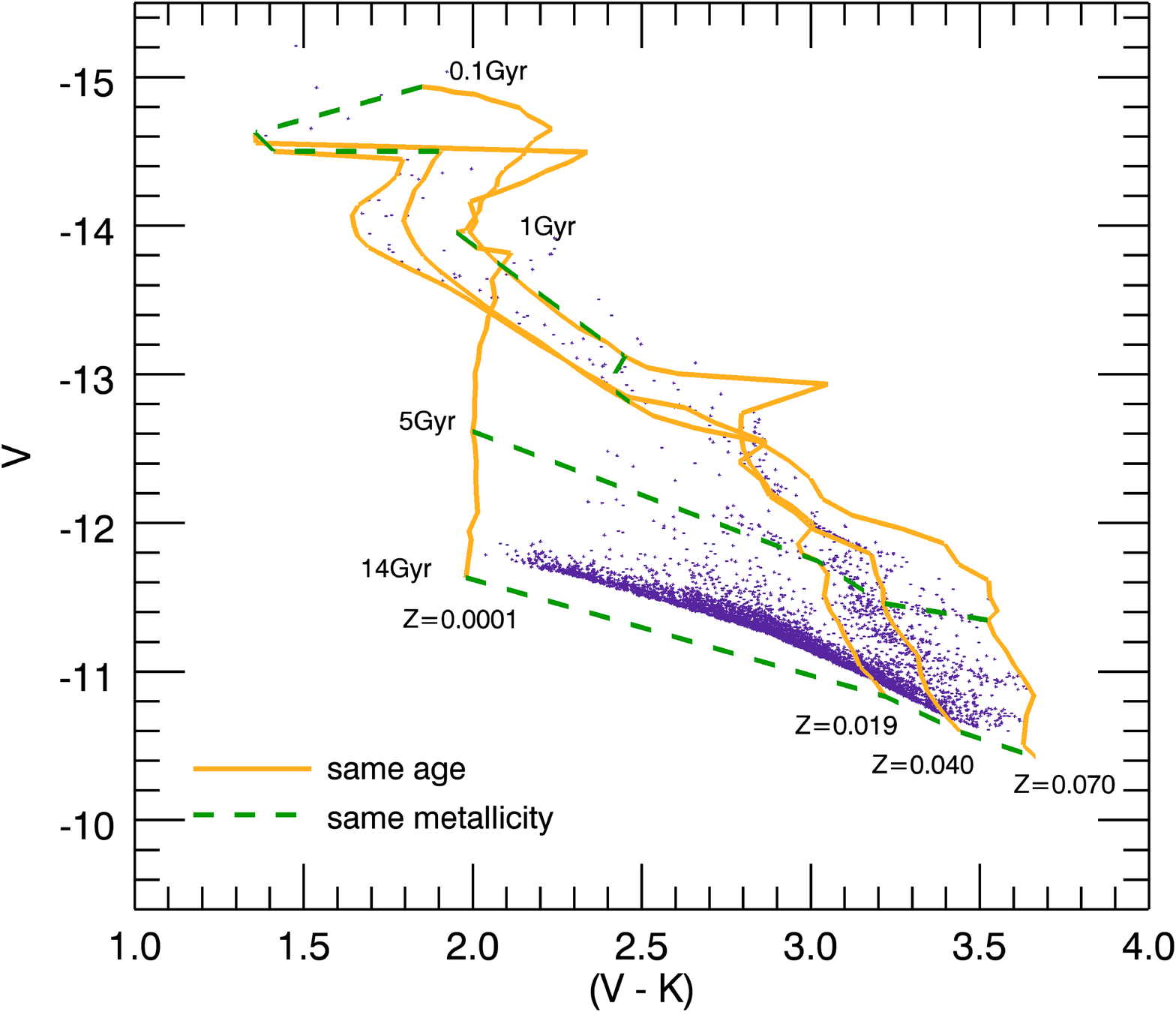}
\includegraphics[width=8.0cm,height=8.0cm]{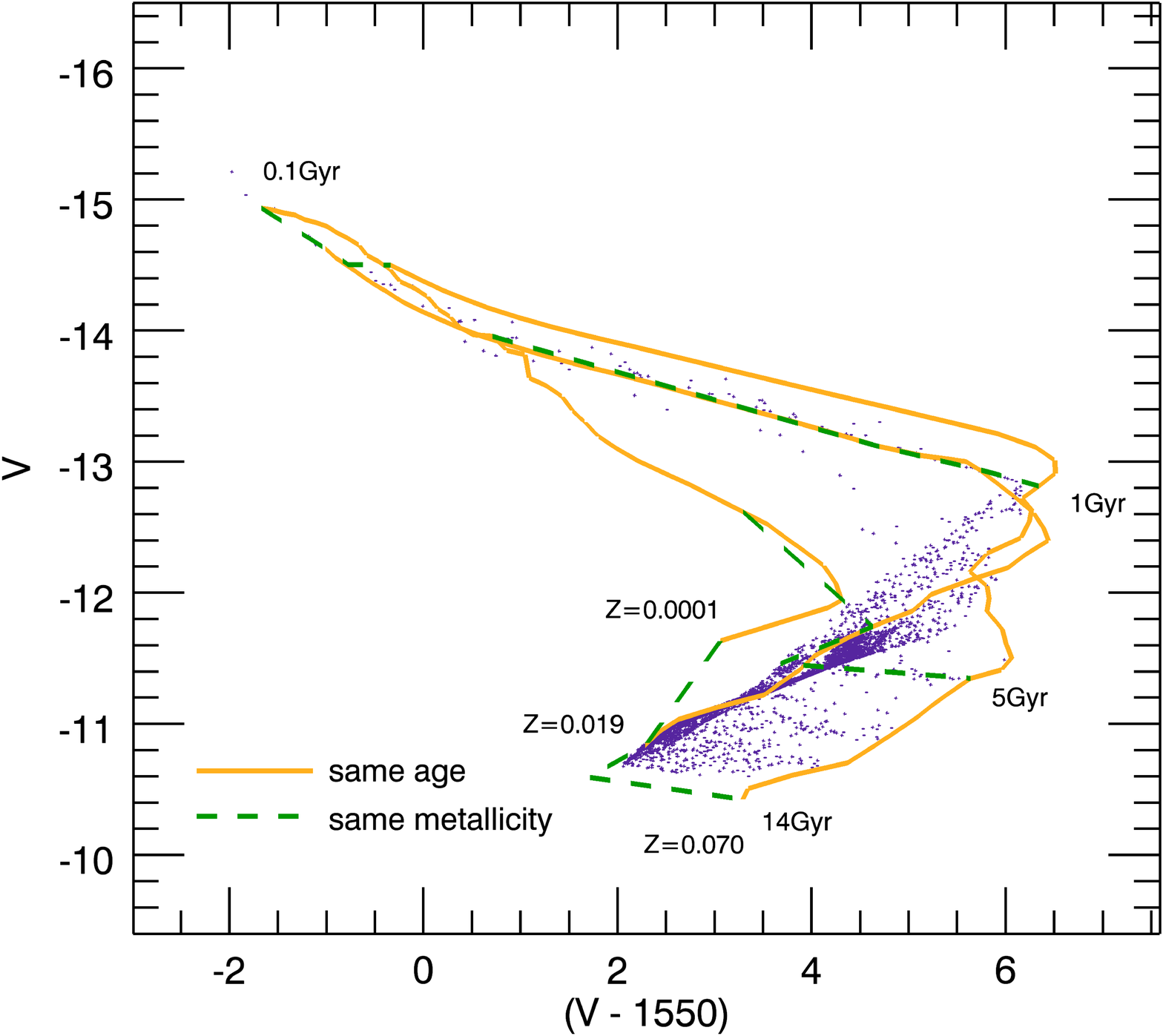}}
\caption[Distribution of stellar populations in the $(V-K)-V$ and
   $(1550-V)-V$ planes] {{\bf Left Panel}: distribution of the
   stellar populations in the $(V-K)-V$ plane for the $SCDM$ model
   at the age of T=13\,Gyr. The SSPs color grid is overlapped for
   interpretation of ages and metallicities. \textbf{Right Panel}: the
   same as in the left panel but for the $(1550-V)-V$ plane. See the text for details.}
\label{cmd1_cmd2}
\end{center}
\end{figure*}

It might be worth of interest to explore the age-metallicity range
spanned by the stellar populations of a galaxy by looking at their
distribution in the color-magnitude diagram (CMD), in analogy to
what currently made for stars in clusters and fields.

\begin{figure*}
\begin{center}
\centerline{
\includegraphics[width=6cm,height=6cm]{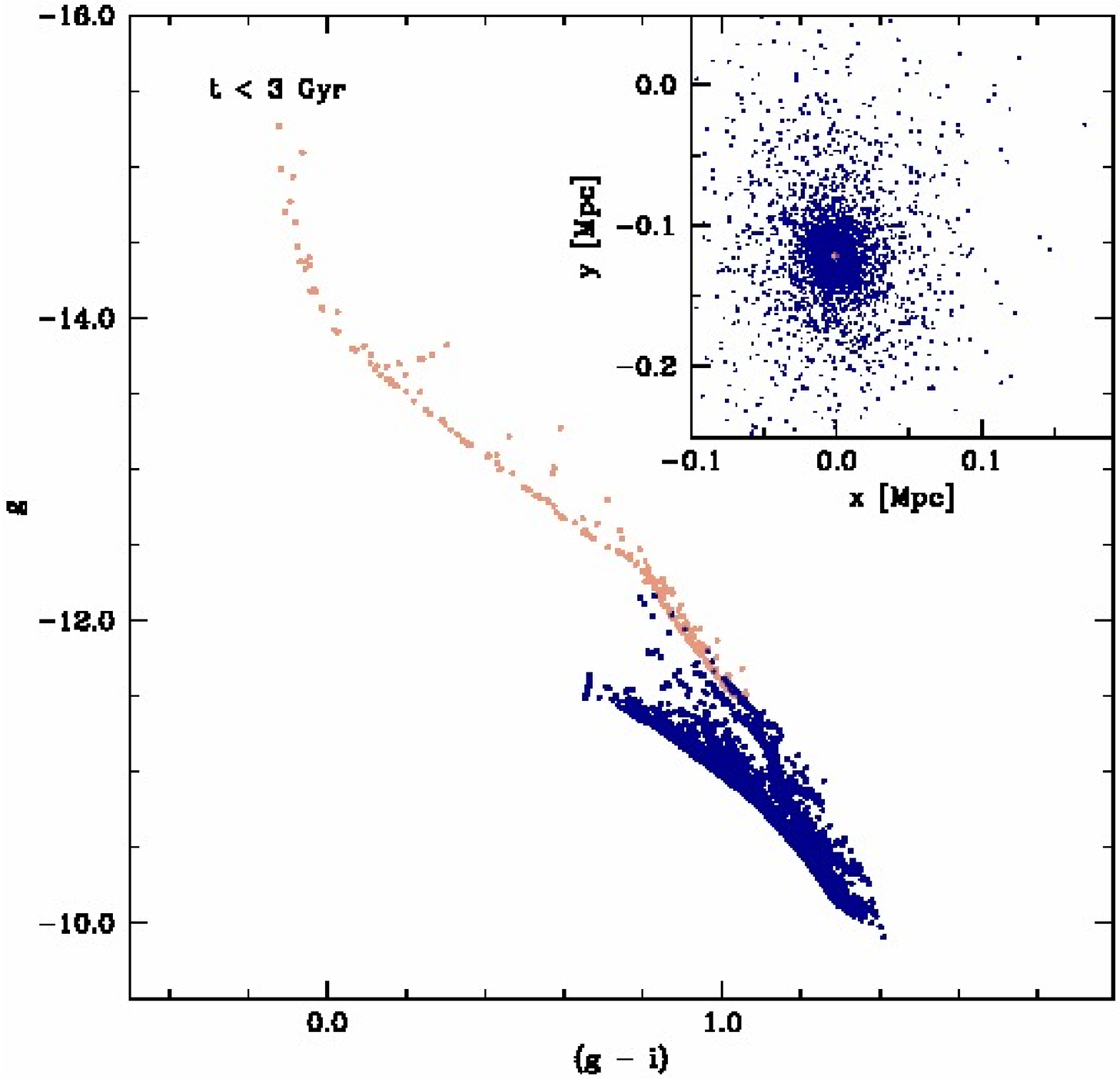}
\includegraphics[width=6cm,height=6cm]{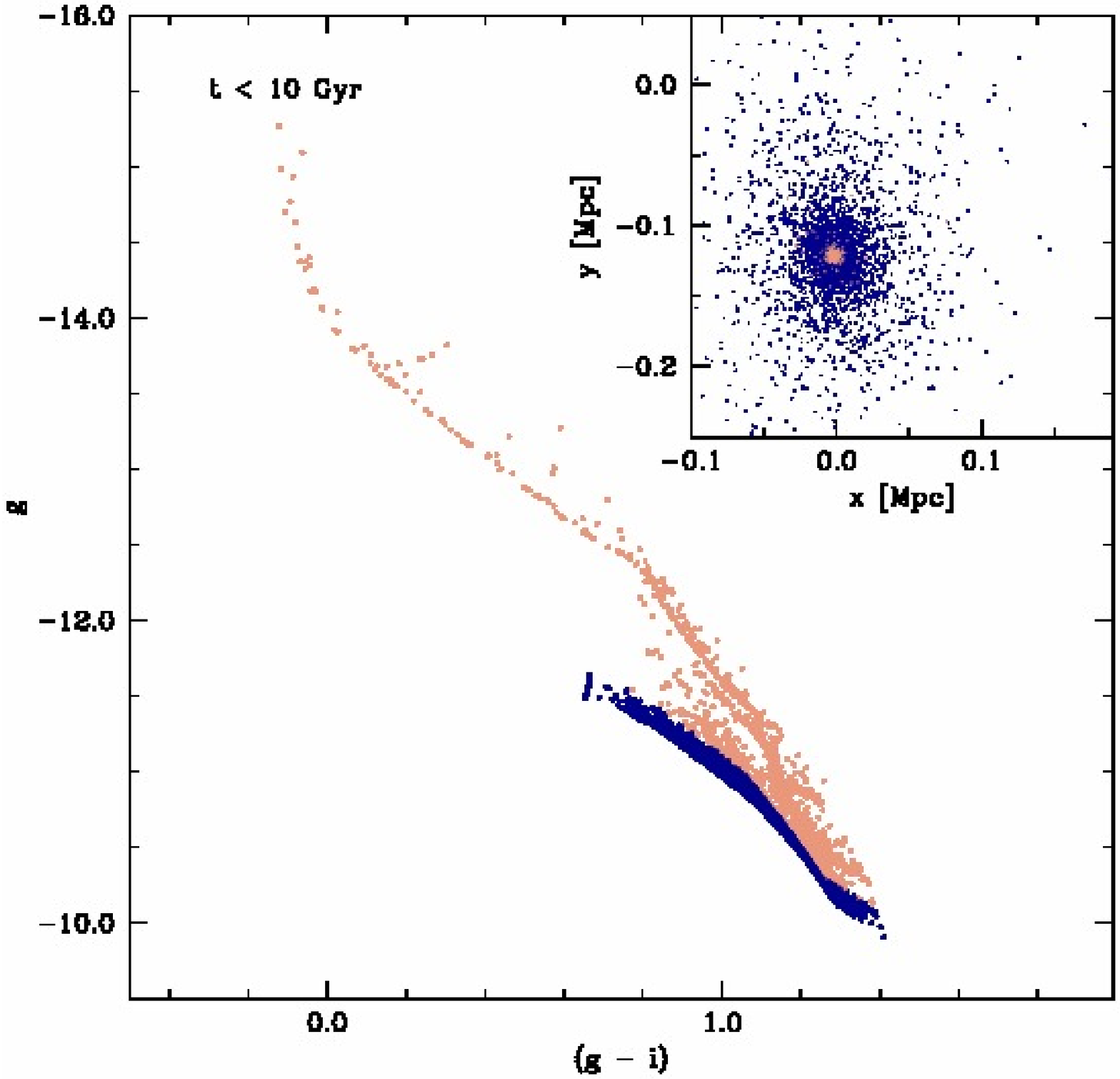}
\includegraphics[width=6cm,height=6cm]{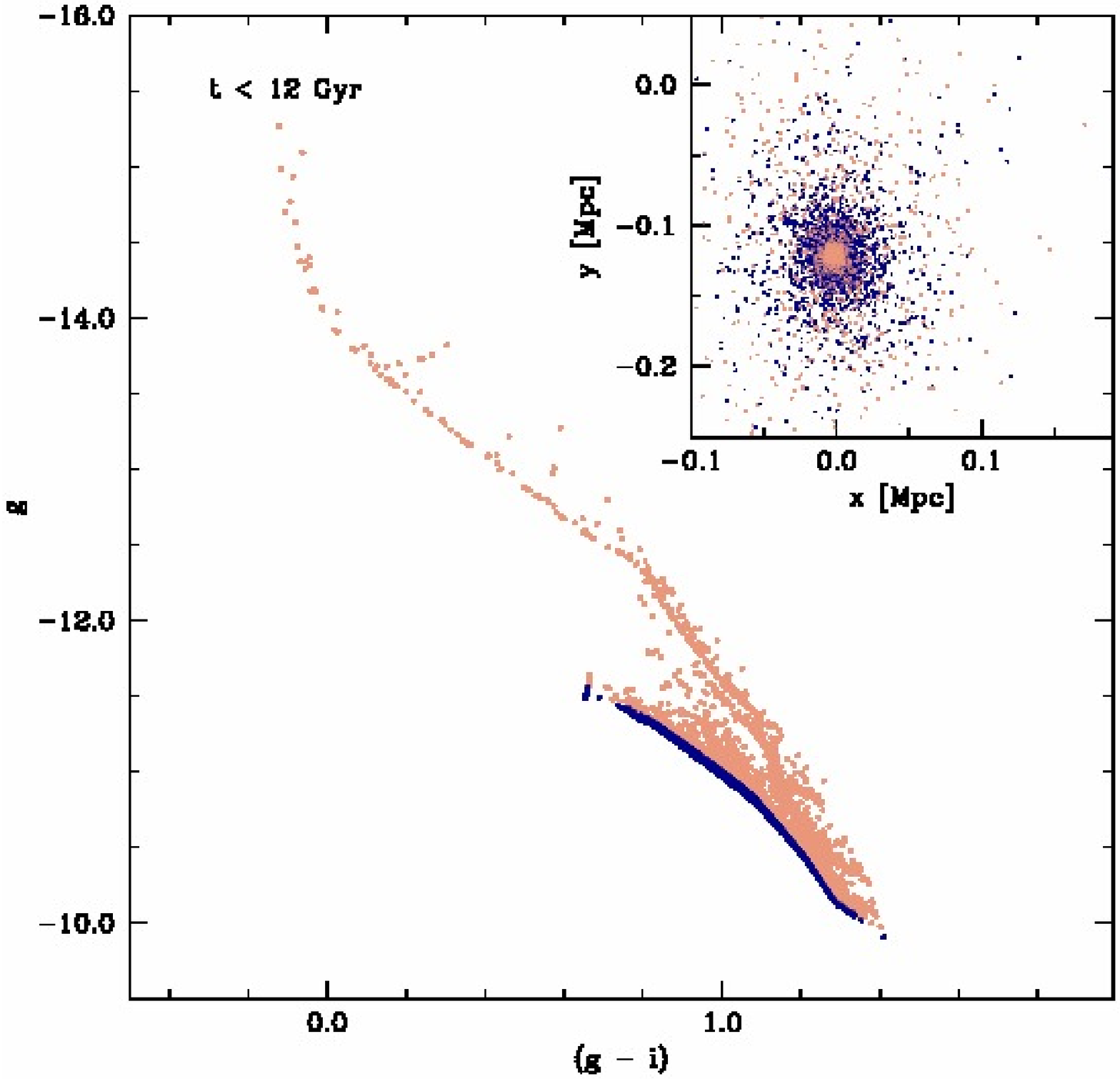}}
\caption[Distribution of stellar populations with different
   age in the $(g-i)-g$ plane.]
   {Distribution of the stellar populations in the $(g-i)$ vs $g$
   plane for the $SCDM$ model at the age of T=13\,Gyr. Stars younger
   than a certain limit (that varies from panel to panel as indicated)
   are plotted as light dots. The insert in the upper right corner in
   each panel shows the position of such stars (with light colors) and
   of the remaining stars (dark dots). As expected, the residual star
   formation (say after the first 5\,Gyr) tends to occur preferentially
   in the very central regions of the galaxy}
\label{cmd_age}
\end{center}
\end{figure*}

\begin{figure*}
\begin{center}
\centerline{
\includegraphics[width=6cm,height=6cm]{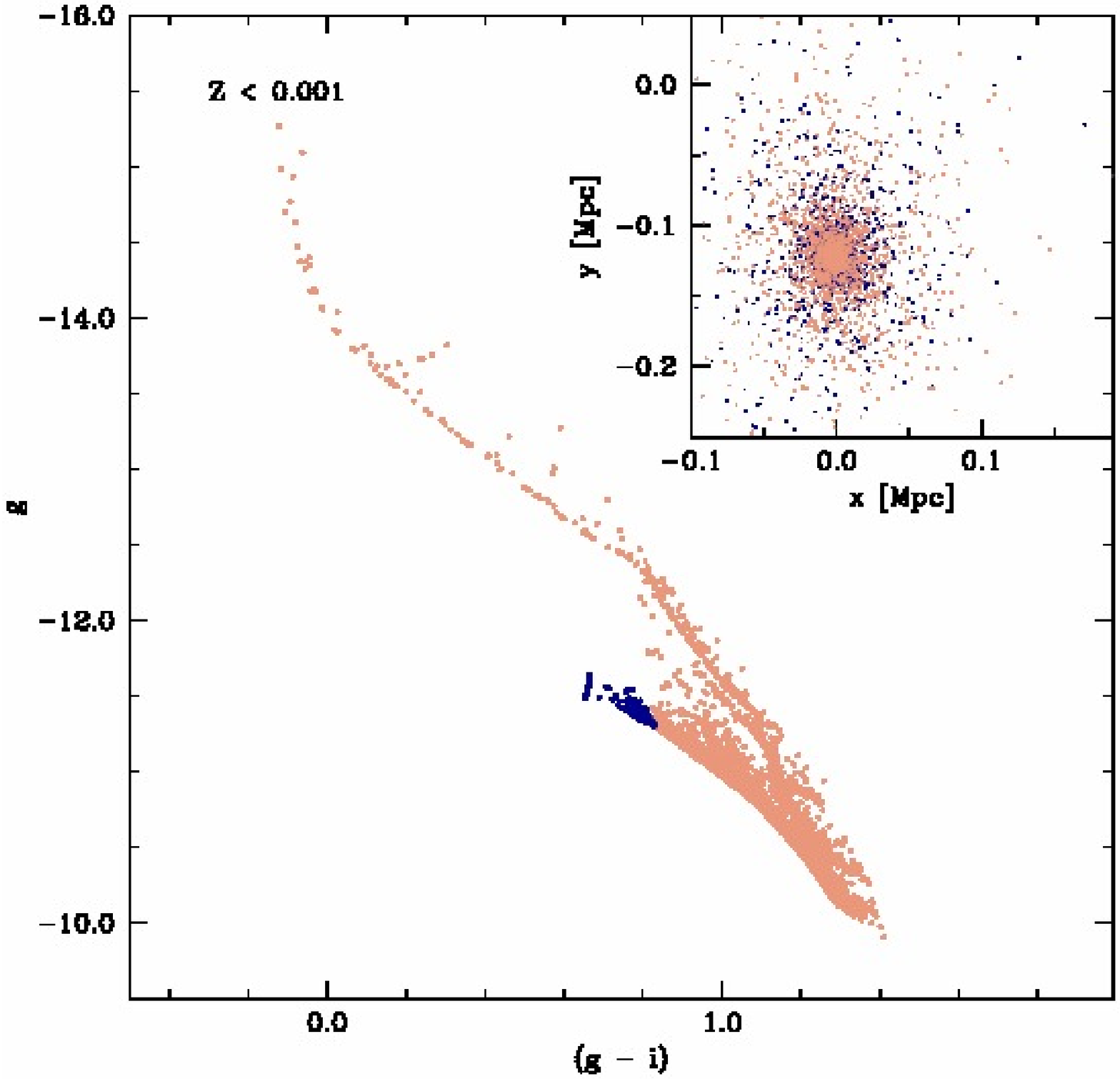}
\includegraphics[width=6cm,height=6cm]{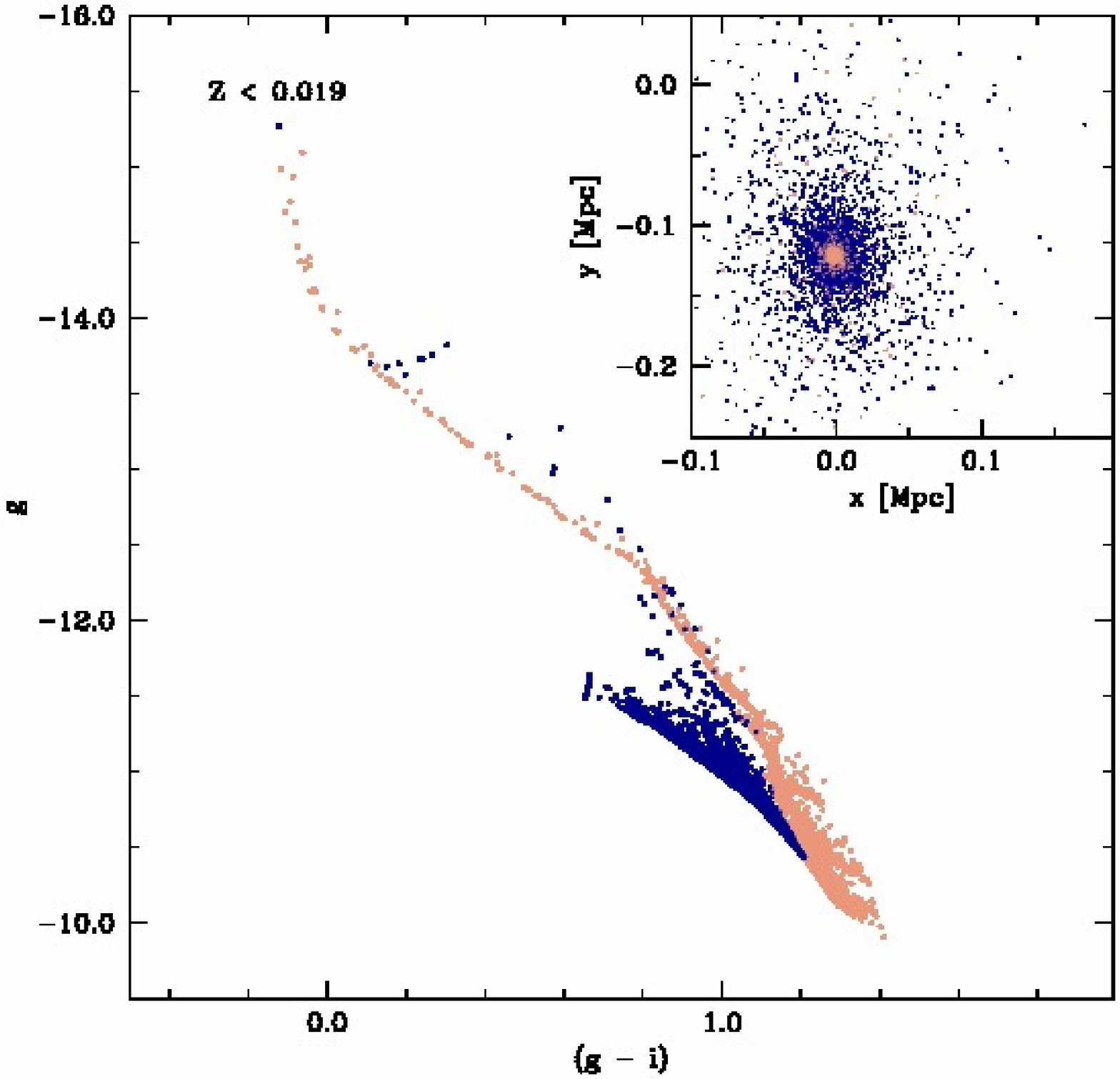}
\includegraphics[width=6cm,height=6cm]{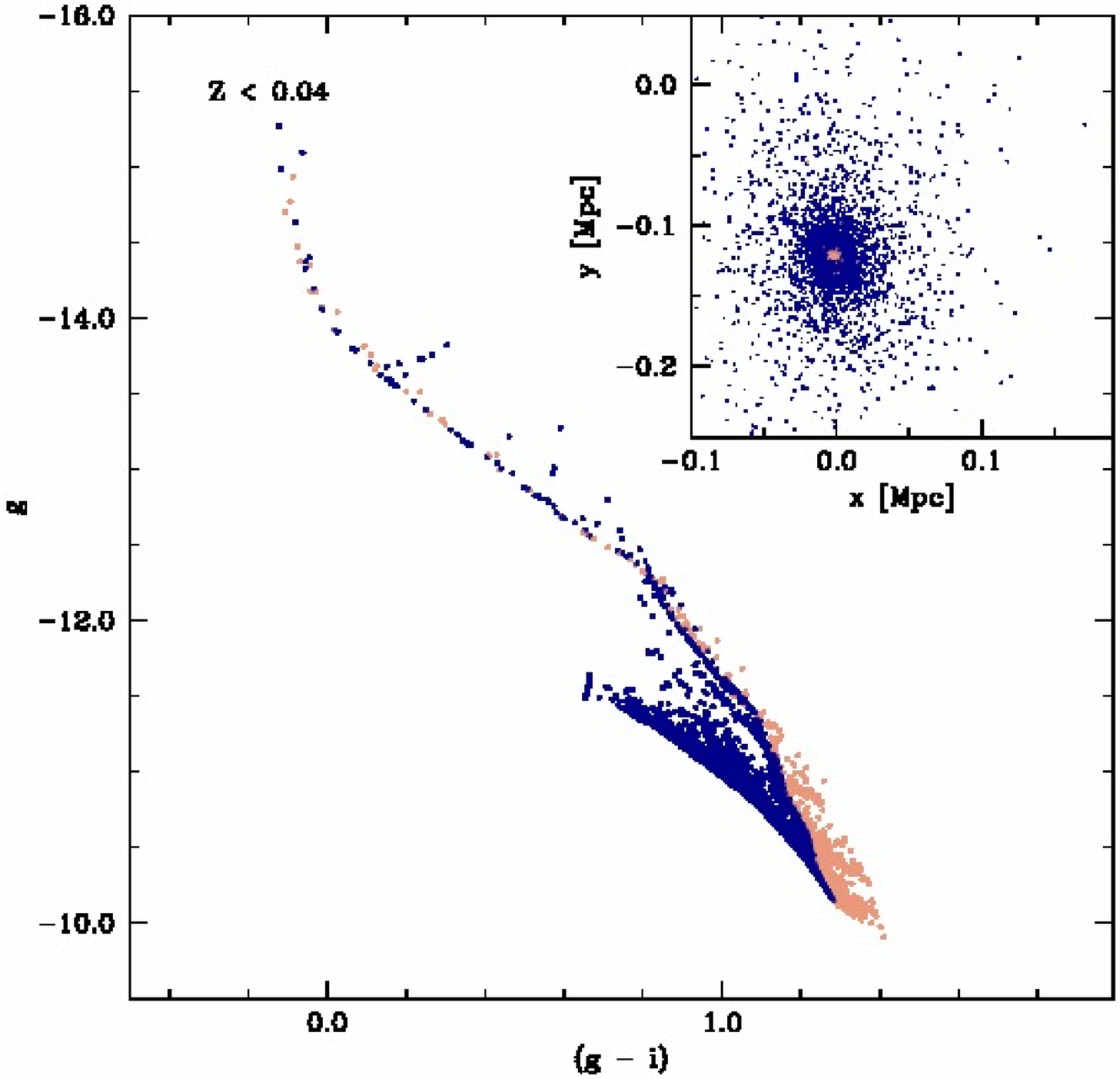}}
\caption[Distribution of stellar populations with different
   metallicity in the $(g-i)-g$ plane.]
   {Same as Fig.~\ref{cmd_age} but for stellar populations with
   different metallicity. As expected the stars of very low
   metallicity are in general very old, whereas at increasing
   metallicity stars of any age are possible.}
\label{cmd_met}
\end{center}
\end{figure*}

\begin{figure*}
\begin{center}
\centerline{
\includegraphics[width=6cm,height=6cm]{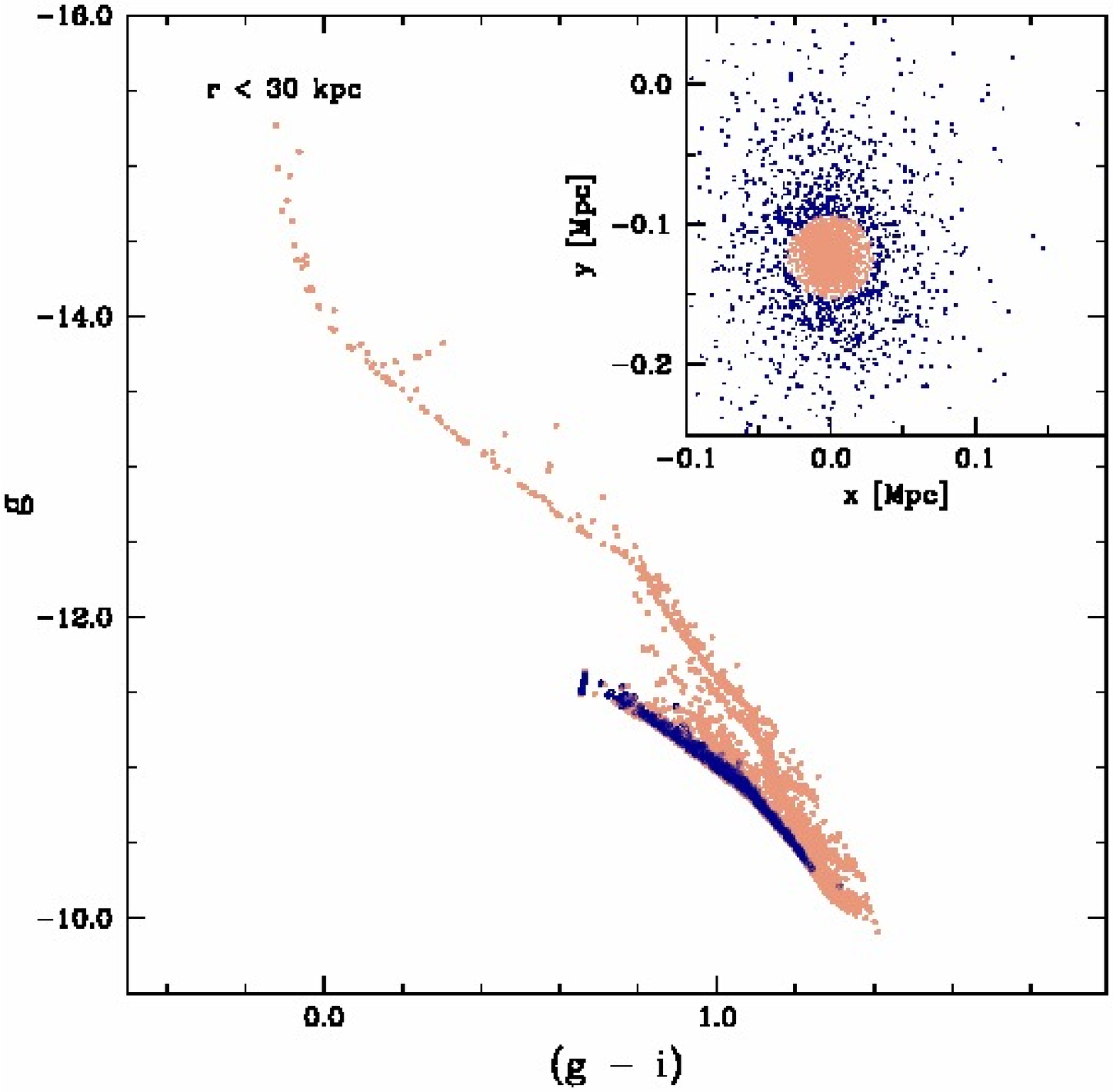}
\includegraphics[width=6cm,height=6cm]{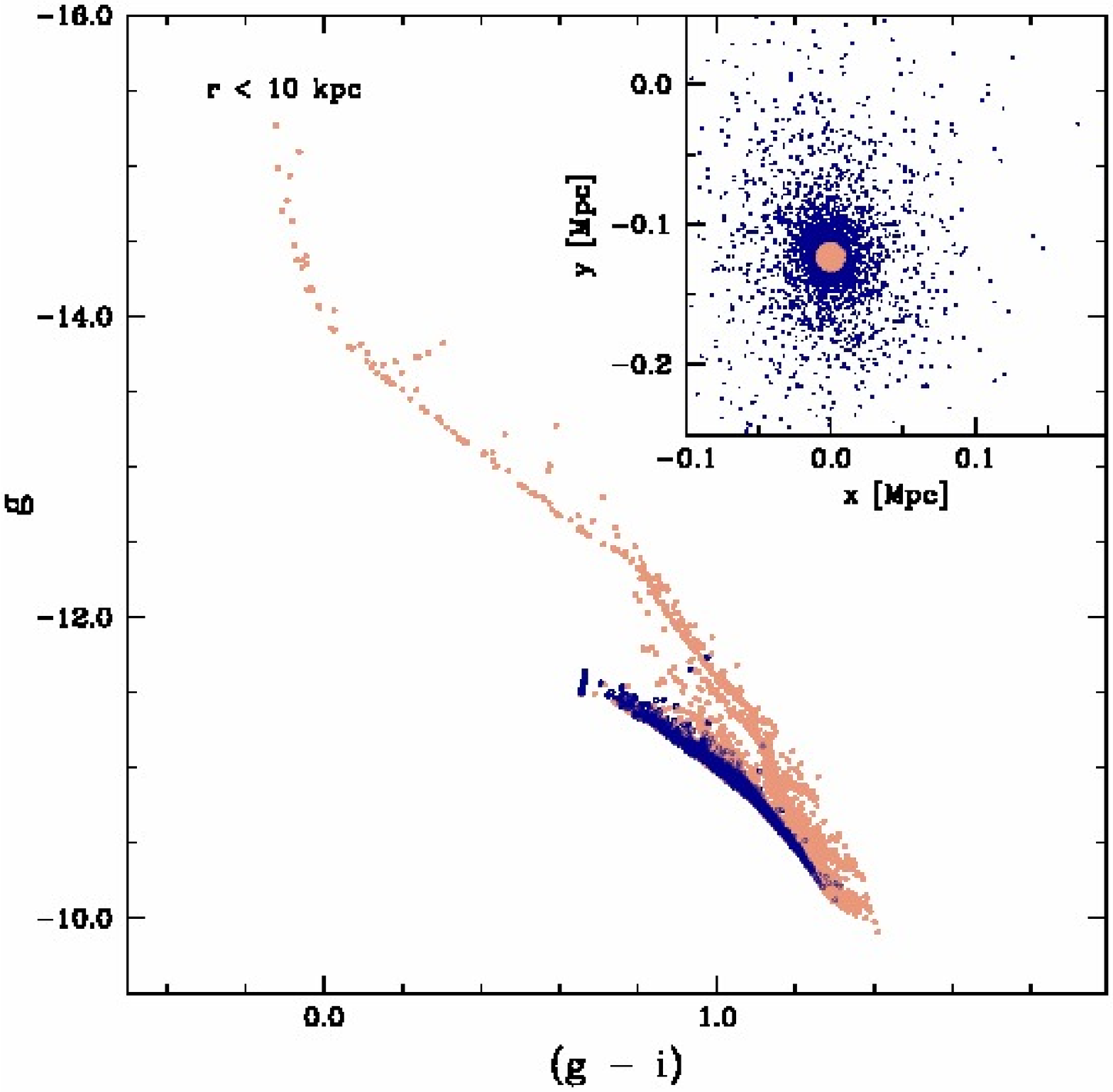}
\includegraphics[width=6cm,height=6cm]{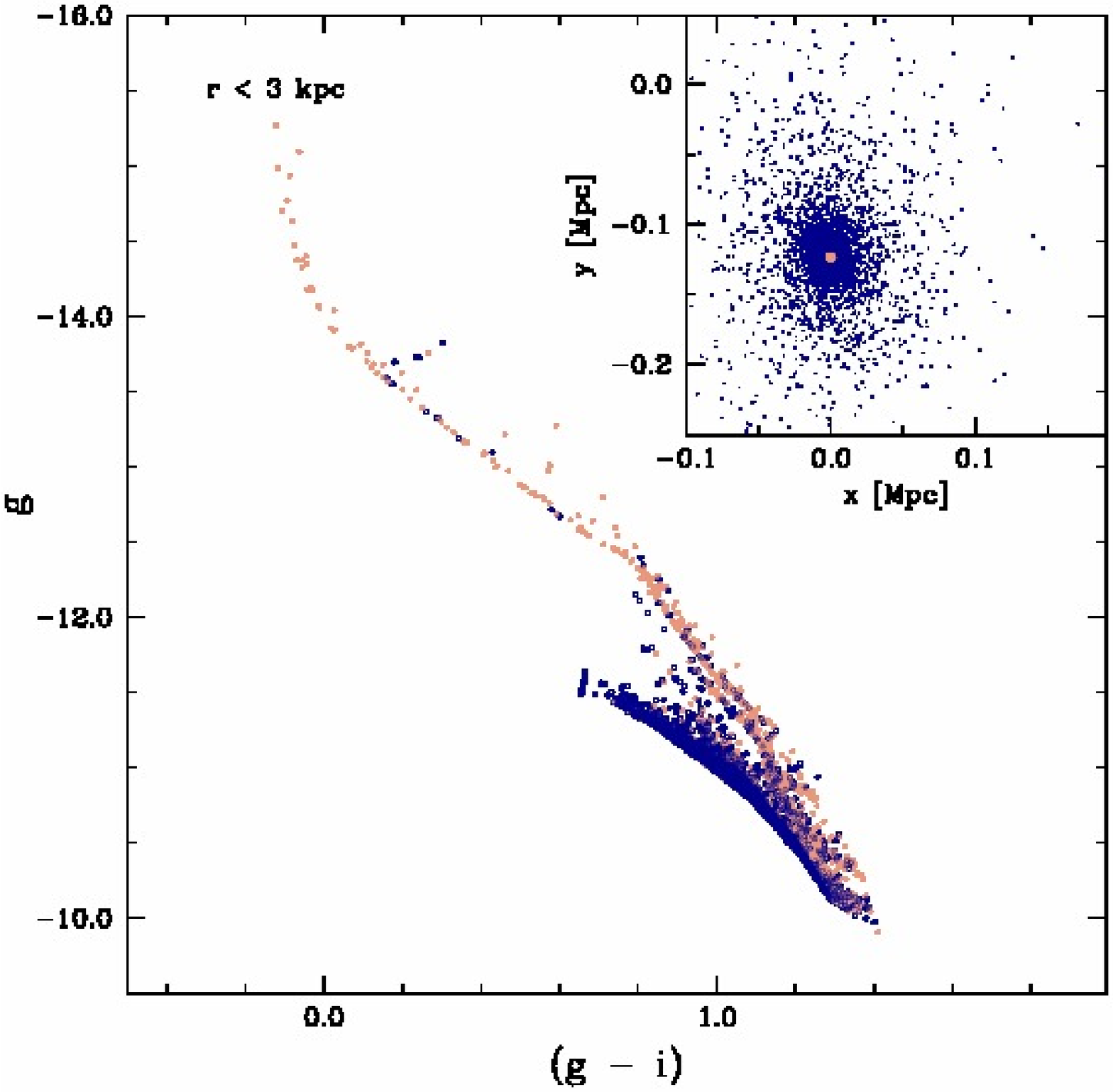}}
\caption[Distribution of stellar populations with different
   position in the galaxy in the $(g-i)-g$ planes]
   {Same as Fig.~\ref{cmd_age} but for stars of any age and
   metallicity but different locations in the galaxy}
\label{cmd_rad}
\end{center}
\end{figure*}

The CMDs of Fig.~\ref{cmd1_cmd2} show SSPs of different age and
metallicity, and the star-particles of the galaxy simulations (the
filled circles). To calculate the magnitude of the SSPs we have
assigned them the same mass of the star-particles in the NB-TSPH
simulations. The solid lines show the evolutionary path followed by
SSPs of different metal content as their integrated luminosity and
color change as a function of the age. Along each line the age goes
from 0.1\,Gyr to 14\,Gyr. This is the analog of the evolutionary
path followed by stars of given mass and chemical composition. The
dashed lines show SSPs of the same age (as indicated) and different
metallicity. Along each line the metallicity goes from Z=0.0001 to
Z=0.07. The CMDs allow us to catch immediately how the stellar
populations of a model galaxy (as represented by its star-particles)
distribute in age and metallicity. The analog of this situation for
real stars would be a CMD built up with the integrated magnitudes
and colors of the stellar clusters of a galaxy, for instance the
clusters of the LMC and SMC, the only difference is that while real
clusters have different mass our star-particles are all with the
same mass. However, this is a point of minor relevance here.

In the (V--K) vs. V diagram we can see how star-particles distribute
at varying the metallicity: in the $SCDM$ galaxy at the age of
13\,Gyr the vast majority of star-particles are very old. The bulk
of stars distribute along the lines of very old ages and span all
the values of metallicity. This means that the (V--K) color tests
the metallicities differences, more than the age. However there is a
fraction of younger stars that tends to crowd the region comprised
between Z=0.019 (Z$_\odot$) and Z=0.070 (3.5Z$_\odot$).

In the (1550--V) vs V diagram, on the other hand, the bulk of the
stars distribute along the Z=0.019 line and have ages going from
very old to very young. Since the metallicity has a lower effect,
this diagram can be used to infer the gross age of the stellar
content of a galaxy.

To get a deeper insight of the whole problem, in Figs.~\ref{cmd_age},
from left to right, we show the distribution of the stellar
populations in the $g-(g-i)$ plane for the $SCDM$ model at the age of
13\,Gyr. In each panel, stars younger than a certain limit (that
varies from panel to panel as indicated) are plotted as light dots, the
remaining ones as dark dots. The insert in the upper right corner in
each panel shows the position on the xy projection plane of such stars
(the light and dark dots). As expected, owing to the residual star
formation activity (say after the first 5\,Gyrs) stars of younger and
younger age tend to concentrate toward the galactic center. Keeping
the same color-code, in Fig.~\ref{cmd_met}, we show how the stars of
different age distribute in metallicity. As expected the stars of very
low metallicity are in general very old, whereas at increasing
metallicity stars of any age are possible. Finally, in
Fig.~\ref{cmd_rad} we show how stars of different age and metallicity
spatially distribute within the galactic volume.

\begin{figure}
\begin{center}
\includegraphics[width=8cm,height=8cm]{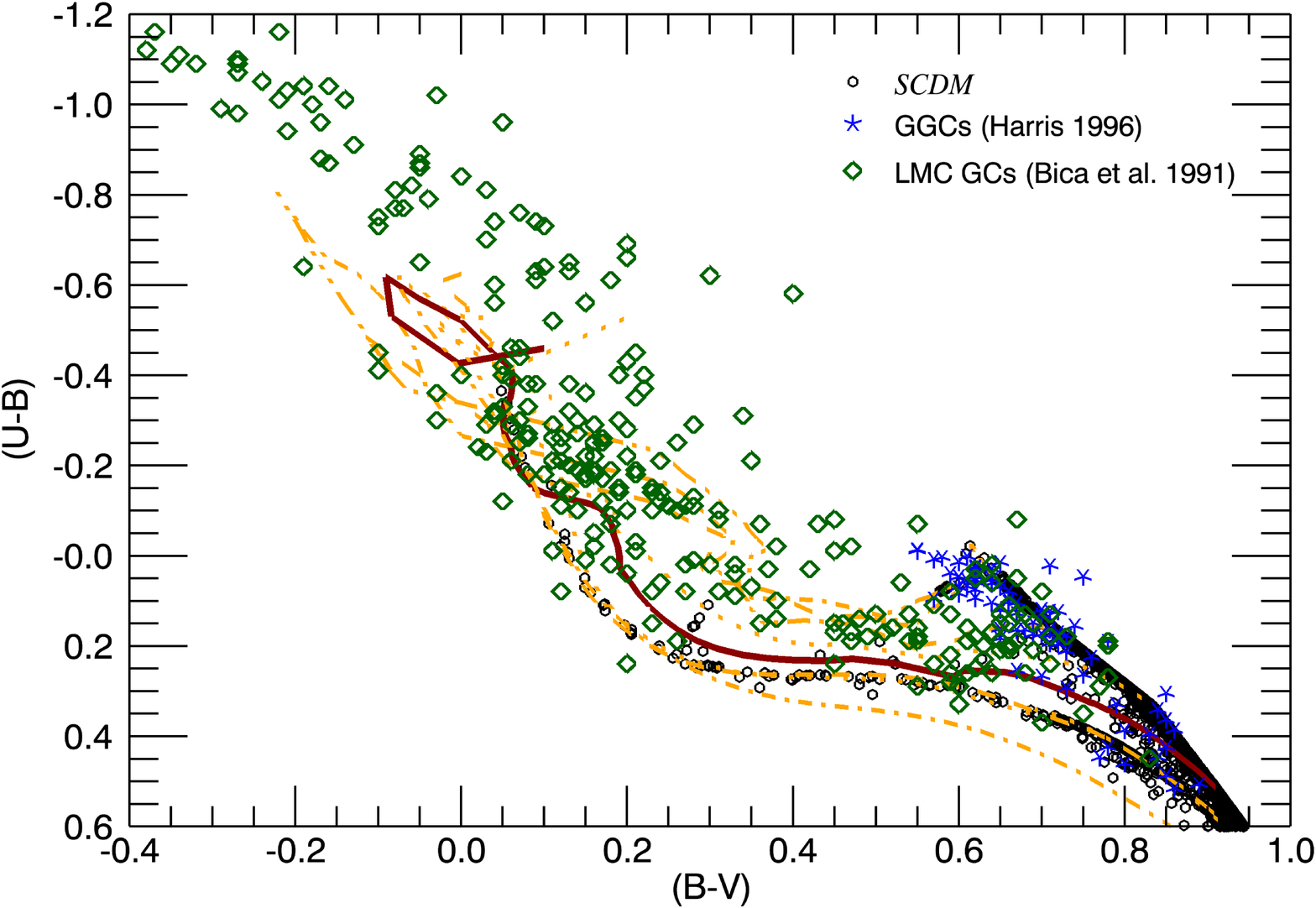}
\caption[Two color plane]
   {Integrated (U--B)$_{0}$ and (B--V)$_{0}$ colors of the LMC
   clusters by \citet{Bica91}, open rhombs; the galactic globular
   clusters by \citet{Harris96}, asterisks; the SSPs with different
   metallicity, dotted-dashed lines, the heavy solid line is the one
   with the metallicity typical of the LMC, i.e. Z=0.008); finally,
   the star-particles of the $SCDM$ galaxy model, the small open
   circles. Data and theoretical predictions seem to agree each other.}
\label{ub_bv}
\end{center}
\end{figure}

In order to validate the quality of our photometry, we compare the
colors of our SSPs and star-particles of the NB-TSPH simulation with
the colors of observed real stellar clusters. This is shown in
Fig.~\ref{ub_bv}, where we display: the integrated (U--B)$_{0}$ and
(B--V)$_{0}$ colors of the LMC clusters by \citet{Bica91} (open
rhombs); the Galactic Globular Clusters by \citet{Harris96} limited to
a few indicative cases (asterisks); the SSPs with different
metallicity indicated by the dotted dashed lines (the heavy solid line
is for Z=0.008, the typical mean metallicity of the LMC); finally the
star-particles of the $SCDM$ galaxy model (small open circles). Data
and theoretical predictions seem to agree each other but for the
youngest clusters of the LMC which tend to scatter above the line for
Z=0.008. However, this is less of a problem as a plausible explanation
has been advanced by \citet{Girardi95}. Therefore, the above
agreement between theory and data secures that our photometry is
carefully calculated.

\section{Cosmological Spectro-Photometric Evolution: Theory and Data}
\label{cosmev}

The advent of the modern giant telescopes has opened a new era in
observational cosmology and galaxy evolution can be traced back to
very early stages. In this context, deep multi-color imaging surveys
provide a powerful tool to access the population of faint galaxies
with relatively high efficiency. These surveys span the whole spectral
range from the UV to the near-IR bands, enabling galaxy evolution to
be followed on a wide range of redshifts. Therefore it is worth
looking at the cosmological evolution of our model galaxies and
compare it with modern data. Since galaxies are observed at different
redshifts in an expanding Universe, we need the so-called
$K$-correction and $E$-corrections that can be easily derived together
with magnitudes and colors from the population synthesis technique
\citep[see][]{Guiderdoni87,Roccavolm88,Guiderdoni88,Bressan94}.
Finally, we compared the colors of our models with those of two
sample of ETGs extracted from the COSMOS and GOODS databases.

\begin{figure}
\includegraphics[width=8.0cm,height=8.0cm]{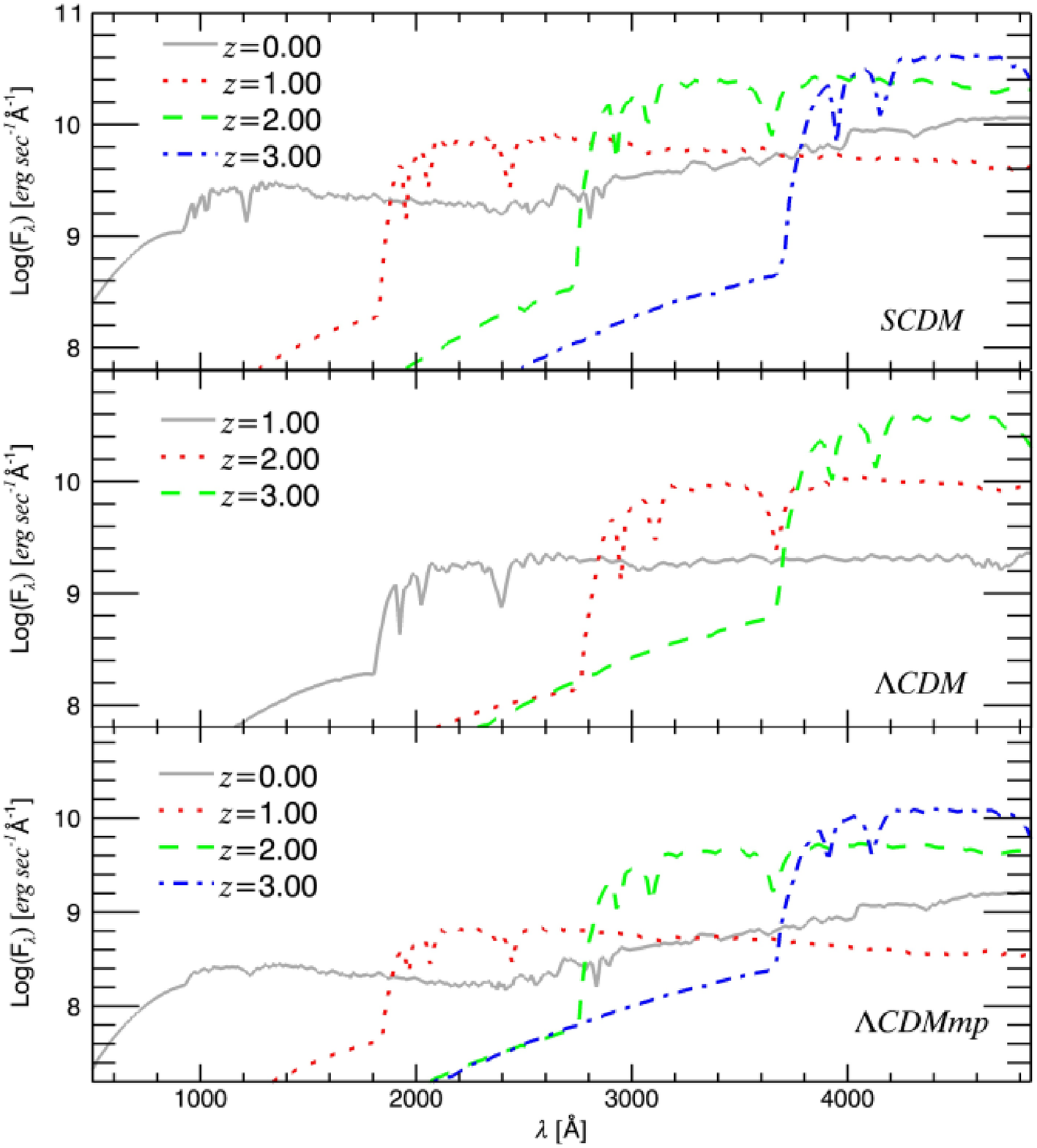}
\caption[Red-shifted spectra]
   {Red-shifted spectra at different redshifts for the three galaxy
   models. Internal extinction is taken into account.}
\label{spred}
\end{figure}

\subsection{Evolutionary and Cosmological Corrections}

When we consider a source observed at redshift $z$, we need to
remember that a photon observed at a wavelength $\lambda_{o}$ has been
emitted at wavelength $\lambda_{e}$. The two wavelengths are related
by

\begin{equation}
\lambda_{e} = \lambda_{o}/(1+z).
\end{equation}

A source with apparent magnitude $m$ measured in a photometric
passband, is related to the absolute magnitude $M$, in the
emission-frame passband, through the cosmological correction,
$K_{corr}$, in the following way \citep{Hubble36}:

\begin{equation}
\label{mR}
m = M + DM + K_{corr},
\end{equation}

\noindent
where $DM$ is the distance modulus, defined by

\begin{equation}
DM =5 \log_{10}\left[ \frac{D_{L}(z)}{10 pc} \right],
\end{equation}

\noindent
being $D_{L}(z)$ the luminosity distance.

The above luminosity distance has been calculated with the same
cosmology of the simulated galaxies ($\Omega_{M}$, $\Omega_{\Lambda}$,
$h_{0}$). In particular, we have adopted the following equation
\citep[see][ for all details]{Weinberg72,Hogg99,Kolb00}:

\begin{equation}
D_{L}(z)= \frac{c}{H_{0}}(1+z)\int_{0}^{z}\frac{dz}
{\bigg[\Omega_{M}(1+z)^{3} + \Omega_{\Lambda} \bigg]^{\frac{1}{2}}}
\end{equation}

If the source is at redshift $z$, then its luminosity is related to
its spectral density flux (energy per unit time, unit area, and unit
wavelength) by

\begin{equation}
L(\lambda_{e}) = 4 \pi (1+z) D_{L}^{2} f(\lambda_{0}),
\end{equation}

\noindent
where $f(\lambda_{0})$ is the monochromatic flux of a galaxy that has
been computed as defined in eqn.~\ref{flamgal}.

\noindent
Finally, the $K_{corr}$ in eqn.~\ref{mR} is

\begin{equation}
K_{corr} = 2.5 \log_{10} (1+z) + 2.5 \log_{10}
\left[ \frac{L(\lambda_{0})}{L(\lambda_{e})} \right]
\end{equation}

\noindent
\citep[see the definition by][]{Oke68}. This means that to
make a fair comparison between objects at different redshifts, we must
derive the rest-frame photometric properties of our observed galaxies
(magnitudes, colors, etc.) by applying $K$-corrections.

In addition, we must also correct these rest-frame quantities for the
expected evolutionary changes over the redshift range studied, by
applying the so called evolutionary corrections, $E_{corr}$. The
$E_{corr}$ are usually derived assuming a model for the galaxy SED and
calculating its evolution with the redshift. In this way we can
recover the evolution of the absolute magnitudes and colors as a
function of the redshift $z$, including the effect of the $K$- and
$E$-corrections on the SED of our models.

Following \citet{Guiderdoni87}, the cosmological $K(z)$ and
evolutionary $E(z)$ corrections are conventionally given in terms of
magnitude differences:

\begin{equation}
K(z) = M(z,t_{0}) - M(0,t_{0}),
\end{equation}

\begin{equation}
E(z) = M(z,t_{z}) - M(z,t_{0}),
\end{equation}

\noindent
where $M(0,t_{0})$ is the absolute magnitude in a passband derived
from the rest frame spectrum of the galaxy at the current time,
$M(z,t_{0})$ is the absolute magnitude derived from the spectrum of
the galaxy at the current time but red-shifted at $z$, and
$M(z,t_{z})$ is the absolute magnitude obtained from the spectrum of
the galaxy at time $t_{z}$ and red-shifted at $z$.

From eqn.~\ref{mR} the apparent magnitude, in some broad-band filter
and at redshift $z$, is given by:

\begin{equation}
m(z) = M(z) + E(z) +  K(z) + DM(z).
\end{equation}

Obviously, the relation $t=t(z)$, between the cosmic time $t$ and the
redshift $z$ of a stellar population formed at a given redshift
$z_{f}$, depends on the cosmology considered and the parameters
adopted. Following \citet{Kolb00}:

\begin{equation}
t(z)= \frac{1}{H_{0}}\int_{z}^{\infty}\frac{dz}{(1+z) \bigg[
\Omega_{M}(1+z)^{3} + \Omega_{\Lambda} \bigg]^{\frac{1}{2}}}
\end{equation}

\subsection{Extinction}

Before calculating the $E_{corr}$, it is worth applying to the
theoretical SEDs the effect of extinction of the stellar luminosity
caused by the presence of a certain amount of metal-rich gas so that
the SEDs get closer to the real ones. Although the task is a
complicate issue requiring a careful analysis
\citep{Piovan03,Piovan06,Piovan06b}, for the purposes of the present
study, the effect of extinction can be evaluated using the relation
proposed long ago by \citet{Guiderdoni87}:

\begin{equation}
\tau_{\lambda} = 3.25(1-\omega_{\lambda})^{0.5}
(A_{\lambda}/A_{V})_{\odot} [Z(t)/Z_{\odot}]^{1.35} G(t),
\end{equation}

\noindent
where $\tau_{\lambda}$, the effective optical thickness of the gaseous
component at a given $\lambda$, is a function of: (i) the albedo
$\omega_{\lambda}$ of the grains for which the  mean values of 0.4
taken from \citet{Draine84} has been used; (ii) the extinction law
$A_{\lambda}/A_{V}$ \citep{Cardelli89}; (iii) and finally the metallicity
$Z(t)$ and gas fraction $G(t)$.

The monochromatic flux of the galaxy with the inclusion of the
effect due to extinction, $F_{\lambda,ext}$, can be expressed in
function of the monochromatic flux of the rest-frame SED of the
model galaxy $F_{\lambda}$ (see eqn.~\ref{flamgal}):

\begin{equation}
F_{\lambda,ext} = F_{\lambda} \frac{1-\exp(-\tau_{\lambda}
\sec i)}{\tau_{\lambda} \sec i},
\label{factor_extin}
\end{equation}

\noindent
where the right-hand part of the expression takes into account the
transmission function for an angle of inclination $i$; we adopt here
$i=45$. Although this relation was originally derived for disk
galaxies, it can be safely used also in our case.

\begin{figure}
\includegraphics[width=8.0cm,height=8.0cm]{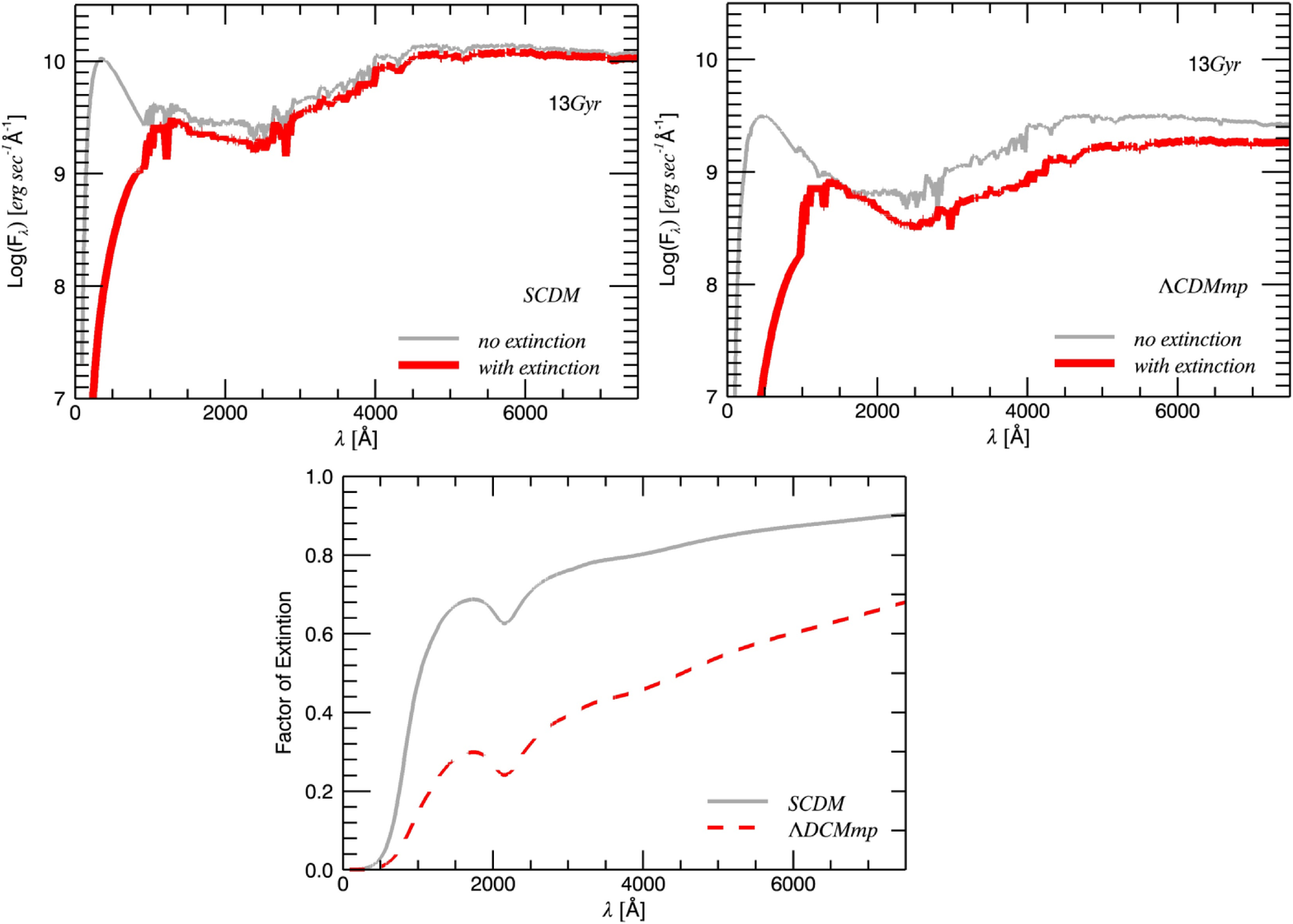}
\caption[Red-shifted spectra]
   {{\bf Top panels:} comparison between the SEDs of the $SCDM$
   and $\Lambda CDM_{mp}$ models at the age of 13\,Gyr with
   extinction (thick line) and without extinction (thin line). {\bf Bottom panel:}
   the extinction factor for the two models as indicated.}
\label{extin}
\end{figure}

The effect of extinction is included in our SEDs using $Z(t)$ and
$G(t)$ obtained from the NB-TSPH simulations. Internal extinction may
significantly redden the colors, the effect being particularly
important on the color-redshift relation. To illustrate the point, in
Fig.~\ref{extin} we compare the SEDs of the $SCDM$ and $\Lambda
CDM_{mp}$ models at the age of 13\,Gyr with and without extinction, as
indicated. First of all the two SEDs are different even neglecting
extinction (they reach indeed different levels of flux), see the top
panels of Fig.~\ref{extin}. This simply reflects the final lower mass
in stars of the $\Lambda CDM_{mp}$ model with respect to $SCDM$ (a factor
of three lower). Second, the effect of extinction is different in the
two models (top panels of Fig.~\ref{extin}). This simply reflects the
different metallicity and gas content at the age of 13 Gyr. These are
$Z(t)= 0.0214$ and $G(t)=0.257$ in $SCDM$ and $Z(t)=0.0513$ and
$G(t)=0.441$ in $\Lambda CDM_{mp}$. The factor
$[{1-\exp(-\tau_{\lambda} \sec i)}/{\tau_{\lambda} \sec i}]$ of
eqn. (\ref{factor_extin}) with $i=45^{\circ}$ is shown in the bottom
panel of Fig.~\ref{extin}.

\begin{figure}
\begin{center}
\includegraphics[width=8cm]{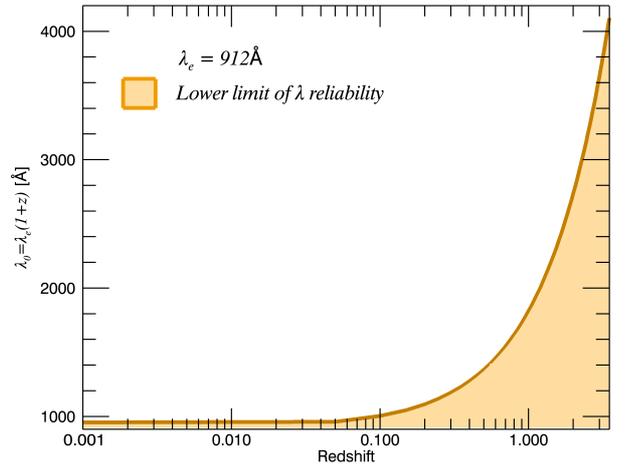}
\caption[912\AA\ Wavelength and its variation]
   {Reliability of magnitudes and colors as function of the redshift
   is because the theoretical spectra in the population
   synthesis algorithm do not extend at wavelengths shorter than
   $\lambda =$912\AA\ and replaced by black-body spectra
   (approximation).}
\label{lambdaev}
\end{center}
\end{figure}

There is a final point to consider, i.e. the intrinsic reliability
of the magnitudes and colors derived from SEDs as a function of the
redshift. To illustrate the point, in Fig.~\ref{spred} we display
the red-shifted spectrum with extinction of the models for some
values of $z$. The spectra show a drastic change in the slope that
occurs at a certain wavelength whose value increases with the
redshift. This effect is because the rest-frame
theoretical spectra have a lower limit of 912\AA\ and that the
extension to $\lambda < $912\AA\ has been made by simply assuming
black-body spectra. The real spectrum short-ward of 912\AA\ could be
different from a pure black-body. The effect of this approximation
should be taken into account in the computation of the colors in
any photometric system. In other words, magnitudes and colors that
contain flux originated in the $\lambda < $912\AA\ interval become
more and more uncertain at increasing redshift. This is illustrated
in Fig.~\ref{lambdaev} which displays the quantity $\lambda_{obs}=
\lambda_{em} (1+z)$. The shaded area shows the region of intrinsic
uncertainty due to the above effect. It emerges from this that at
redshift of $z=3$ (observational data in the surveys considered reach
this redshift) magnitudes are considered accurate at wavelengths
$\lambda > $4100\AA.

\subsection{Comparison with Data}

The advent of large-scale space and ground-based surveys in a wide
range of wavelengths is giving us unprecedented access to
statistically large populations of galaxies at different redshifts
(and also environments). The practical use of these immense databases
requires some caution as far as galaxy detections, redshift
assignment, galaxy classification, and galaxy selection are
concerned. Prior to anything else it is worth recalling that owing to
the enormous amounts of data to handle, the data acquisition process
is usually made following automatic procedures that deserve some
remarks.

\noindent
\textsf{Detection}.
At $z>5$, traditional optical bands, e.g. UBVR, fall below the
rest-frame wavelength that corresponds to the Lyman-break spectral
feature (1216\AA), where most of the stellar radiation is extinguished
by interstellar or intergalactic hydrogen. Because of this, galaxies
at $z>5$ are practically invisible at those photometric bands, and
even if they were detected, their colors would provide very little
information about their stellar population. The color selection
technique, e.g. the UGR selection of Lyman-Break Galaxy (LBGs) by
\citet{Steidel96,Steidel99} and \citet{Steidel03}, has been used in
some surveys to identify galaxies at high redshift, dramatically
improving the efficiency of spectroscopic surveys at $z>3$.

\noindent
\textsf{Redshift assignment}.
Photometric redshifts are the logical extension of color selection by
estimating redshifts and SEDs from many photometric bands. Unlike
color selection, photometric redshifts take advantage of all
available information, enabling redshift estimates along with the age,
star formation rate and mass.

\noindent
\textsf{Morphological classification}.
One of the main characteristics of deep photometric surveys is the
richness of detected objects, where a significant fraction of them
appear as point sources that cannot be neither easily distinguished
from real stars nor morphologically classified. Therefore, the
classification by means of morphological and photometric criteria is a
crucial issue. In this paper, we consider two deep catalogues that
allow us to select good samples of ETGs.

\noindent
\textsf{Selection}.
Morphological selection of ETGs can be made using automated pipelines
that isolate objects on the basis of their two-dimensional light
distributions: this is the case of the COSMOS survey (see below). On
the other hand, in the case of GOODS, it is possible to select more
accurately these objects by correlating a catalogue of photometric and
spectroscopic redshifts with a morphological one.

\subsection{COSMOS Survey}
\label{cosmsurv}

In this analysis, we use the Cosmic Evolution Survey - COSMOS official
photometric redshift catalogue \citep{Scoville07}, designed to probe
the evolution of galaxies in the context of their large scale
structure out to moderate redshift. Details of the COSMOS catalogue
are described in \citet{Capak07} and \citet{Mobasher07}. It covers a 2
square degree area with deep panchromatic data and includes objects
whose total $i$ magnitudes ($i^{+}$ or $i^{\ast}$) are brighter than
25.

The COSMOS multi-band catalogue embraces data from different
telescopes, as listed in Table~\ref{tabcosmos}, and presents imaging
data and photometry that cover various photometric bands between
0.3$\mu m$ and 2.4$\mu m$. The catalogue was generated using
SExtractor \citep{Bertin96} and contains photometry measured over 3
arcsec diameter apertures for all the bands. All magnitudes are in the
AB system. The cosmology adopted is: $H_{0}=70${\it km/s/Mpc},
$\Omega_{M}=0.3$ and $\Omega_{\Lambda}=0.7$. It also contains
photometric redshifts, 68 and 95 percent confidence intervals, and
spectral types calculated with two different packages: the
\citet{Mobasher07} and the Bayesian Photometric Redshift
\citep[BPZ by][]{Benitez00}.

Such a large database needs an automated and objective morphological
classification procedure separating ETGs from other objects. The goal
is achieved using the parameter $T_{phot}$, which is based on the
spectral type. Objects with $T_{phot} \leq 1.5$ correspond to ETGs,
those with $T_{phot} > 1.5$ to all the remaining types.

\begin{table*}
\begin{center}
\caption[COSMOS Survey dataset]{{COSMOS Survey: Telescopes and optical/IR bands}}
\label{tabcosmos}
\vspace{1mm}
\footnotesize{
\begin{tabular*}{110mm}{l l c}
\hline
\multicolumn{1}{c}{Telescope} &
\multicolumn{1}{c}{Filters} &
\multicolumn{1}{c}{Band Width (in \AA) }\\
\hline
CFHT    & $u^{\ast}$, $i^{\ast}$                                               & 3200-1000  \\
CTIO    & $Ks$                                                                 & 9000-25000 \\
HST-ACS & F814W (i-band)                                                       & 4000-11000 \\
KPNO    & $Ks$                                                                 & 9000-25000 \\
SDSS    & $u$, $g$, $r$, $i$, $z$                                              & 3200-11000 \\
Subaru  & $B_{J}$, $V_{J}$, $g^{+}$, $r^{+}$, $i^{+}$, $z^{+}$, NB816 (i-band) & 4000-11000 \\
\hline
\end{tabular*}}
\end{center}
\end{table*}

\begin{figure}
\begin{center}
\includegraphics[width=8cm,height=8cm]{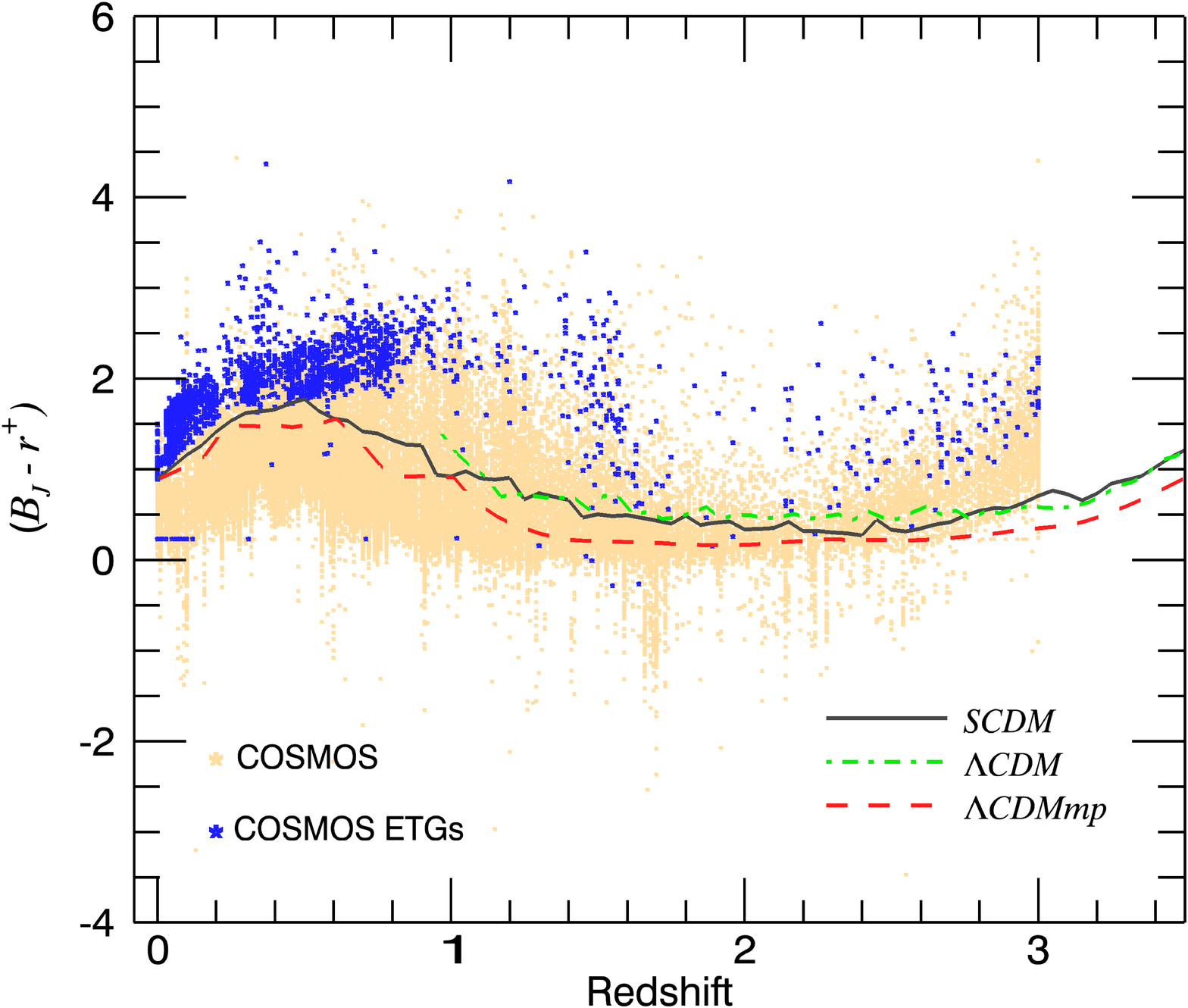}
\caption[Evolution with redshift: COSMOS $B_{J}-r^{+}$]
   {Cosmological evolution with redshift for the ($B_{J}-r^{+}$)
   color of the COSMOS survey. Both passbands are those of the
   Subaru Telescope. All galaxies of the catalog are shown in light stars.
   The ETGs selected with the pipeline morphological $T_{phot}<1.1$
   parameter are marked in dark stars. The galaxy models for the three
   different cosmological scenarios are shown superimposed to the
   data, continuous and dotted lines as labelled. The $\Lambda CDM$
   case is shown for $z > 1$.}
\label{cosmbr}
\end{center}
\end{figure}

\begin{figure}
\begin{center}
\includegraphics[width=8cm,height=8cm]{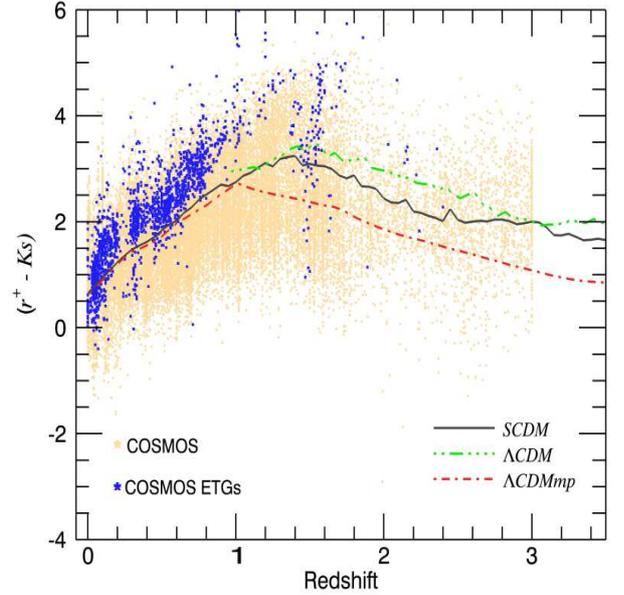}
\caption[Evolution with redshift: COSMOS $r^{+}-Ks$]{Same as
     Fig.~\ref{cosmbr} but for the ($r^{+}-Ks$) colors ($r^{+}$ band
     from Subaru and $Ks$ band from KPNO).}
\label{cosmrk}
\end{center}
\end{figure}

\begin{figure}
\begin{center}
\includegraphics[width=8cm,height=8cm]{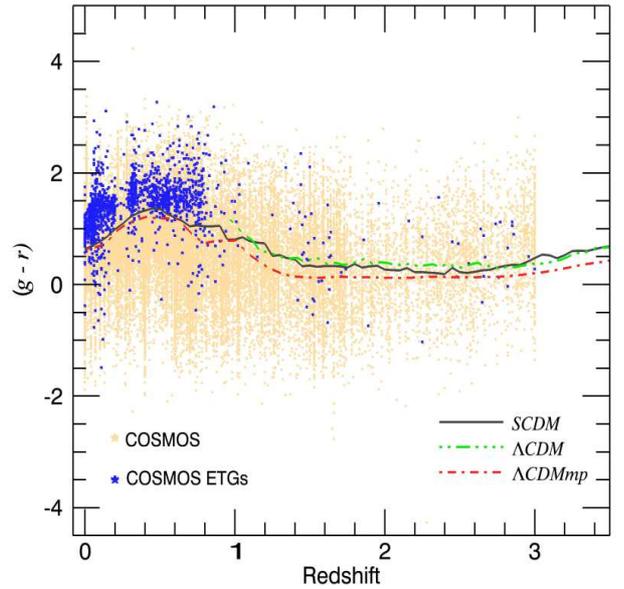}
\caption[Evolution with redshift: COSMOS $g-r$]
     {Same as Fig.~\ref{cosmbr} but for the ($g-r$) color (SDSS
     survey passbands).}
\label{cosmgr}
\end{center}
\end{figure}

For our model galaxies we generate magnitudes and colors in the
same photometric system of COSMOS, and calculate the cosmological
and evolutionary corrections suited to the cosmological background
in use. Colors and their cosmological evolution in different bands
are shown in Figs.~\ref{cosmbr}, \ref{cosmrk}, and \ref{cosmgr},
where the $B_{J}-r^{+}$ (Subaru), $r^{+}-Ks$ (Subaru/KPNO), and the
$g-r$ (SDSS) colors are plotted as an example. It is worth calling
attention that the model $\Lambda CDM$ is plotted from $z_{ini}$
down to $z\sim 1$ because it has been stopped at the age of 7\,Gyr.
For the sake of comparison, first we display all the galaxies from
the survey, independently of their classification (light stars) and,
superposed to them, we mark in dark stars the sub-sample of ETGs selected
following the classification suggested by the automated pipeline
($T_{phot} \leq 1.5$). The photometric evolution for our three model
galaxies is also shown: the solid line is for the $SCDM$, the
dotted-dashed line is for the $\Lambda$CDM model, and the dotted
line is for the $\Lambda CDM_{mp}$ case.

The models follow the general trend of the observations and, in
particular, are marginally consistent with the group of ellipticals up
to $z\sim 1$ beyond which the data are too poor to say anything.  The
observed ellipticals are indeed redder than the mean value of the data
and theoretical predictions. Concerning the theoretical values, their
bluer colors can be ascribed to the tail of star formation extending
to the present. Although this minor stellar activity does not
significantly affect the gross features of the models (structure, mass
distributions etc.) it certainly affect the colors making them bluer
than desired and expected. This secondary star formation activity is
likely a spurious effect (work is progress to cope with this). Another
point of uncertainty could reside in the selection criteria to
identify ETGs.

\subsection{GOODS Database}
\label{sectgoods}

For the sake of comparison, we consider now the Great Observatories
Origins Deep Survey - GOODS database \citep{Giavalisco04}. The survey
is based on the observations of two separate fields centered on the
Hubble Deep Field North (HDFN) and Chandra Deep Field South (CDFS) and
includes ultra-deep images from ACS on HST, from mid-IR satellite
Spitzer, as well as from a number of ground-based facilities (see
Table~\ref{tabgoods}).

Galaxies exhibit a range of morphologies that are difficult to
determine automatically, so a manual classification is often used to
test the efficacy of automated classifiers. For this reason, in
order to select ETGs from the database we have cross-correlated two
catalogues: the GOODS - Multi-wavelength Southern Infrared Catalogue
(GOODS-MUSIC) by \citet{Grazian06} to determine the redshift and
the one by \citet{Bundy05} to fix the morphology. These are good
catalogues to rely on.  The first one indeed contains redshifts of
high precision. In fact the GOODS-MUSIC database \citep{Grazian06}
comprises photometric and spectroscopic information for galaxies in
the GOODS Southern Field. For these objects they find excellent
agreement between photometric and spectroscopic redshifts over the
range $0<z<6$ \citep[see Fig.~12 in][ for the $z_{spec}-z_{phot}$
relation]{Grazian06}.

\begin{table*}
\begin{center}
\caption[GOODS database]{{GOODS database: Telescopes and optical/IR bands}}
\label{tabgoods}
\vspace{1mm}
\footnotesize{
\begin{tabular*}{145mm}{l l c}
\hline
\multicolumn{1}{c}{Telescope} &
\multicolumn{1}{c}{Filters used} &
\multicolumn{1}{c}{Band Width (in \AA)}\\
\hline
ESO-WFI                   & $U_{38}$                                       & 3100-4000    \\
VLT-VIMOS                 & $U_{VIMOS}$                                    & 3300-4000    \\
ACS-HST                   & $B(F435W)$ $V(F606W)$, $i(F775W)$, $z(F850LP)$ & 3400-11000   \\
VLT-ISAAC                 & $J_{ISAAC}$, $H_{ISAAC}$, $Ks_{ISAAC}$         & 11000-24000  \\
Spitzer - IRAC instrument & 3.6$\mu$, 4.5$\mu$, 5.8$\mu$, 8$\mu$           & 30000-100000 \\
\hline
\end{tabular*}}
\end{center}
\end{table*}

In the second one, \citet{Bundy05} present a morphological catalogue
of galaxies in the GOODS North and South Fields, for which the
morphological classification has been made by hand, therefore more
being reliable than the one derived from automated procedure. The
study relies on the combination of many different data sets in the
GOODS fields including infrared observations as listed in
Table~\ref{tabgoods}, spectroscopic and photometric redshifts, and
HST morphologies. The catalogue contains objects with a magnitude
limit based on HST-ACS imaging data released by the GOODS team
\citep{Giavalisco04}. A $z$-band selected catalogue was constructed
running SExtractor \citep{Bertin96} and considering a magnitude
limit of $z_{AB}<22.5$, where reliable visual morphological
classification was deemed possible. All magnitudes are defined in
the ABmag system and they assume a cosmology with $\Omega_{M}=0.3$,
$\Omega_{\Lambda}=0.7$, and $H_{0}=70$. The resulting sample of
objects over both GOODS fields was inspected visually by
\citet{Bundy05} who classified each one of them, by using the
technique discussed in \citet{Brinchmann00}, according to a scale
that separates stars from compact objects and galaxies of different
morphological type.

We cross-correlated the two catalogues described above to recover a
data set of galaxies with reliable morphological classification and
precise redshift determination. Therefore these samples of galaxies
should be considered much better selected than the sample derived from
COSMOS. Standing on those arguments, a final sub-sample of ETGs is
selected from the complete database. The total amounts to $118$
objects.

\begin{figure}
\begin{center}
\includegraphics[width=8cm,height=8cm]{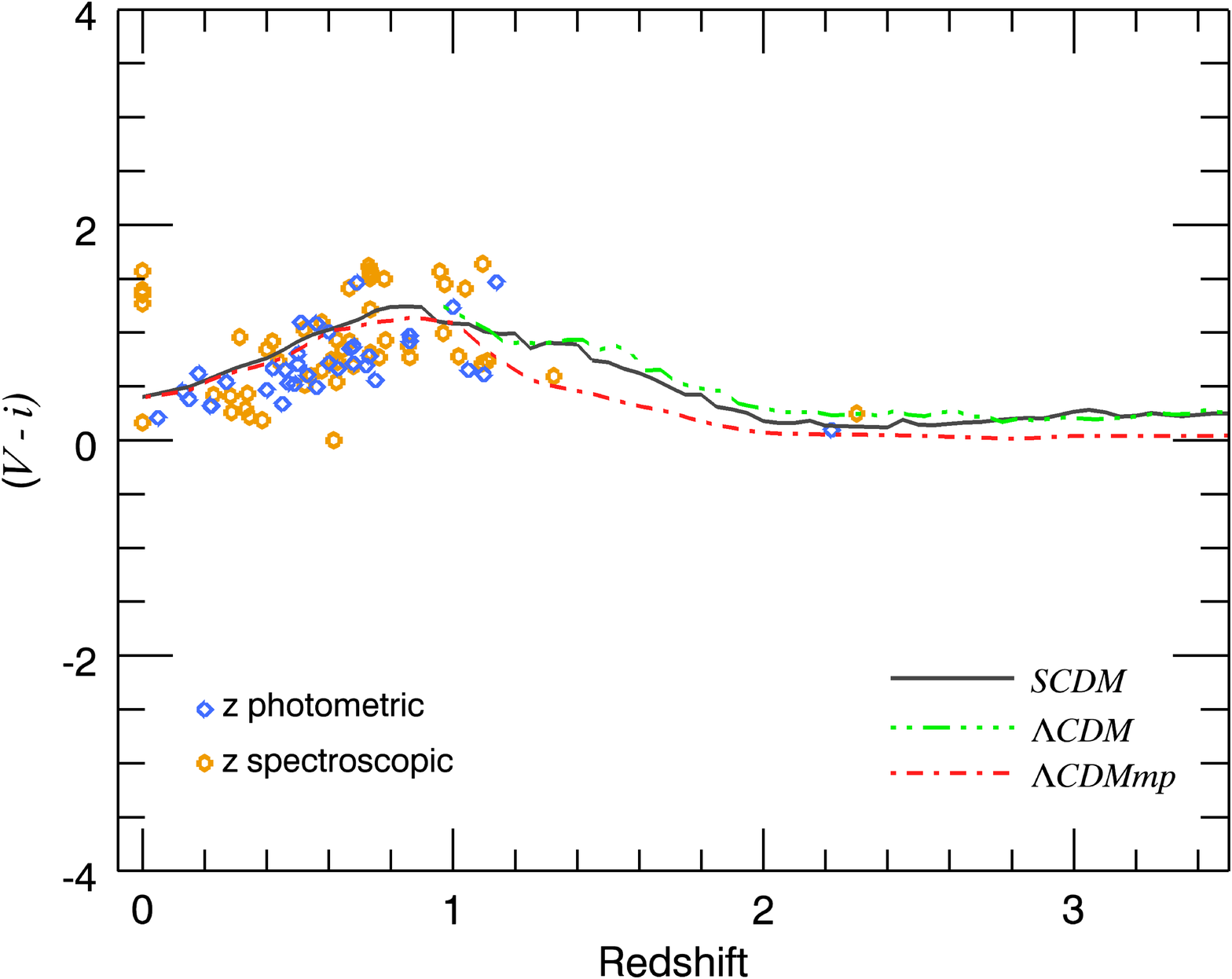}
\caption[Evolution with redshift: GOODS $V-i$]{Cosmological evolution with
   redshift of the $V(F606W)-i(F775W)$ colors of the GOODS survey
   (ACS-HST passbands) for early-type galaxies with spectroscopic
   (empty circle) and photometric (filled circles)
   redshift determination as indicated. The galaxy models for the
   three different cosmological scenarios are shown superimposed to
   the data, continuous and dotted lines as labelled. The $\Lambda
   CDM$ case is shown for $z > 1$.}
\label{goodsvi}
\end{center}
\end{figure}

\begin{figure}
\begin{center}
\includegraphics[width=8cm,height=8cm]{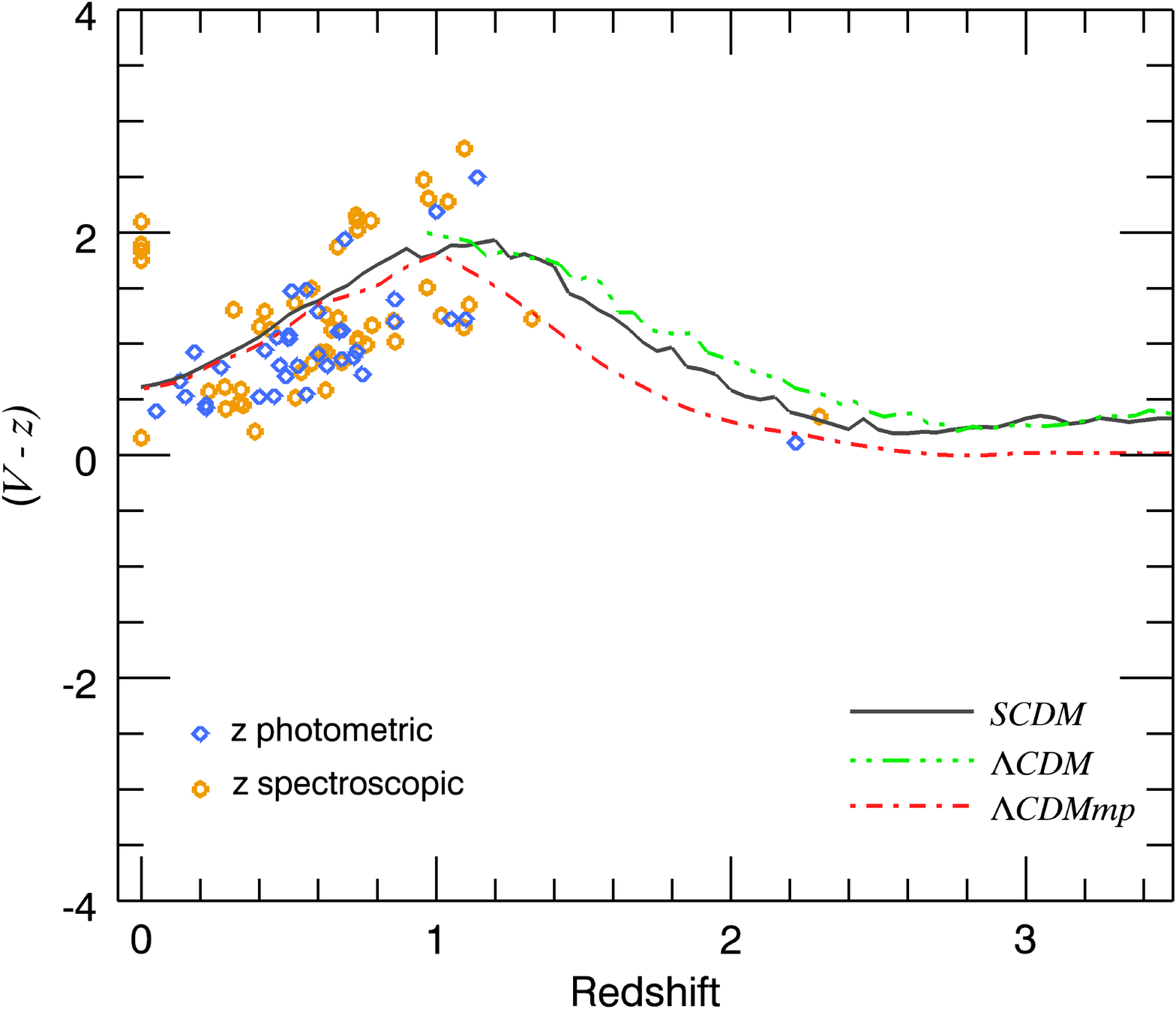}
\caption[Evolution with redshift: DS $V-z$]
   {Same as Fig.~\ref{goodsvi} but for the $V(F606W)-z(F850LP)$
   colors (ACS-HST passbands).}
\label{goodsvz}
\end{center}
\end{figure}

\begin{figure}
\begin{center}
\includegraphics[width=8cm,height=8cm]{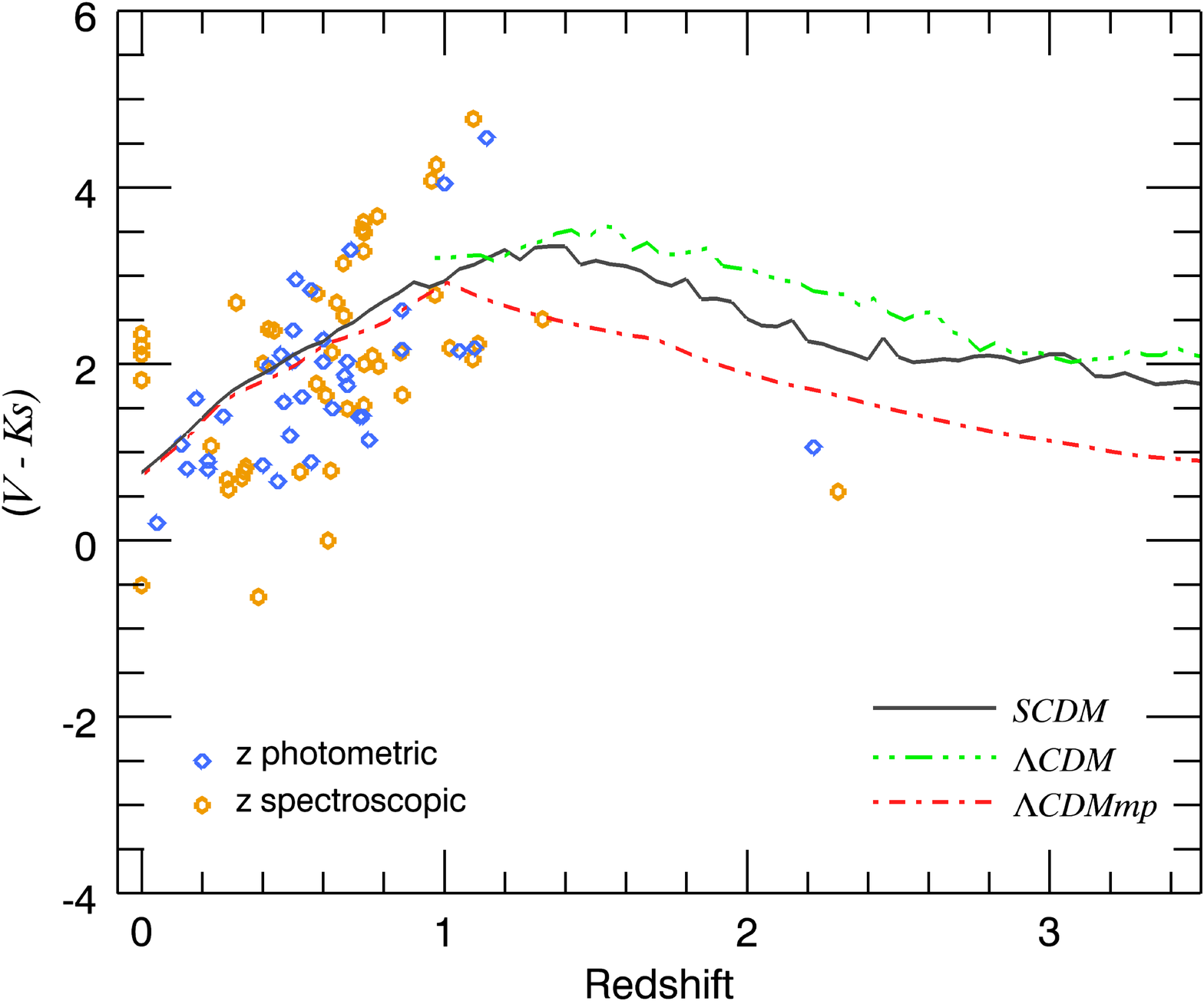}
\caption[Evolution with redshift: GOODS $V-Ks$]
   {Same as Fig.~\ref{goodsvi} but for the $V(F606W)-Ks_{ISAAC}$
   colors (V passband from ACS-HST and $Ks_{ISAAC}$ passband from VLT-ISAAC).}
\label{goodsvk}
\end{center}
\end{figure}

In Figs.~\ref{goodsvi}, \ref{goodsvz}, and \ref{goodsvk}, the
($V(F606W)-i(F775W)$), ($V(F606W)-z(F850LP)$), and
($V(F606W)-Ks_{ISAAC}$) colors of this selected sample are shown
together with the theoretical predictions. The same remark on the
redshift spanned by the model $\Lambda CDM$ made above applies also
here. For all of the colors, the agreement between data and theory
concerning the photometric, cosmological evolution with redshift is
remarkably good. Furthermore, the theoretical colors seem to match
the GOODS data much better than COSMOS values.

This clearly shows that the reliability of the morphology classifier
plays an important role in these matters. The selection in GOODS is
made by correlating two catalogues, one containing photometric and
spectroscopic redshifts, the other listing galaxies with good
morphological classification (made by hand). In contrast, in COSMOS
classification and selection are fully automated with no any
cross-correlation with other criteria.

Another possible explanation for the differences between the results
from COSMOS and GOODS, could be an intrinsic difference in the
samples of observed galaxies. First, the COSMOS database has a lower
magnitude limit that makes the morphological classification more
difficult. Second, two different packages are used to calculate the
redshift with a mean dispersion of 0.0315. Finally, there are some
problems with the using the SExtractor \citep{Bertin96} star/galaxy
separator parameter that is not stable with a variable seeing. All
this goes in favor of the GOODS database of ETGs.

\section{Surface Photometry of ETGs}
\label{chapim}

Surface photometry is one of the most powerful tools to study the
properties and history of ETGs. The analysis is based on fitting
ellipses to the isophotes of a galaxy. The oldest fit of the radial
intensity profile is the empirical {\it de Vaucouleurs law}
\citep{deVaucouleurs48} that has now been extended by the more
flexible {\it S\'ersic profile} \citep{Sersic68}. The derived
properties, of which the most important ones are the intensity
distribution, the radial ellipticity, and the position angle
profiles, provide basic information such as effective radius,
deviations from the ellipses, isophote twisting, triaxiality, and
absolute magnitudes. The associated higher order Fourier
coefficients from the fits reveal the intrinsic "boxy" or "disky"
appearance of the isophotes which can be used to uncover the
underlying stellar components.

In this section we will show how, starting from a 3-D numerical
simulation of a galaxy, we can recover "artificial" images projected
on a plane, from which photometric and structural parameters, such as
luminosity, magnitudes, colors, and effective radius, can be
calculated. These images are analyzed as if they were realistic images
of galaxies taken with a telescope. In this way we can derive
morphological and structural parameters of the models that can be
compared with those of real ETGs. In particular we derive the Kormendy
relation (the projection of the Fundamental Plane on the
luminosity-radius plane). Finally, we make use of the SDSS photometric
system.

For the aims of this analysis, we did not include the effects of
extinction. This can be justified as follows. The artificial images we
are using refer to the present epoch, when SFR has dropped by orders
of magnitudes with respect to the past and the galaxies are nearly
passively evolving. Furthermore, as a consequence of galactic winds,
all interstellar gas and dust have been expelled from the central
regions of galaxies. The issue has been discussed in some detail by
\citet{Galletta07} who made use of the same galaxy models and to whom
we refer. In brief, to investigate the spectro-photometric evolution
of the NB-TSPH models by \citet{Merlin06,Merlin07}, \citet{Galletta07}
studied the relative spatial distribution of stars and gas and found
that while most of the stars are located inside about half viral
radius, most of the gas falls outside. This implies that effects of
dust (extinction) become important only in the very outer regions of
ETGs. Of course, a completely different situation is expected to occur
at high redshifts, when the evolutionary stage reached by a galaxy is
such that most of gas is likely to fall in the innermost regions
whereby star formation is still active. In such a case, the effects of
extinctions cannot be neglected. In any case, work is in progress to
include extinction in the SEDs of simulations as observed along a
given line of sight, by considering the geometrical structure of the
galaxies.

\subsection{Artificial Images}

Starting from the 3-D NB-TSPH simulations, we construct 2-D images
projecting the volume with $|z| < 100$ kpc onto the $xy$ plane. The
information we need are the spatial coordinates of the star-particles
and their SEDs. With these projections on a plane we construct
discrete grids at the nodes of which we calculate the total flux given
by all the particles along the line of sight. These fluxes feed the
photometric code to compute the magnitude and/or color in a given
band for each mesh point of the grid. The flux of each star-particle
of given age and metallicity is the one of the associated
SSPs. Therefore there is full consistency between the SEDs of the star
particles and those required by the photometric code in use.

To create good artificial images, the  grids must be large enough to
encompass all the star-particles of the model galaxy, and must
contain a large enough number of grid points to eventually get
smooth images. A scale length of 80-100 kpc is long enough to
include all the star-particles of our simulations and a grid with
100-200 points along each direction is good enough to secure smooth
images; the best combination to obtain satisfactory results is a
scale of 0.4 to 0.8 kpc per grid point and direction.
Table~\ref{pixmesh} gives the resolution under which images have
been created: the choice of the scale depends on the model galaxy
and its diameter.

\begin{table}
\begin{center}
\caption[Grid values]{Construction of artificial images: dimensions and grid scales}
\label{pixmesh}
\vspace{1mm}
\begin{tabular}{r c c c c c}   
\hline \multicolumn{1}{c}{} & \multicolumn{1}{c}{40 kpc} &
\multicolumn{1}{c}{60 kpc} & \multicolumn{1}{c}{80 kpc} &
\multicolumn{1}{c}{100 kpc} &
\multicolumn{1}{c}{200 kpc} \\
\hline
50{\it gp}  & 0.8 {\footnotesize $\frac{kpc}{gp}$} & 1.2 & 1.6 & 2.0 & 4.0 \\
100{\it gp} & 0.4 & 0.6 & 0.8 & 1.0 & 2.0 \\
200{\it gp} & 0.2 & 0.3 & 0.4 & 0.5 & 1.0 \\
500{\it gp} & 0.08& 0.12& 0.16& 0.2 & 0.4 \\
\hline
\end{tabular}
\end{center}
\end{table}

\begin{figure*}
\begin{center}
{\includegraphics[width=5cm,height=5cm]{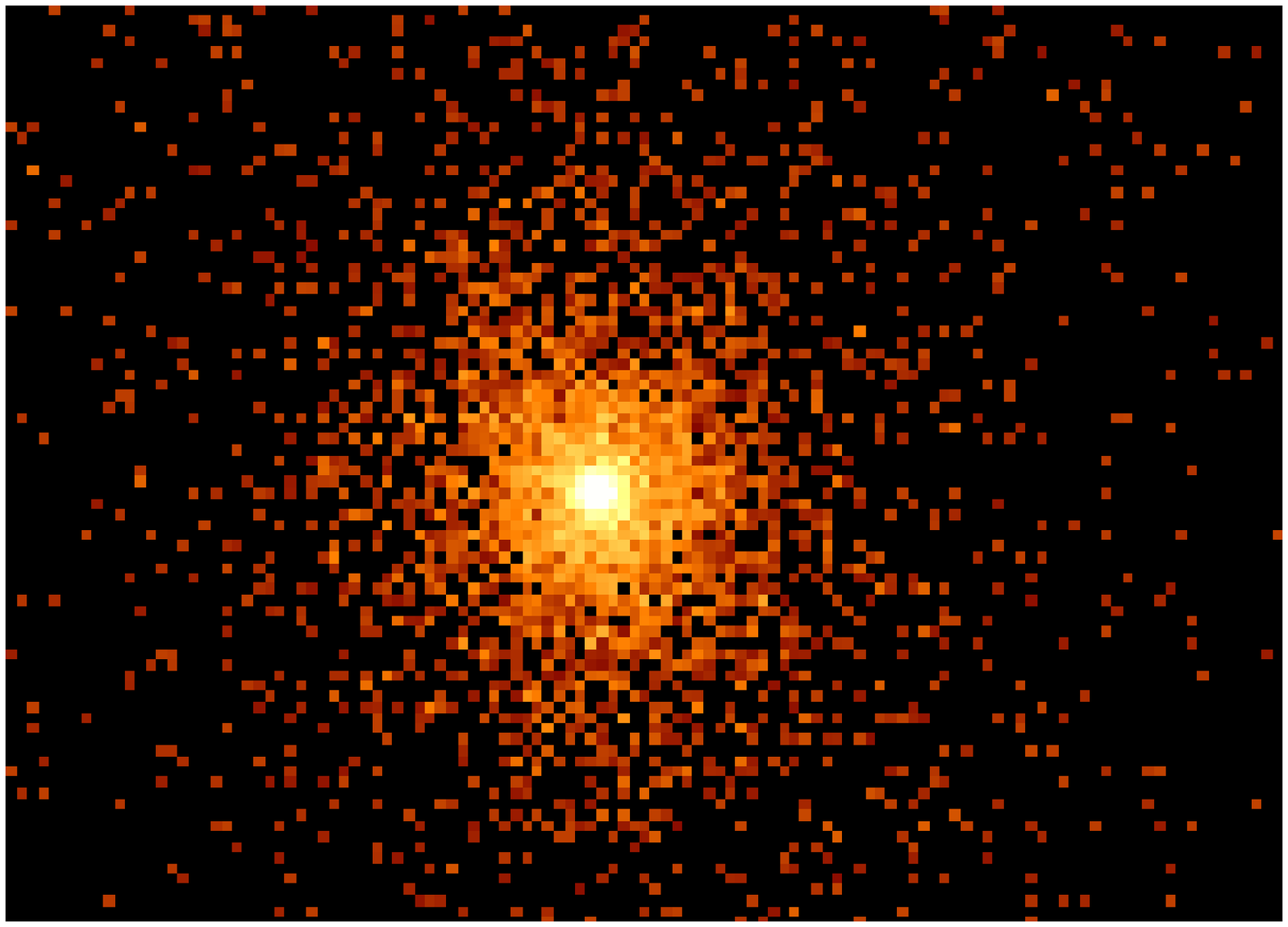}
\includegraphics[width=5cm,height=5cm]{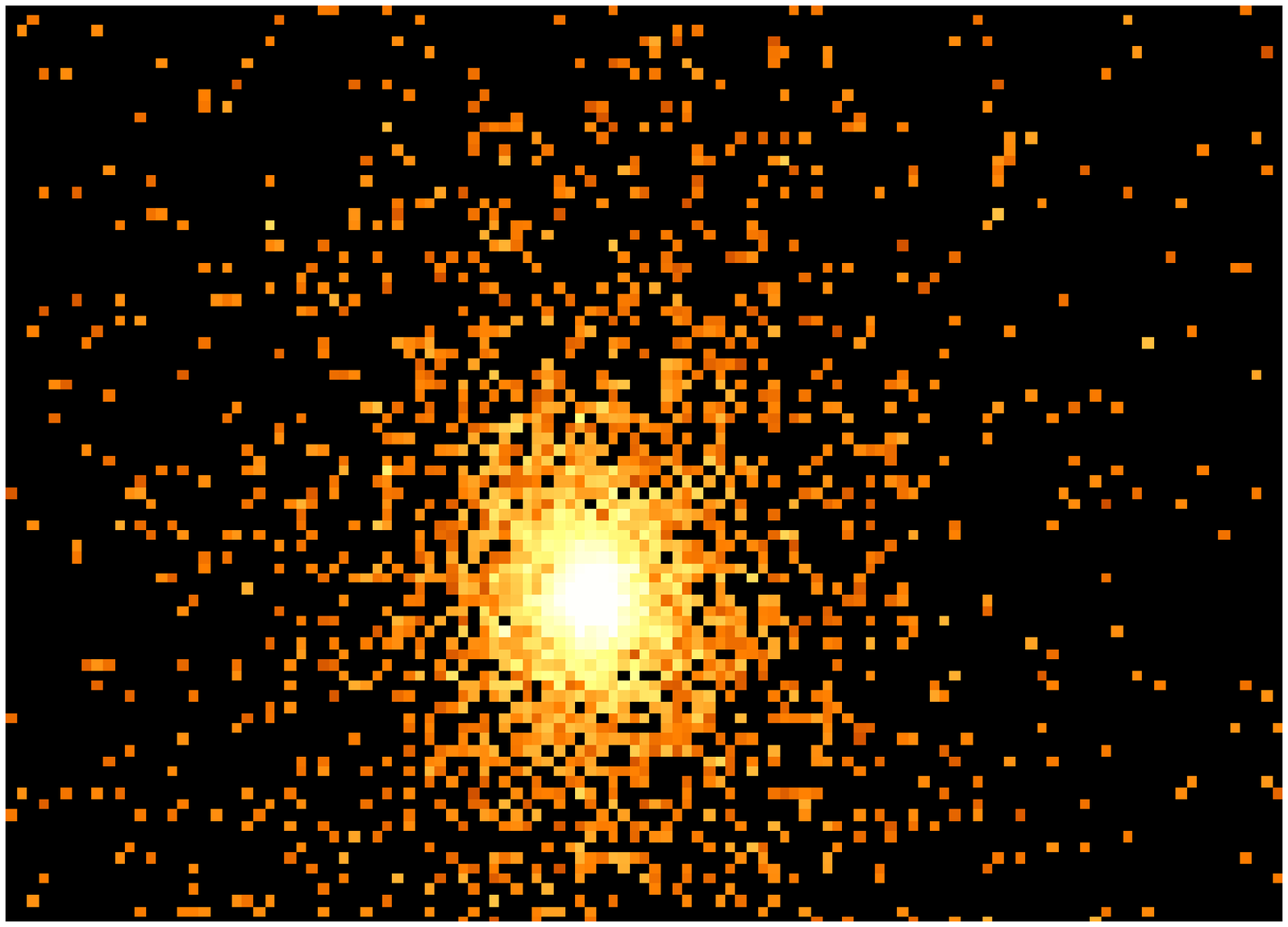}
\includegraphics[width=5cm,height=5cm]{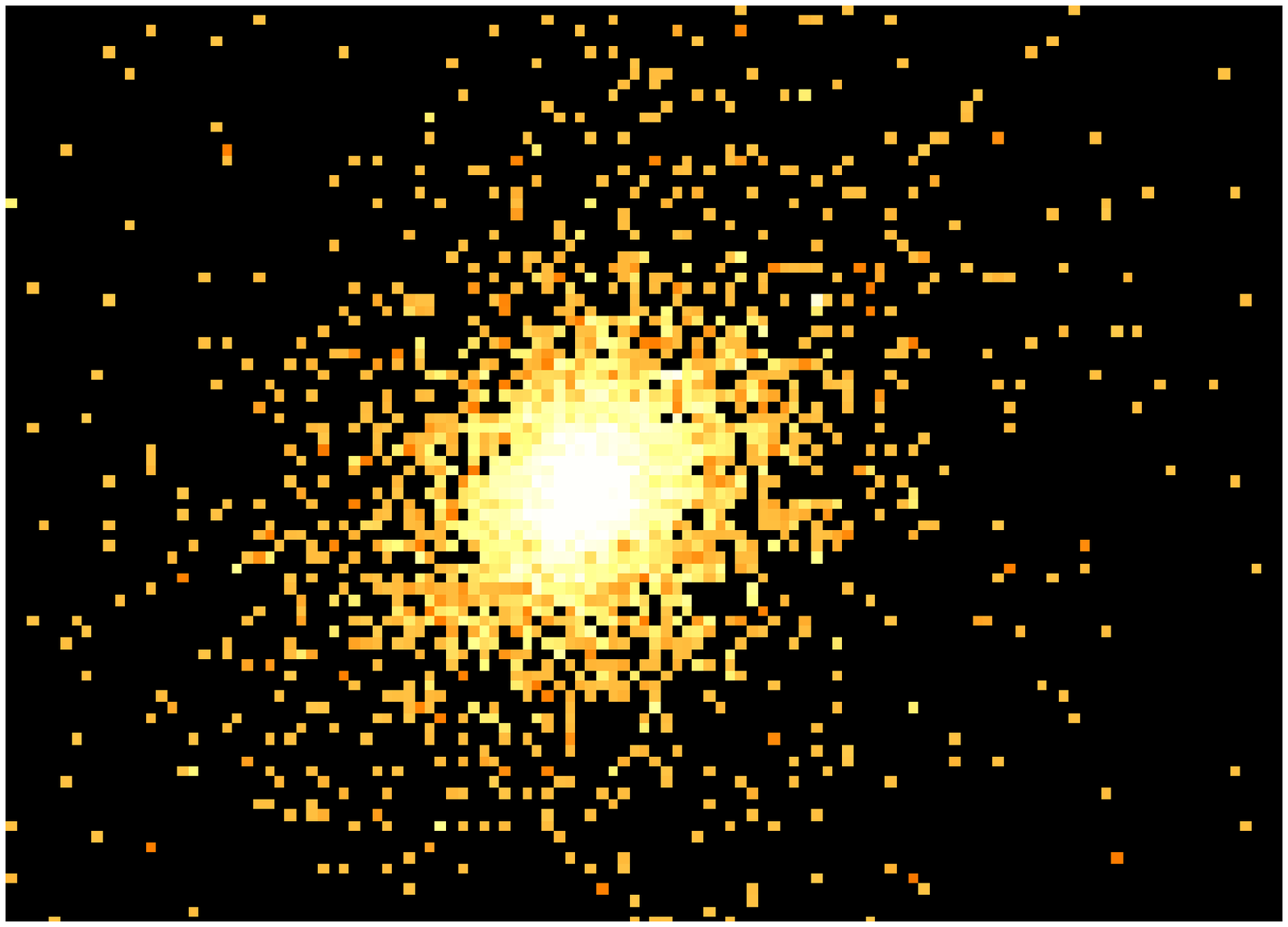}}
\includegraphics[width=5cm,height=1cm]{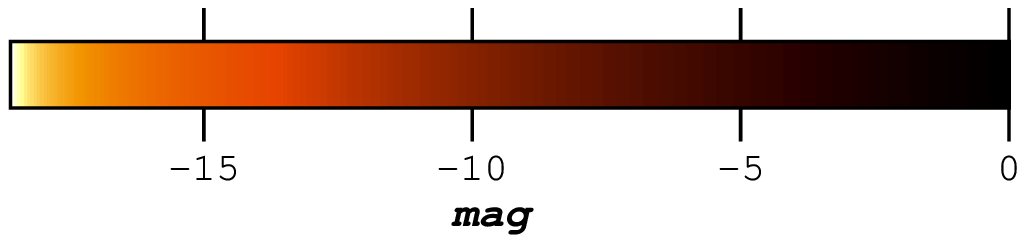}
\caption[Artificial images in the $r$-band]
  {$r$-band artificial image of a 80$\times$80 kpc$^{2}$ region with a
  200$\times$200 mesh points for the last age of the evolution of the
  model galaxies: $SCDM$ left, $\Lambda CDM$ middle, $\Lambda
  CDM_{mp}$ right.}
\label{80200b}
\end{center}
\end{figure*}

Images are built up for the last computed stage of the model galaxies
at the ages listed in Table~\ref{endpar}. Figure~\ref{80200b} shows
the 2-D distribution of the star-particles of the models as traced by
the star-particle $r$-magnitudes of the SDSS system. A grid of
200$\times$200 mesh points is used to span a square region of 80 kpc
side. In all the models, the old star-particles appear to be spatially
distributed mimicking the morphology of an elliptical galaxy. The left
panel refers to the $SCDM$ model, the middle to the $\Lambda CDM$, and
the right to the $\Lambda CDM_{mp}$. In displaying these frames the
intensity contrast between the bright central regions and the low
surface brightness of the outer parts is very difficult to portray
using a linear relation, so that the logarithmic scale is more
convenient. It is worth noting that the inclination of the major axis
of the projected distribution with respect to the coordinate axes
simply reflect the spatial distribution of the star-particles in the
3-D space and in the 2-D projection. No other physical meaning in
involved.

\subsection{Isophotal Analysis}

The images used in this section are similar to those of
Fig.~\ref{80200b} but are based on a grid of 100$\times$100 mesh
points (enough to describe a square region of 80 kpc side). These
images are analysed with the IRAF packages {\it Stsdas.Isophot}
(designed to deal with elliptical isophotes) and {\it Ellipse}
(designed to derive the brightness distribution of a galaxy).

{\it Stsdas.Isophot} (using all the data of the galaxy image) extracts
the intensity levels, which are nicely fitted by ellipses, and
determines the center, the ellipticity, the position angle, and the
length of the semi-major axis. However, small but significant
deviations from pure ellipses are also present. For each isophote,
these deviations are measured along the contour and are recovered by
the Fourier analysis on the distance of the ellipsoidal contour as a
function of the azimuthal angle.

{\it Ellipse} fits the elliptical isophotes and measures the isophotal
parameters of a galaxy's image in whatsoever passband (SDSS in our
case).  The parameters are determined for each isophote. The algorithm
follows the method described by \citet{Jedrze87}. In brief, assumed a
provisional guess for an isophote's center coordinates ($X_c, Y_c$),
ellipticity ($\epsilon$) and semi-major axis position angle ($\phi$),
at each point along the semi-major axis, the intensity $I(\phi)$ of
the image is azimuthally sampled along an elliptical path. $I(\phi)$
is then expanded into a Fourier series as

\begin{equation}
I(\phi) = I_0 + \sum_k[A_{k}{\rm sin}(k\phi) + B_{k}{\rm cos}(k\phi)].
\end{equation}

\noindent A best fit procedure (minimization of the sum of the
squares of the residuals between the real distribution of data and an
elliptical one) fixes the parameters $X_c$, $Y_c$, $\epsilon$, and $
\theta$. The expansion is truncated at the first two moments that are
sufficient to describe an ellipse. The intensity profile as a function
of the position angle is then fitted by the weighted least-squares

\begin{small}
\begin{equation}
I = I_{0} + A_{1}{\rm sin}(\phi) + B_{1}{\rm cos}(\phi) +
A_{2}{\rm sin}(2\phi) + B_{2}{\rm cos}(2\phi)
\end{equation}
\end{small}

\noindent
The amplitudes $A_{1}$, $B_{1}$, $A_{2}$, $B_{2}$ give information on
how much the estimated intensity profile deviates from the real one,
so they give the errors in the fitting procedure. The image data is
fitted by the function

\begin{small}
\begin{equation}
I = I_{0} + A_{3}{\rm sin}(3\phi) + B_{3}{\rm cos}(3\phi) +
A_{4}{\rm sin}(4\phi) + B_{4}{\rm cos}(4\phi)
\label{a4b4}
\end{equation}
\end{small}

\noindent
where the amplitudes $A_{3}$, $B_{3}$, $A_{4}$, $B_{4}$ measure the
isophote's deviations from perfect ellipticity.

Higher order moments ($k \geq 3$) define deviations of the isophotes
from ellipses. In practice, moments beyond the fourth cannot be
measured accurately; third and fourth-order moments are calculated
from the equation above by fixing the first and second-order moments
to their best-fit values.  The third-order moments ($A_{3}$ and
$B_{3}$) represent isophotes with three-fold deviations from ellipses
(e.g., egg-shaped or heart-shaped) while the fourth-order moments
($A_{4}$ and $B_{4}$) represent four-fold deviations.  Rhomboidal or
diamond-shaped isophotes translate into a non-zero $A_{4}$. For
galaxies which are not distorted by interactions, $B_{4}$ is the most
meaningful moment: a positive $B_{4}$ indicates "disky" isophotes
(i.e., with semi-major axis $B_{4}\times$100 percent longer than the
best fitting ellipse), whereas a negative $B_{4}$ indicates "boxy"
isophotes \citep[i.e., with semi-major axis $B_{4}\times$100 percent
shorter than the best fitting ellipse][]{Jedrze87}.

When running {\it Ellipse} on model galaxies, we adopt the following
guidelines. The length of the semi-major axis is increased on a
logarithmic scale (10\% longer in our case). With a new semi-major
axis, a new isophote is calculated using the best-fit parameters
obtained from the previous isophote. In general, all the parameters
can vary freely, even though the routine may fail to
converge on isophotes with large deviations from ellipses. In such
cases to achieve convergence it is necessary to fix the value of one
or more parameters. The isophotal center is not let wander around by
more than 2 grid points between consecutive isophotes; in practice the
center is found to be rather stable.

Given the isophotal center and the semi-major axis, at any position
$a$ along the semi-major axis we determine the radius, $r_{gm}$, of a
circle corresponding to the local ellipse as

\begin{equation}
r_{geo} = a \sqrt{1-\epsilon (a)}
\label{rgeo}
\end{equation}

\noindent
where $\epsilon(a)$ is the ellipticity at the position $a$. This is
equivalent to take the geometric mean of the local semi-major and
semi-minor axes.  In this way all the profiles are reduced to
equivalent circular profiles and the integrated intensity is measured
within this equivalent circle. With this procedure we recover, as a
function of the semi-major axis length, the radial profiles of the
mean isophotal intensity, ellipticity, position angle, local radial
intensity gradient, mean isophotal magnitude, $3^{rd}$ and $4^{th}$
order deviations from ellipse. The errors on intensity, magnitude and
gradient are obtained from the rms scatter of the intensity along the
fitted ellipse. The errors on the geometrical parameters are obtained
from the internal errors in the harmonic fit, after removal of the
first and second fitted harmonics. Errors on the harmonic amplitudes
are obtained from the fit error after removal of all the harmonics up
to including the one being considered.

\begin{figure}
\begin{center}
\includegraphics[width=0.4\textwidth]{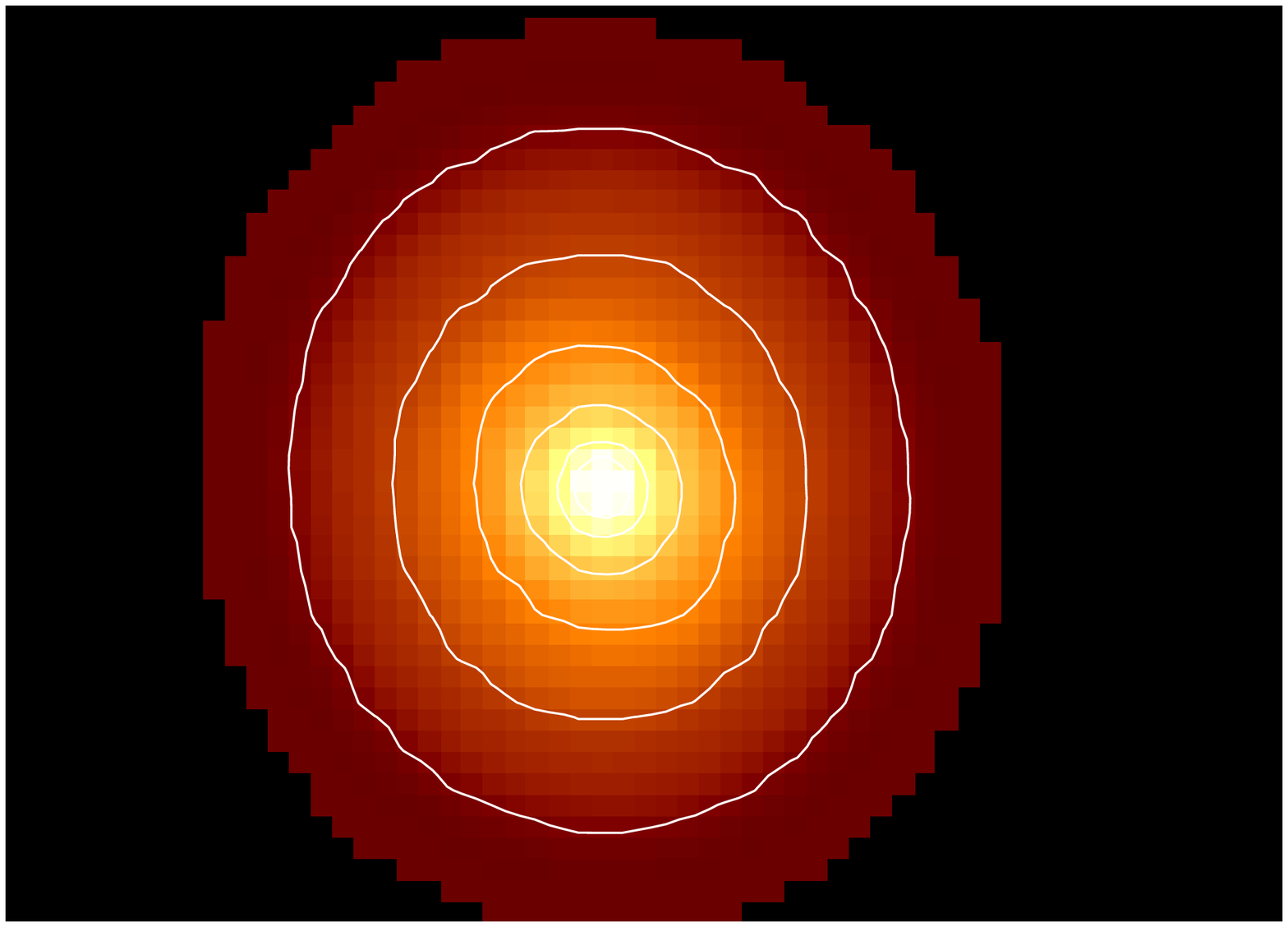}
\includegraphics[width=0.4\textwidth]{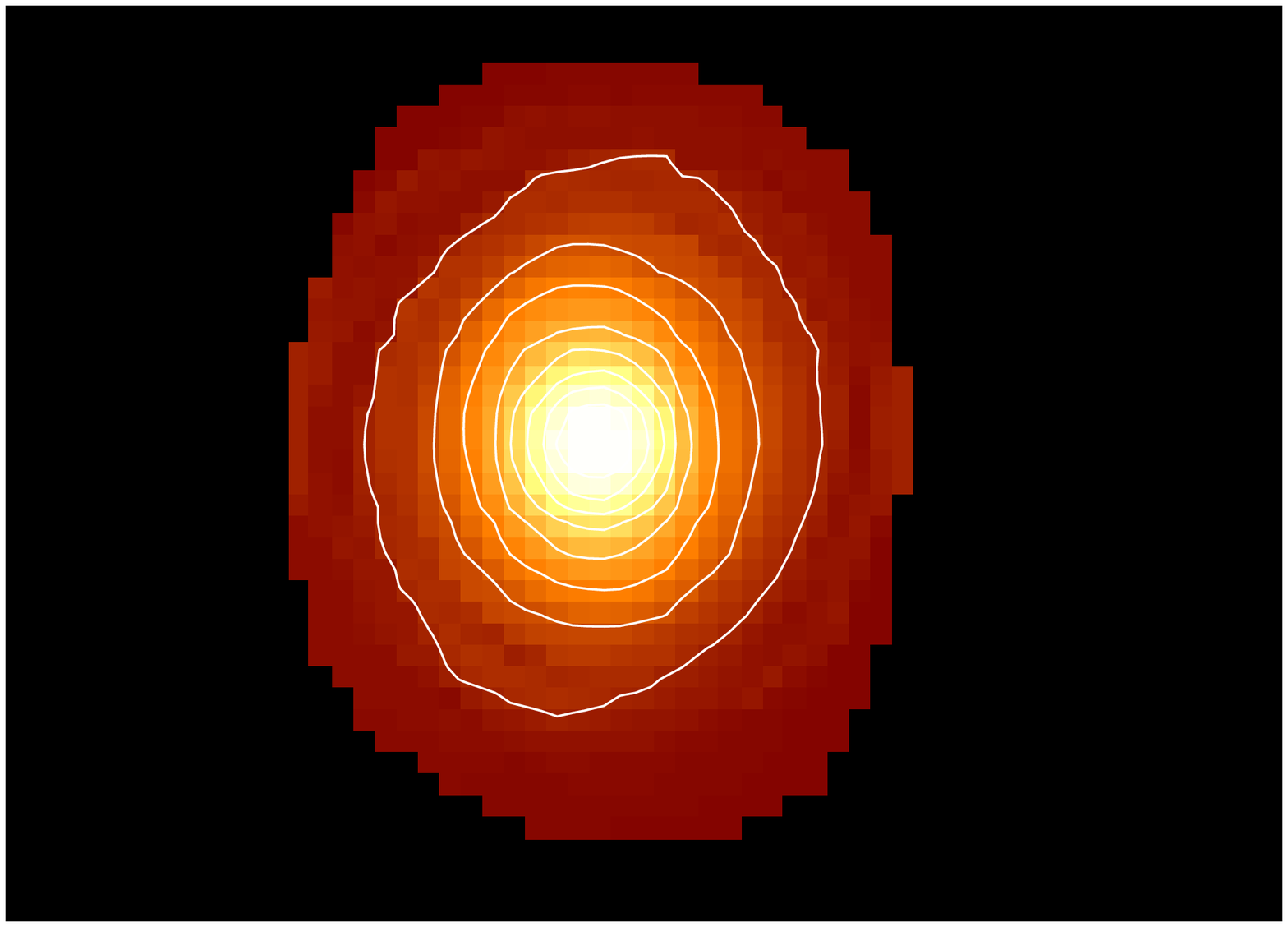}
\includegraphics[width=0.4\textwidth]{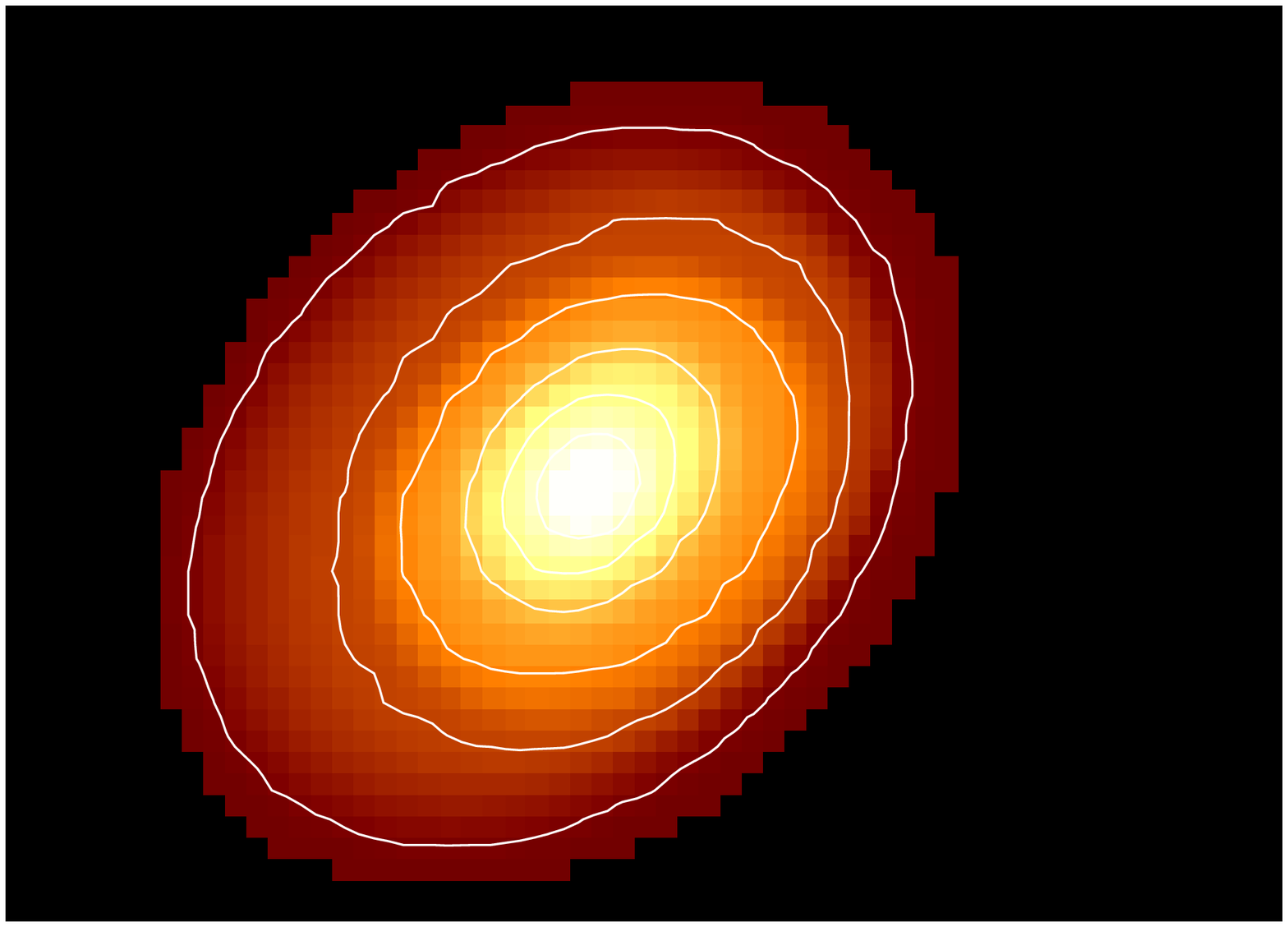}
\caption[Models for the artificial images in the $r$-band and isophotes]
   {Two-dimensional models of the optical $r$-band magnitude in the
   xy-plane of the images shown in Fig.~\ref{80200b}. Superposed are
   elliptical isophotes (white solid lines). The top panel is for $SCDM$,
   the middle panel for  $\Lambda CDM$, and the bottom panel for $\Lambda
  CDM_{mp}$.}
\label{modsimcont}
\end{center}
\end{figure}

After using the {\it ellipse} task to measure the mean radial
intensity profiles and fit ellipses to the image, the program {\it
Bimodel} is used to reconstruct a model image from the results of
isophotal analysis. {\it Bimodel} creates a 2-D smooth image of the
source image.  In Fig.~\ref{modsimcont} are shown the smooth images of
the three simulations corresponding to the discrete images of
Fig.~\ref{80200b} with the elliptical isophotes overlapped. The same
remark made for the inclination of the semi-major axis with respect to
the X-axis in Fig.~\ref{80200b} applies also here.

The $3^{rd}$ and $4^{th}$ harmonics from the photometry are added to
the model. This option is most useful when working close to the
central intensity peak. As explained in \citet{Jedrze87}, the sampling
at small radii may introduce a "boxy" component. More accurate
modelling of the central region is generally achieved when including
this component.

\begin{figure}
\begin{center}
\includegraphics[width=0.5\textwidth]{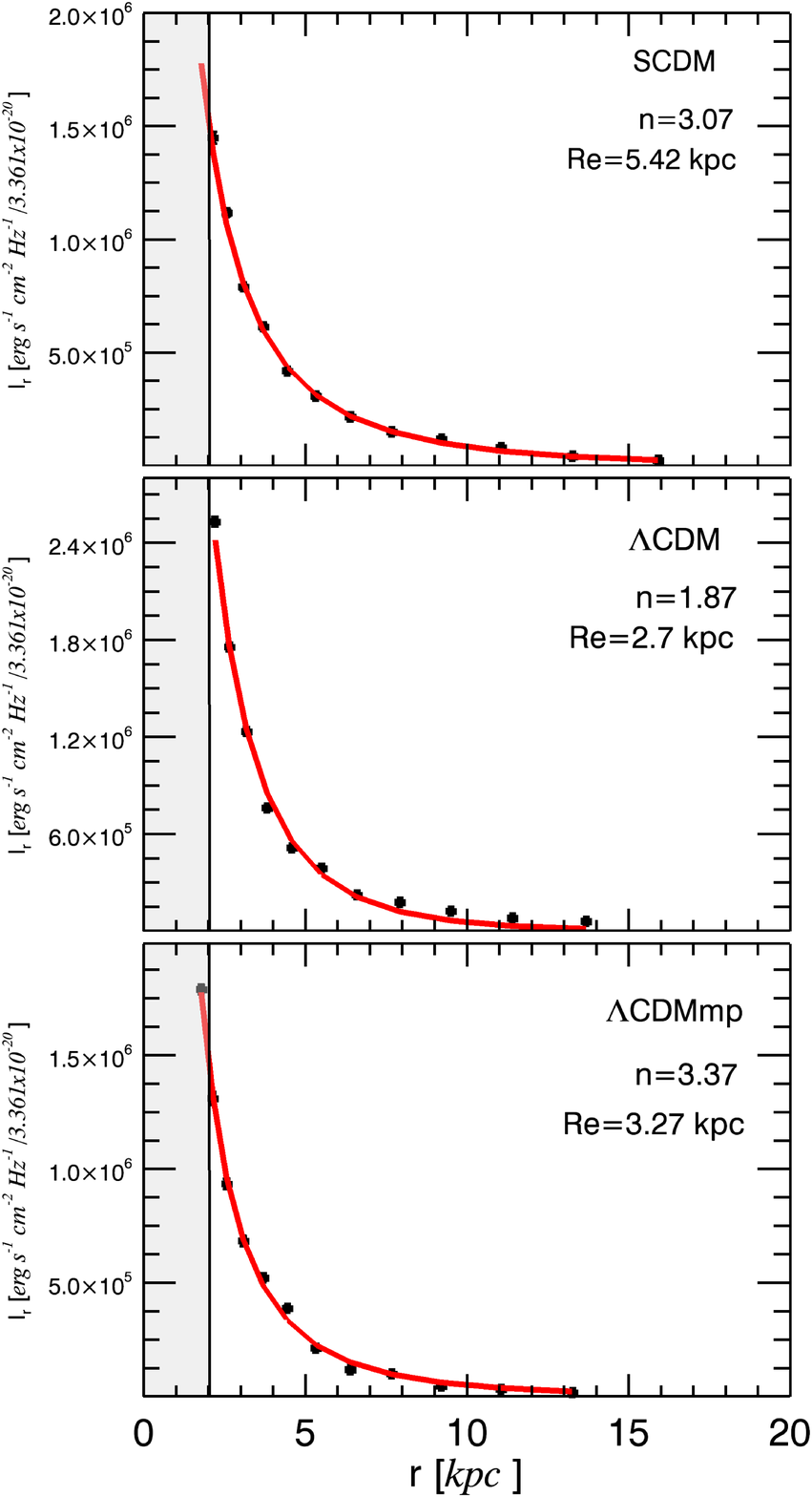}
\caption[Intensity profiles] {Intensity profiles fitted with a S\'ersic law of
   variable index. The $r$-band S\'ersic index $n$ and the effective
   radius $R_e$ are specified in the panel for each models. We
   exclude the central region within $R \sim 1-2$ kpc (shaded area),
   because the intensity profile is smeared out by the gravitational
   softening technique.}
\label{serfit}
\end{center}
\end{figure}

\subsection{Parametrization of the Intensity Profile}
\label{secintprof}

The intensity $I(R)$ and corresponding surface brightness $\mu(R)$ are
given by the S\'ersic laws

\begin{equation}
I(R) = I_e\exp \left\{ -b_{n} \biggl[ \biggl( \frac{R}{R_{e}}
\biggl)^{1/n} - 1 \biggr] \right\},
\end{equation}

\begin{equation}
\mu(R) = \mu_{e} - 2.5b_{n} \biggl[ \biggl( \frac{R}{R_{e}}
\biggl)^{1/n} - 1 \biggl] \log e,
 \end{equation}

\noindent
where $r$ is the radius from the center in kpc, $R_{e}$ is the
effective radius, and $I_{e}$ and $\mu_{e}$ are the effective
intensity and effective surface brightness, respectively, within
$R_{e}$. $b_n$ is a positive parameter that, for a given $n$, can be
determined from the definition of $R_{e}$ and $\mu_{e}$.

For a S\'ersic model with 1 $\lesssim n \lesssim$ 10, the effective
radius $R_{e}$ contains roughly half the integrated light if
$b_{n}=2n-0.324$ \citep{Trujillo01}. The parameter $n$ controls the
overall shape of a S\'ersic profile, with low $n$ values producing
curved profiles with logarithmic slopes which are shallow in the inner
regions, and steep in the outer parts, while high $n$ values produce
extended profiles with less overall curvature.

By definition, the parameters are chosen in such a way that the model
reduces to the de Vaucouleurs law for $n=4$, and the consistency with
the usual interpretation of $R_{e}$ as the radius enclosing half-light
is secured.

The S\'ersic model offers significant advantages. First and foremost,
it provides a good description of the inner (100 $pc$ scale) profiles,
and a significantly better description when the profiles are extended
to the kpc-scale \citep{Graham03,Trujillo04}.

The intensity profile is derived using the {\it nfit1d} algorithm
which provides 1-D, non-linear fits to the image (chi-square
minimization). As {\it nfit1d} supports any analytical fitting
function, given the initial guesses for the function coefficients, we
have obtained the intensity profile of the model galaxy with a
S\'ersic law with suitable index $n$, together with the effective
radius $R_{e}$ and the effective intensity $I_{e}$.
Figure~\ref{serfit} shows the best fitting S\'ersic function for the
$r$-band in the particular sample. The profile inside the central
region whose radius is equal to 1-2 times the softening length
(1-2\,kpc in our case), is excluded from the fit as it is likely to be
influenced by force softening. To derive the effective radius, we
reach large enough distances such as 15-20\,kpc. The solid curve
represents the best-fit S\'ersic model to the final profile (shown as
solid symbols).

From the fitted intensity profiles and the best-fit parameters, we
calculate the effective radii $R_e$ and the total magnitudes in all
bands for the three galaxy models. The S\'ersic index and $R_e$ are
listed in Table~\ref{valnr} for the various passbands.

\begin{table}
\begin{center}
\caption[S\'ersic indexes and effective radii]
{S\'ersic indexes and effective radii expressed in kpc for the SDSS
photometric passbands and for all model galaxies.}
\label{valnr}
\vspace{1mm}
\begin{tabular*}{70mm}{c| c c| c c| c c}
\hline
\multicolumn{1}{c|}{} &
\multicolumn{2}{c|}{$SCDM$} &
\multicolumn{2}{c|}{$\Lambda CDM$} &
\multicolumn{2}{c}{$\Lambda CDM_{mp}$} \\
\hline
\multicolumn{1}{c|}{band} &
\multicolumn{1}{c}{$n$} &
\multicolumn{1}{c|}{$R_{e}$} &
\multicolumn{1}{c}{$n$} &
\multicolumn{1}{c|}{$R_{e}$} &
\multicolumn{1}{c}{$n$} &
\multicolumn{1}{c}{$R_{e}$} \\
\hline
$u$ & 3.22 & 6.42 &  2.2  & 2.38 & 3.97 & 4.31 \\
$g$ & 3.08 & 5.78 &  2.1  & 2.43 & 3.69 & 3.67 \\
$r$ & 3.07 & 5.42 &  1.87 & 2.7  & 3.37 & 3.27 \\
$i$ & 3.86 & 5.83 &  1.99 & 2.73 & 3.14 & 3.07 \\
$z$ & 4.42 & 5.98 &  2.21 & 2.6  & 2.94 & 2.95 \\
\hline
\end{tabular*}
\end{center}
\end{table}

The effective radius varies with the band from which is derived,
roughly decreasing as the band moves toward the shorter wavelengths.
\citet{Bernardi03a} finds that half-light angular sizes of the
galaxies in their sample are indeed larger in the blue bands and
they show how the effective physical radii in their sample changes
in the four\footnote{They exclude the $u$-band as the measurements
are affected by errors larger than the others} bands. This trend is
followed by our simulations: the agreement is particularly good for
the $\Lambda CDM_{mp}$ model. For the other two, the  trend is
recovered only in the bluer bands, whereas the radius increases
again as the band shifts toward longer wavelengths.

\subsection{Structural Properties}

\begin{figure*}
\begin{center}
{\includegraphics[width=0.30\textwidth]{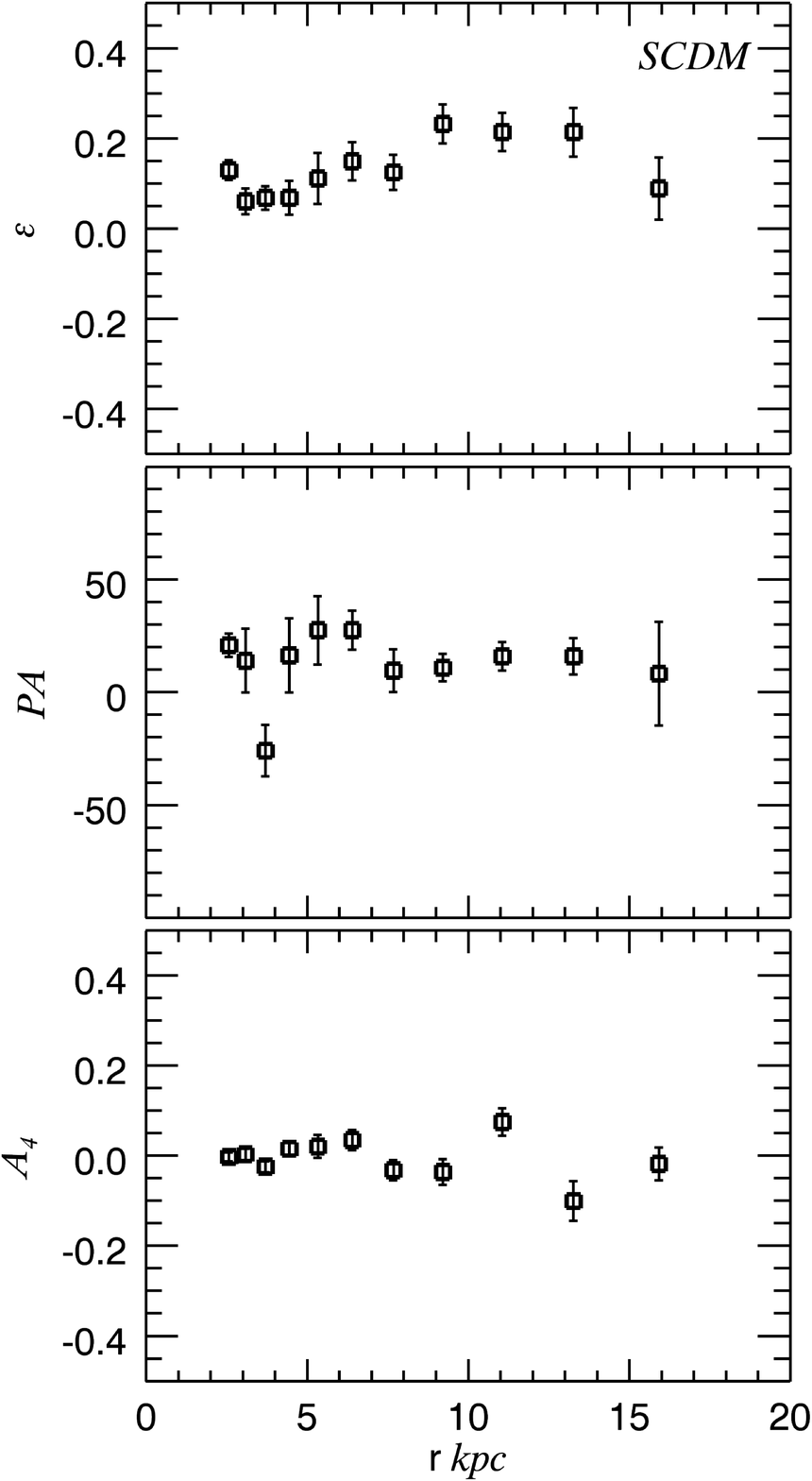}
\includegraphics[width=0.30\textwidth]{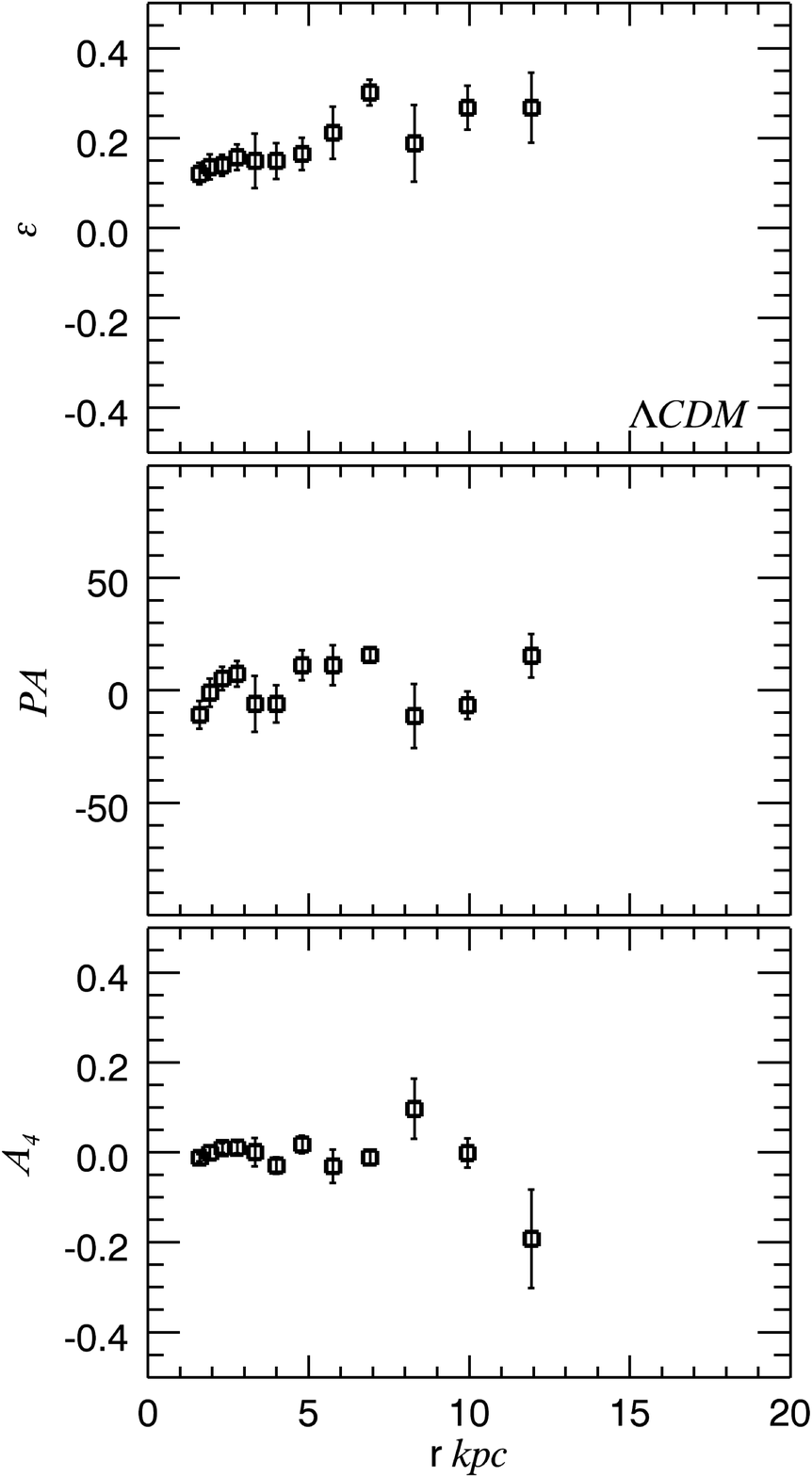}
\includegraphics[width=0.30\textwidth]{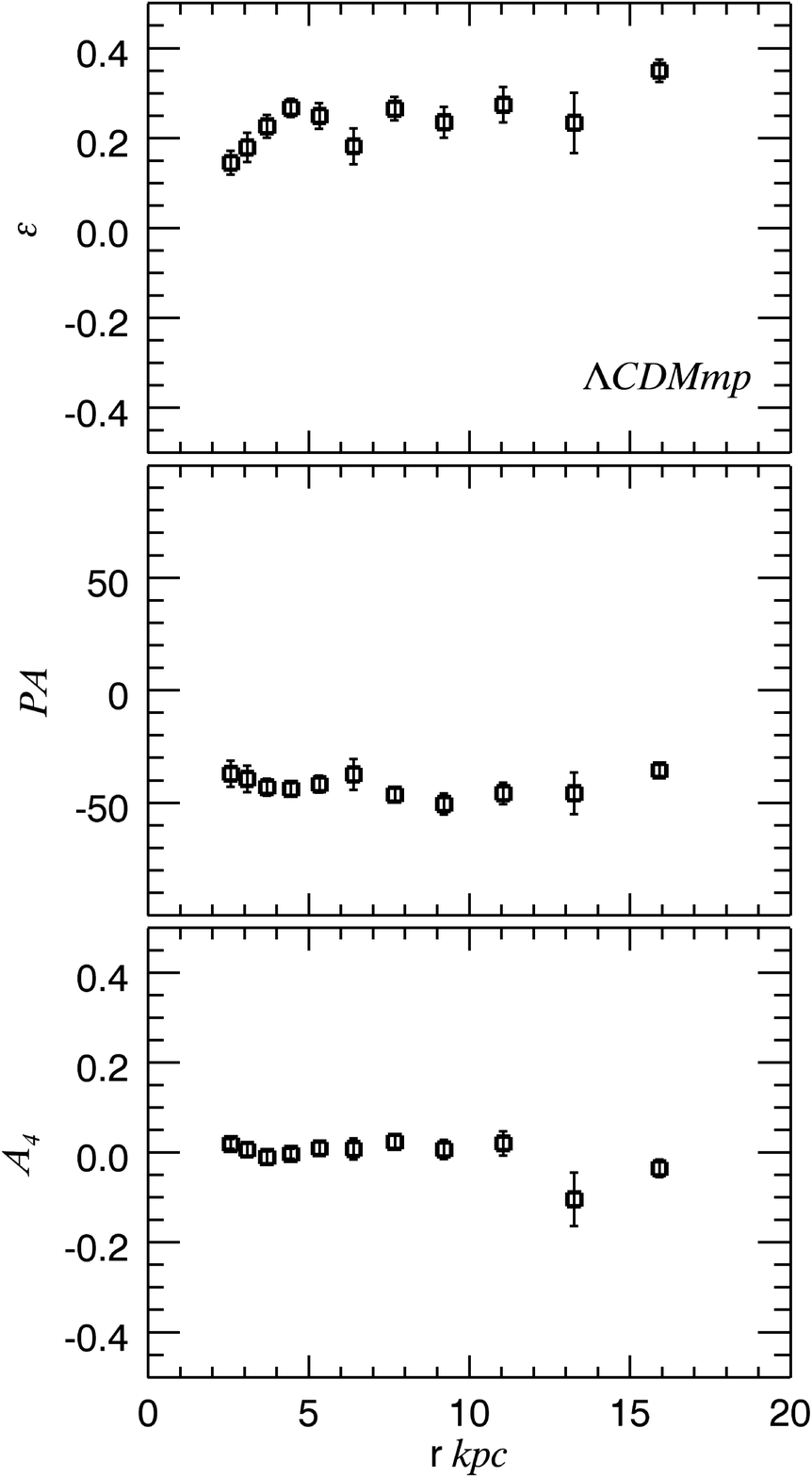}}
\caption[Morphological profiles for the $SCDM$ model]{Ellipticity $\epsilon$,
   position angle $PA$, and $A_{4}$ radial profiles
   for the three models as indicated.}
\label{ellprof1pr}
\end{center}
\end{figure*}

In the various panels of Fig.~\ref{ellprof1pr} we show the basic
structural parameters of the three models. In brief, the ellipticity
$\epsilon$ (top panels), position angle $PA$ (middle panels), and
deviation $A_4$ of the isophotes from pure ellipses (bottom panels)
are shown as a function of the "geometric mean" radius $r_{gm}$, as
defined in eqn.~\ref{rgeo} for all the galaxy models as indicated. To
the sake of illustration, we show the results limited to the $r$-band
image.

Although the isophotes shown in Fig.~\ref{modsimcont} are well
approximated by ellipses, small but significant deviations from
perfect ellipsoidal shapes are measured.  Of particular interest is
the $A_{4}$ parameter\footnote{$A_{4}$ corresponds to the fourth
order moment $B_{4}$ of eqn.~\ref{a4b4}}, which measures the
deviations from perfect ellipses: $A_{4} < 0$ corresponds to "boxy"
isophotes, whereas $A_{4} > 0$ implies "disky" isophotes.

The radial profile of the $A_{4}$ parameter indicates that the
deviations from perfect ellipses are generally negligibly small for
the $SCDM$ and $\Lambda CDM_{mp}$ models. In the case of the $\Lambda
CDM$ model, the negative values of $A_{4}$ in the outer regions tell
us that there is a "boxy" structure.

\subsection{Color Profiles}

As mentioned before, \citet{Bernardi03a} found that half-light angular
sizes of the galaxies change in function of the band in which they are
measured, finding larger radii with bluer bands. On average, the fact
that ETGs have this trend implies that they present color
gradients. It is known from observations that these gradients are such
that ETGs are redder in their cores and bluer in the outskirts. This
fact is thought to originate from variations in age or metallicity of
the underlying stellar populations \citep{Worthey94,Tantalo98a}.

\begin{figure*}
\begin{center}
\centerline{
\includegraphics[width=0.30\textwidth]{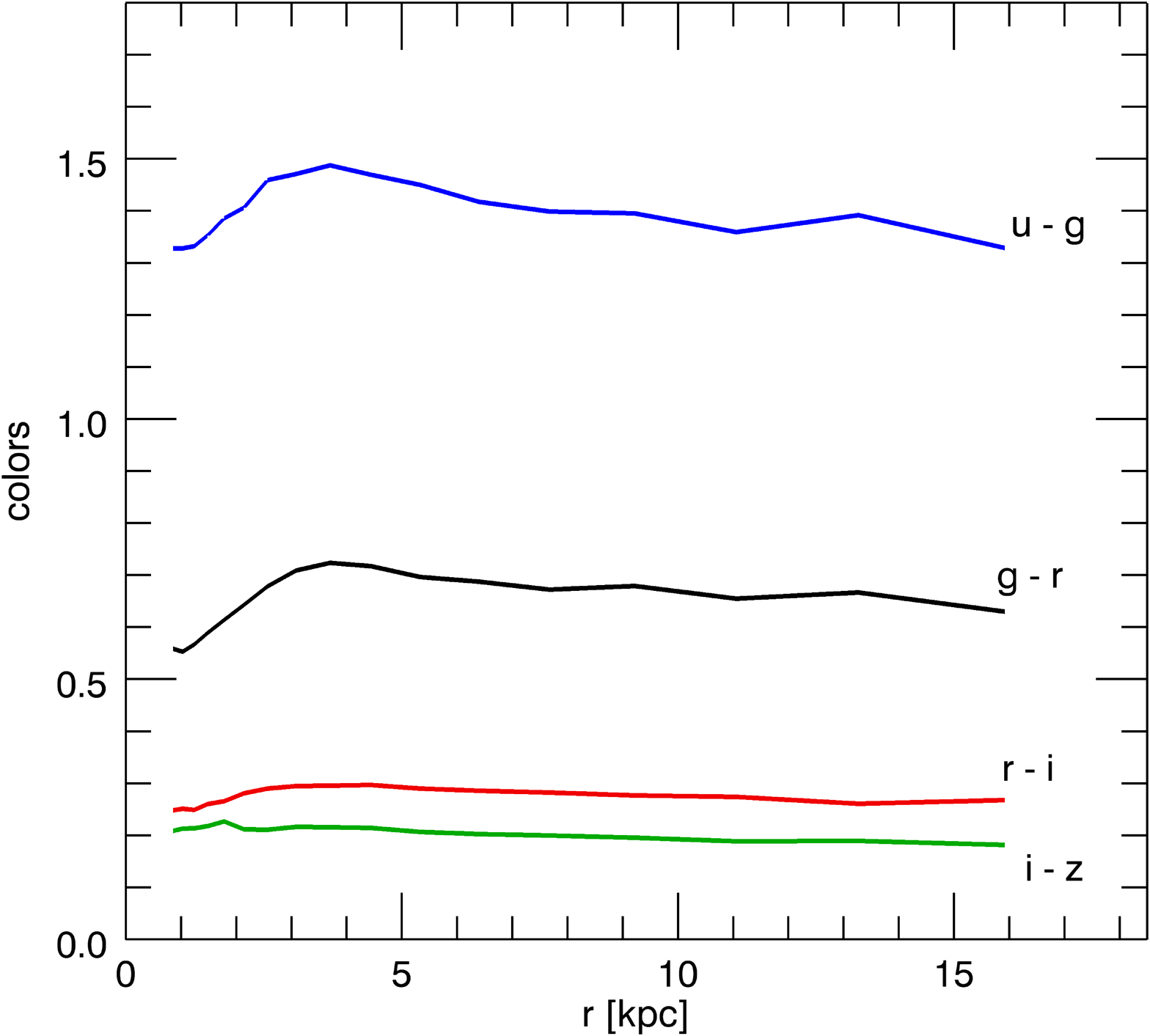}
\includegraphics[width=0.30\textwidth]{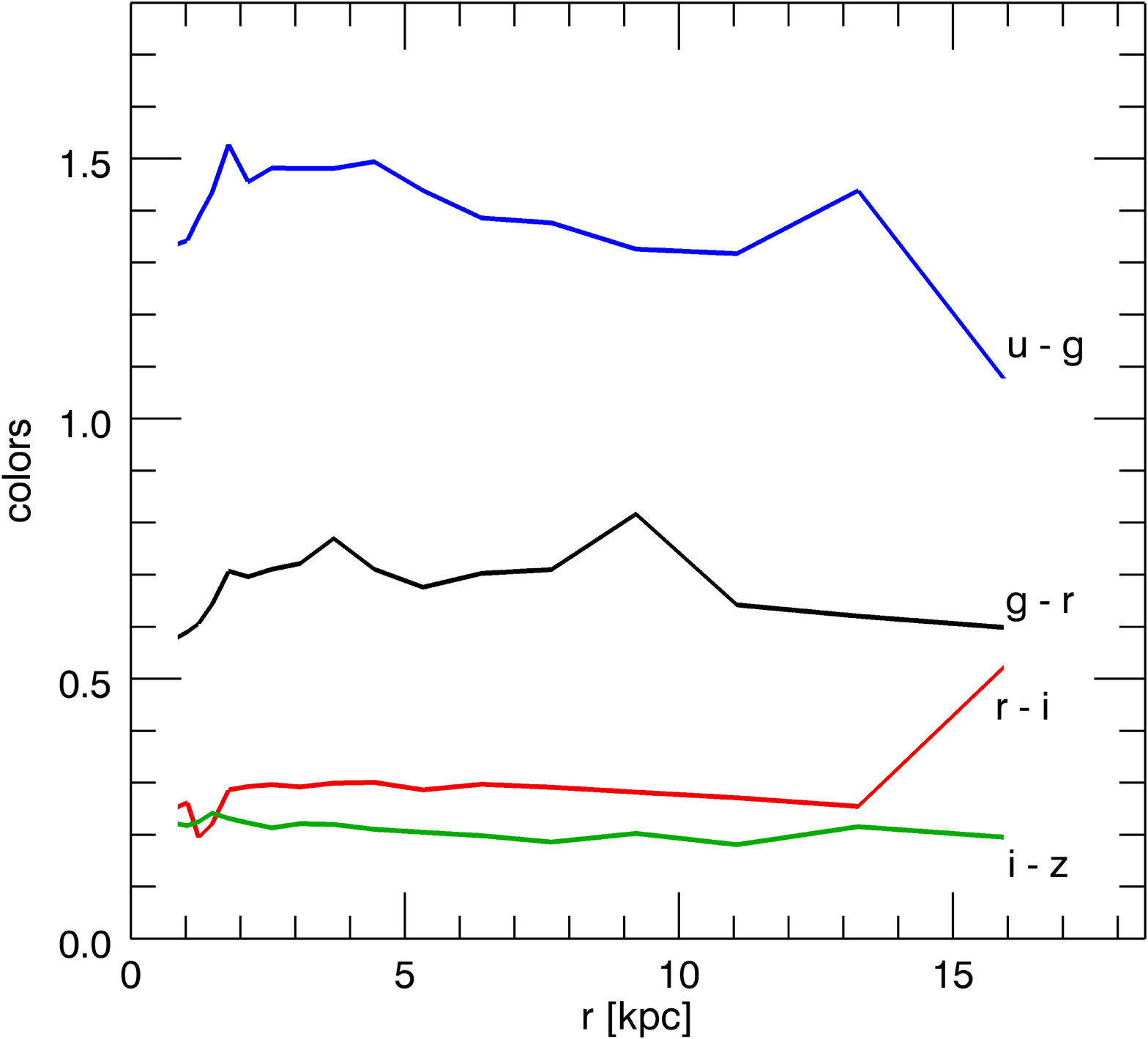}
\includegraphics[width=0.30\textwidth]{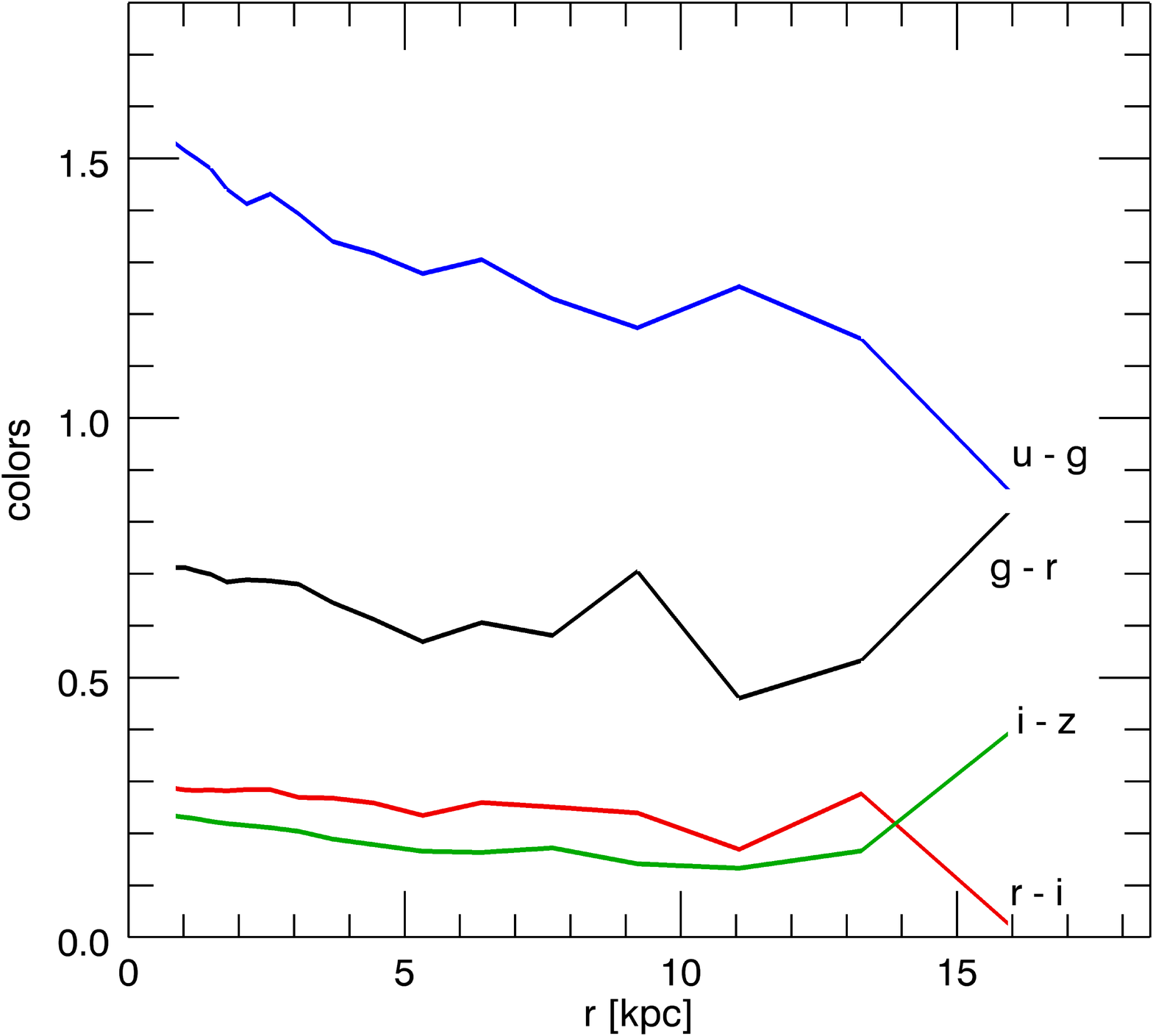}}
\caption[Color profiles for the $SCDM$, $\Lambda CDM$,
   and $\Lambda CDM_{mp}$ model]
   {{\bf Left Panel}: radial profiles of the $SDSS$ colors for the
   $SCDM$ model derived from the simulated image 100$\times$100 grid
   points shown in Fig.~\ref{80200b}. {\bf Middle Panel}: the same as
   in the left panel but for the $\Lambda CDM$ model. {\bf Right
   Panel}: the same as in the previous panel but for the $\Lambda
   CDM_{mp}$ model.}
\label{radgradcolscdm_lcdm}
\end{center}
\end{figure*}

The radial color profiles of our models are shown in the three panels
of Fig.~\ref{radgradcolscdm_lcdm} which display the $u-g$, $g-r$,
$r-i$, and $i-z$ color profiles in the SDSS photometric system. With
the physical processes we have considered, colors are bluer in the
central cores and the color gradients are very small for the $SCDM$
and $\Lambda CDM$ models (left and middle panels of
Figs.~\ref{radgradcolscdm_lcdm}). In contrast, the central regions are
redder and the color gradients are significant for the $\Lambda
CDM_{mp}$ model (right panel of Fig.~\ref{radgradcolscdm_lcdm}).
Therefore, the typical color gradients of ETGs are not strictly
reproduced by our models, except for the $\Lambda CDM_{mp}$ case.

The main reason for the central regions being bluer than expected and
observed resides in the prolonged stellar activity present mainly in
the central regions of all the models together with insufficient
increase of the mean metallicity in those regions. To cast light on
the effect of the long tail of star formation, we have artificially
stopped star formation at different epochs. The details of these
experiments are not shown here for the sake of brevity. As expected,
stopping star formations immediately yields redder colors in the
center, for instance by about 0.1 mag in (B-V). In addition, we have
applied the cut in star formation to star-particles enclosed within a
given galacto-centric distance going from 50 kpc to $R_e/2$. The
result is that the effect on the color gets stronger moving
outward. In all cases, the color difference amounts to 0.1 to 0.2 mag
in the typical (B-V). What we learn from this is that further
investigation of other star formation prescriptions and effects of
feedback with better resolution in the central regions is required.

\section{Scaling Laws: the Kormendy Relation}
\label{scalrel}

As long known, ETGs are similar in their structural and dynamical
properties and obey to empirically relationships among colors,
luminosities, half-light radii, surface brightness profiles, and
velocity dispersions that are ultimately related to their stellar
content and dynamics. They are known as the \emph{Scaling Laws}, among
which particularly important are the {\it Fundamental Plane} and the
{\it Kormendy Relation}.

\noindent
\textsf{ Fundamental Plane (FP)}. Long ago \citet{Djorgovski87} and
\citet{Dressler87} found that

\begin{equation}
\log R_{e} = \alpha \log \sigma_{0} + \beta <\mu>_{e} + \gamma
\end{equation}

\noindent
where $R_{e}$ is the effective radius, $\sigma_{0}$ the central
velocity dispersion, $<\mu>_{e}$ the mean surface brightness within
the effective radius ($<\mu>_{e} = -2.5 \log <I_{e}> + const$), and
$I_{e}$ is the effective luminosity. The coefficients change with the
passband in use. Typical values for the Johnson $B$-band are: $\alpha
= 1.25$, $\beta = 0.32$, and $\gamma = -8.895$ \citep{Bender98}. A
remarkable property is the variation of the coefficient $\alpha$ in
the different filter passbands, and the almost constant value of
$\beta$ \citep{Bernardi03c}.

The existence of the FP has strong implications on galaxy formation
and evolution theories. The small scatter in the FP \citep[see
e.g.][]{Jorgensen96} and its apparent lack of evolution with redshift
\citep{Franx95,VanDokkum96,Bender97,Ellis97,VanDokkum00}, the
homogeneity \citep{Bower92b}, and the evidence for short ($<$1\,Gyr)
star formation timescales in these systems, all indicate that the bulk
of stellar population in ETGs indeed formed at high redshift
($z>2$). Measuring the three parameters entering the FP for ETGs at
varying redshift highlights some important questions concerning their
age, formation history, and internal properties. One can answer
questions such as how far in the past does the FP apply and whether
its parameters evolved significantly with time. A study of the galaxy
properties as a function of look-back time provides a good probe of
the possible evolutionary differences. In this sense, we plan to study
this scaling law at intermediate redshift by deriving its key
parameters as function of time, i.e. considering the evolution with
redshift of the ETGs.

In order to compare our models with real galaxies on the FP, we need
the theoretical central velocity dispersion, $\sigma_0$.
Observationally, this  is evaluated within an aperture typically
less than about half the effective radius. Unfortunately, velocities
within such small radii in the simulations are significantly
affected by the softening parameter  that in our case amounts to
1\,kpc. As accurate predictions for $\sigma_0$ are not possible with
our models,  we have to leave aside the analysis of the FP.

\noindent
\textsf{Kormendy Relation (KR)}. Better chances are possible with this relation.
The KR is the projection of the FP onto the luminosity-radius
plane. It relates $<\mu>_{e}$ to $R_{e}$. Once the dependence on $I_e$
is made explicit we get

\begin{equation}
R_{e} \propto I_{e}^{-0.83}.
\end{equation}

\noindent
Many studies have confirmed that the luminous ETGs in clusters
approximately follow the relation $<\mu>_e = \log R_e + const$ found
by Kormendy with slope of $\sim 3$ and intrinsic scatter of $0.3-0.4$.

The KR for galaxies in clusters at increasing redshift has been
claimed to be consistent with passively evolving stellar populations
\citep{Bower92a,Aragon93,Bender96,VanDokkum96,Jorgensen97,Ziegler97,Bender98,VanDokkum98}.
On the other hand, some studies have also claimed that the data are
consistent with the hierarchical evolutionary scenario
\citep{White78}. \citet{LaBarbera03}, working with cluster ETGs  at
different redshift, found that the slope of the KR  is almost
invariant up to $z \sim 0.64$ with value of $\sim 2.91 \pm 0.08$.
The homogeneity and the invariance with redshift of these
distributions is also suggested by the analysis of the SDSS data by
\citet{Bernardi03b}.

To compare our models with observational data, we have chosen a
sub-sample of ETGs from the Sloan Digital Sky Survey
\citep[SDSS;][]{York00,Stoughton02} database. The SDSS survey has
mapped one-quarter of the entire sky, producing a detailed
image of it and determining the positions and photometric properties
of more than 100 million celestial objects. The SDSS obtained
high-resolution images in five different bands, namely $u$, $g$,
$r$, $i$, and $z$ \citep{Fukugita96}, thus allowing for a reliable
identification of ETGs and precise measurements of their photometric
properties.

Galaxies can be selected using automated pipelines that isolate
objects on the basis of their 2-D light distributions. We consider
ETGs from the DR2 release selected following the criteria described in
\citet{Bernardi03a} who have produced a catalogue of $\sim 9000$
low-redshift ETGs, selected using a combination of SDSS pipeline
parameters. This catalogue contains galaxies with a high $i$-band
concentration index $(r_{50}/r_{90})>2.5$ and in which a
\citet{deVaucouleurs48} fit to the surface brightness profile is
significantly more likely than an exponential fit. Details of the
selection can be found in \citet{Bernardi03a}.

\begin{figure}
\begin{center}
\includegraphics[width=8.0cm,height=8.0cm]{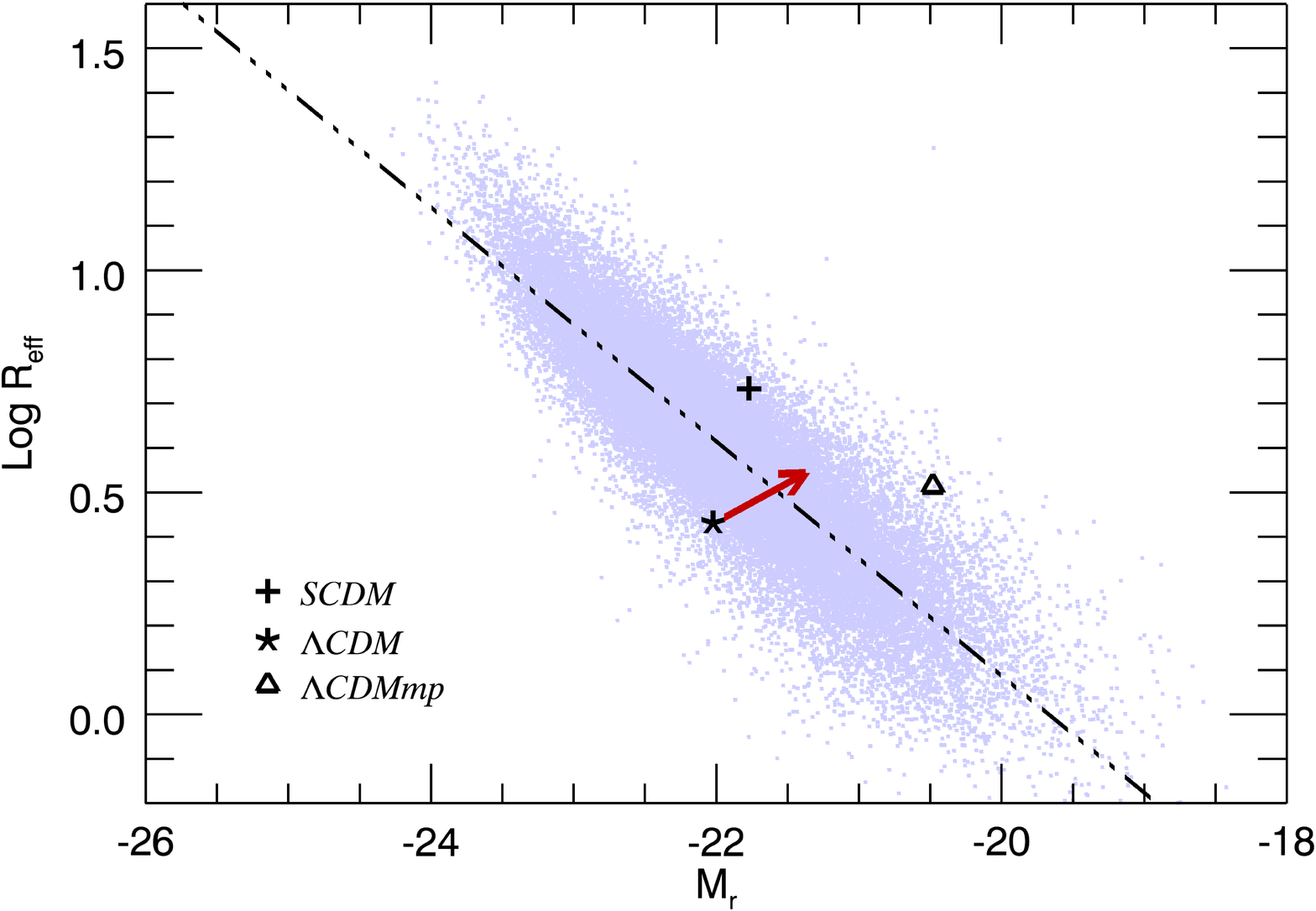}
\caption[Magnitude-radius relation]{Magnitude-radius relation in
   the $r$-band for the ETGs selected from the SDSS DR2 database
   following the criteria from \citet{Bernardi03a}. Results for our
   three model galaxies are shown for comparison. See the text for all
   details.}
\label{korm}
\end{center}
\end{figure}

In Fig.~\ref{korm} we show the $M_{r}$-$R_{e}$ relation found by
\citet{Bernardi03c} for the sample of elliptical galaxies selected
as mentioned above. The  luminosity-size relation for $r$-band is
$R_e \propto I^{-0.75 \pm 0.01}$. For comparison, we show the
results for our three model galaxies. The  $R_{e}$ and  $M_{r}$
magnitude within it of the models are: $R_{e}$=5.42 kpc and
$M_{r}$=--21.77 for the $SCDM$ model; $R_{e}$=2.7 kpc and
$M_{r}$=--22.02 for the $\Lambda CDM$ case; $R_{e}$=3.27 and
$M_{r}$=--20.48 for the $\Lambda CDM_{mp}$ galaxy.

The $SCDM$ and $\Lambda CDM_{mp}$ lie above the mean relation but fall
in the data crowd. The $\Lambda CDM$ model falls below the mean line
but still compatible with data. No case lies close to the mean
line. There are several reasons to account for the marginal
discrepancy. The discussion is slightly different for the $SCDM$ and
$\Lambda CDM_{mp}$ models evolved up to the age of 13\,Gyr and the
$\Lambda CDM$ stopped at 7\,Gyr.

To bring the position of the $SCDM$ and $\Lambda CDM_{mp}$ models down
to the mean line one should increase the total star mass and $R_{e}$ by
a factor of 1.5 or so. The solution is viable in the sense that other
models of the same type with better tuned parameters could reach the
agreement. The same reasoning cannot be applied to the model $\Lambda
CDM$ because its evolution terminated at 7\,Gyr ($z \sim 1$). Looking
at the fading lines (magnitudes versus age) of Fig.~\ref{magevsb} a
shift of the $M_{r}$ magnitude of about +1 mag in 6\,Gyr (to get the age
of 13\,Gyr) is possible without changing the star mass, thus bringing
the model onto the mean line without changing $R_{e}$. Most likely, an
increase of $R_{e}$ of the same entity as in the previous cases is also
possible.  The arrow in Fig.~\ref{korm} shows the expected shift. Also
in this case, agreement can be easily achieved. To conclude, all the
three models are marginally consistent with the data. No better
comparison is possible at this stage.

\section{General Discussion and Concluding Remarks}
\label{concl}

We have presented a package of numerical codes to compute the
spectroscopic and photometric properties of model galaxies by
combining the evolutionary population synthesis technique with the 3-D
geometrical structure of the galaxy and its history of star formation
and chemical enrichment, provided by NB-TSPH simulations. The tool is
very flexible in the way input libraries of evolutionary tracks,
isochrones, SSPs and important physical laws such as the initial mass
function, star formation rate, and metallicity enrichment can be
changed, tested, and added to the database for future use. Finally, it
can be adapted to any photometric system currently in use. The method
has been tested so far on three models of ETGs evolved within
different cosmological backgrounds and the analysis has been done in
the following photometric systems: Bessell-Brett, SDSS, COSMOS, and
GOODS filters among all the ones we have at disposal. Our main results
may be summarized as follows:

\textsf{1.} The application of the  tool to three model galaxies at
our disposal, allow us to compute the SED, magnitudes, and colors as
function of time and redshift, together with the evolutionary and
cosmological corrections. With the aid of it, we analyzed samples of
ETGs taken from the COSMOS and GOODS databases, and made a qualitative
and quantitative comparison between theoretical results and
observational data. For both data-sets, we find that the simulated
colors for the different cosmological scenarios follow the general
trend over the entire range of redshift considered and are in good
agreement with the data up to $z \sim 1$, above which the number of
observed ETGs falls abruptly.  In conclusion, within the redshift
range considered, all the simulated colors reproduce quite well the
observational data.

\textsf{2.} We have also generated synthetic 2-D images of the
galaxy models in a given photometric system. These synthetic images
can be processed with the same algorithms used to analyze real images
to derive the structural and morphological parameters, e.g. the
galaxy's $R_e$ and the luminosity within this, the shape indices
through Fourier and S\'ersic analysis, the color profiles, and the
radial profiles of most of the parameters that define the structure of
galaxies. We find that the luminosity profiles of the model galaxies
at $z = 0$ can be reasonably fitted with a S\'ersic $R^{1/n}$ law. The
evaluation of the $R_e$ in the photometric bands of SDSS shows the
same dependence on the passband wavelength range of the observational
data.  Furthermore, the isophotes are well approximated by ellipses,
with only a weak radial variation in position angle and
ellipticity. Small but significant deviations from perfect ellipses
are also measured. In general, we can recover properties that resemble
those of observed galaxies.

\textsf{3.} In addition to that, we looked at the Kormendy
relation, one of the Scaling Laws of ETGs, for which all theoretical
counterparts of observational data were available. The theoretical
luminosities and effective radii of the models are consistent with the
archival data from the SDSS for a sample of ETGs.

\noindent From the above results we can conclude that the package
provides good results that permit to reasonable recover the
observational properties of ETGs. Owing to the small number of galaxy
models to our disposal, neither statistical generalization of the
results is possible, nor the underlying cosmological background can be
tested. We have to wait for a more complete library of model galaxies
in different cosmological scenarios to better elucidate the evolution
of photometric properties and scaling laws, and to provide insight
into the meaning of the correlations (or lack thereof) between shape,
kinematics, and photometry of ETGs.

\begin{acknowledgements}
We like to thank Dr. Mariangela Bernardi for kindly providing the
data from the SDSS database.
\end{acknowledgements}

\bibliographystyle{aa}
\bibliography{mnemonic,biblio2009}
\end{document}